\documentclass[10pt,aps,showpacs]{revtex4}
\usepackage{epsfig}
\usepackage{amssymb,amsmath,amsthm,epsfig,graphicx}
\begin{document}

\title{{\Large Fluctuations in Ideal and Interacting Bose-Einstein
Condensates:}\\
From the laser phase transition analogy \\
to squeezed states and Bogoliubov quasiparticles
\footnote{It is a pleasure to dedicate this review to Prof.\ Herbert Walther,
our guide in so many fields of physics. His contributions to atomic, molecular
and optical physics are enlightened by the deep insights he has given us into
the foundations of quantum mechanics, statistical physics, nonlinear dynamics
and much more. }}

\author{Vitaly V. Kocharovsky$^{1,2}$, Vladimir V. Kocharovsky$^2$,
Martin Holthaus$^3$, C. H. Raymond Ooi$^{1}$, Anatoly Svidzinsky$^{1}$,
Wolfgang Ketterle$^{4}$, and Marlan O. Scully$^{1,5}$}

\affiliation{$^{1}${Institute for Quantum Studies and Dept.\ of Physics,
Texas A\&M Univ.,TX 77843-4242}\\
$^{2}${Institute of Applied Physics, Russian Academy of Science,
600950 Nizhny Novgorod, Russia}\\
$^{3}${Institut f\"ur Physik, Carl von Ossietzky Universitat,
D-2611 Oldenburg, Germany}\\
$^{4}${MIT-Harvard Center for Ultracold Atoms, and Dept.\ of Physics,
MIT, Cambridge, Mass. 02139}\\
$^{5}${Princeton Institute for Materials Science and Technology,
Princeton Univ., NJ 08544-1009}}

\date{\today}

\begin{abstract}
We review the phenomenon of equilibrium fluctuations in the number of
condensed atoms $n_0$ in a trap containing $N$ atoms total. We start with a
history of the Bose-Einstein distribution, a similar grand canonical problem
with an indefinite total number of particles, the Einstein-Uhlenbeck debate
concerning the rounding of the mean number of condensed atoms $\bar n_0$
near a critical temperature $T_c$, and a discussion of the relations between
statistics of BEC fluctuations in the grand canonical, canonical, and
microcanonical ensembles.

First, we study BEC fluctuations in the ideal Bose gas in a trap and explain
why the grand canonical description goes very wrong for all moments $\langle
(n_0-\bar n_0)^m\rangle$, except of the mean value. We discuss
different approaches capable of providing approximate analytical results and
physical insight into this very complicated problem. In particular, we
describe at length the master equation and canonical-ensemble quasiparticle
approaches which give the most accurate and physically transparent picture
of the BEC fluctuations. The master equation approach, that perfectly
describes even the mesoscopic effects due to the finite number $N$ of the
atoms in the trap, is quite similar to the quantum theory of the laser. That
is, we calculate a steady-state probability distribution of the number of
condensed atoms $p_{n_0}(t=\infty )$ from a dynamical master equation and
thus get the moments of fluctuations. We present analytical formulas for the
moments of the ground-state occupation fluctuations in the ideal Bose gas in
the harmonic trap and arbitrary power-law traps.

In the last part of the review, we include particle interaction via a
generalized Bogoliubov formalism and describe condensate fluctuations in the
interacting Bose gas. In particular, we show that the canonical-ensemble
quasiparticle approach works very well for the interacting gases and
find analytical formulas for the characteristic function and all cumulants,
i.e., all moments, of the condensate fluctuations. The surprising conclusion
is that in most cases the ground-state occupation fluctuations are
anomalously large and are not Gaussian even in the thermodynamic limit. We
also resolve the Giorgini, Pitaevskii and Stringari (GPS) vs.\ Idziaszek et
al.\ debate on the variance of the condensate fluctuations in the interacting
gas in the thermodynamic limit in favor of GPS. Furthermore, we clarify a
crossover between the ideal-gas and weakly-interacting-gas statistics which
is governed by a pair-correlation, squeezing mechanism and show how, with an
increase of the interaction strength, the fluctuations can now be understood
as being essentially 1/2 that of an ideal Bose gas. We also explain the
crucial fact that the condensate fluctuations are governed by a singular
contribution of the lowest energy quasiparticles. This is a sort of infrared
anomaly which is universal for constrained systems below the critical
temperature of a second order phase transition.
\end{abstract}

\maketitle
\tableofcontents

\section{INTRODUCTION}

Professor Herbert Walther has taught us that good physics unifies and unites
seemingly different fields. Nowhere is this more apparent than in the
current studies of Bose-Einstein condensation (BEC) and coherent atom optics
which draw from and contribute to the general subject of coherence effects
in many-body physics and quantum optics. It is in this spirit that the
present paper presents the recent application of techniques, ideas, and
theorems which have been developed in understanding lasers and squeezed
states to the condensation of $N$~bosons. Highlights of these studies, and
related points of BEC history, are described in the following paragraphs.

1. Bose~\cite{Bose,TheimerRam76} got the ball rolling by deriving the Planck
distribution without using classical electrodynamics, as Planck~\cite{Planck}
and Einstein~\cite{Einstein17} had done. Instead, he took the extreme
photon-as-a-particle point of view, and by regarding these particles as
indistinguishable obtained, among other things, Planck's result,
\begin{equation}
\bar{n}_{\mathbf{k}}=\frac{1}{e^{\beta \varepsilon _{\mathbf{k}}}-1}
\end{equation}
where $\bar{n}_{\mathbf{k}}$ is the mean number of photons with energy $%
\varepsilon _{\mathbf{k}}$ and wavevector $\mathbf{k}$, $\beta
=(k_{B}T)^{-1} $, $T$ is the blackbody temperature, and $k_{B}$ is
Boltzmann's constant.

However, his paper was rejected by the \emph{Philosophical Magazine} and so
he sent it to Einstein, who recognized its value. Einstein translated it into
German and got it published in the \emph{Zeitschrift f\"{u}r
Physik}~\cite{Bose}. He then applied Bose's method to atoms and predicted that
the atoms would ``condense'' into the lowest energy level when the temperature
was low enough~\cite{Einstein,Einstein25,Einstein25b}.

Time has not dealt as kindly with Bose as did Einstein. As is often the case
in the opening of a new field, things were presented and understood
imperfectly at first. Indeed Bose did his ``counting'' of photon states in
cells of phase space in an unorthodox fashion. So much so that the famous Max
Delbr\"{u}ck wrote an interesting article~\cite{Delbruck} in which he
concluded that Bose made a mistake, and only got the Planck distribution by
serendipity. We here discuss this opinion, and retrace the steps that led
Bose to his result. Sure, he enjoyed a measure of luck, but his mathematics
and his derivation were correct.

2. Einstein's treatment of BEC of atoms in a large box showed a cusp in the
number of atoms in the ground state, $\bar{n}_{0}$, as a function of
temperature,
\begin{equation}
\bar{n}_{0}=N\left(1-\left(\frac{T}{T_{c}}\right)^{3/2}\right)
\end{equation}
for $T \le T_{c}$, where $N$ is the total number of atoms, and $T_{c}$ is the
(critical) transition temperature.

Uhlenbeck~\cite{Uhlenbeck} criticized this aspect of Einstein's work,
claiming that the cusp at $T=T_{c}$ is unphysical. Einstein agreed with the
Uhlenbeck criticism but argued that in the limit of large numbers of atoms
(the thermodynamic limit) everything would be okay. Later, Uhlenbeck and his
student Kahn showed~\cite{Kahn} that Einstein was right and put the matter
to rest (for a while).

\begin{figure}[tbp]
\bigskip
\centerline{\epsfxsize=1.05\textwidth\epsfysize=0.5\textwidth
\epsfbox{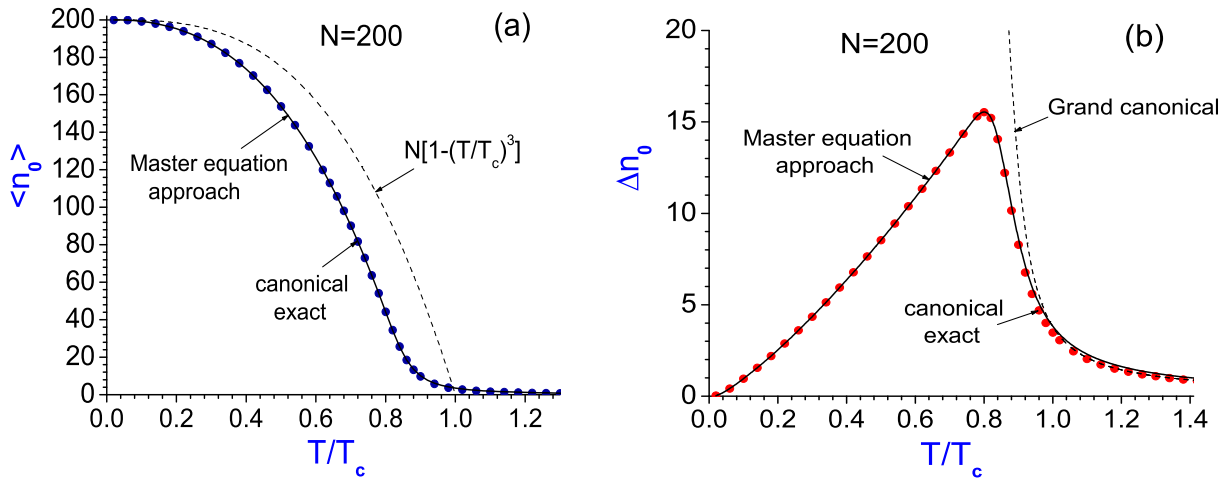}}
\caption{(a) Mean value $\langle n_0 \rangle$ and (b) variance $\Delta n_{0}=\protect\sqrt{\langle n_{0}^{2}\rangle -\langle n_{0}\rangle^{2}}$ of the number of condensed atoms as a function of temperature for $N=200$ atoms in a harmonic trap calculated via the solution of the condensate master equation (solid line). Large dots are the exact numerical results obtained in the canonical ensemble. Dashed line for $\langle n_0 \rangle$ is a plot of
$N[1-(T/T_c)^3]$ which is valid in the thermodynamic limit. Dashed line for $\Delta n_{0}$ is the grand canonical answer $\sqrt{\bar n_0(\bar n_0+1)}$ which gives catastrophically large fluctuations below $T_c$.}
\label{n0n2me}
\end{figure}

Fast forward to the present era of mesoscopic BEC physics with only
thousands (or even hundreds) of atoms in a condensate. What do we now do
with this Uhlenbeck dilemma? As one of us (W.~K.) showed some time ago
\cite{kd}, all that is needed is a better treatment of the problem. Einstein
took the chemical potential to be zero, which is correct for the ideal Bose
gas in the thermodynamic limit. However, when the chemical potential is
treated more carefully, the cusp goes away, as we discuss in detail,\;
see, e.g., Figs.~\ref{n0n2me}a and~\ref{n0}.

3. So far everything we have been talking about concerns the average number
of particles in the condensate. Now we turn to the central focus of this
review: fluctuations in the condensate particle number. As the reader will
recall, Einstein used the fluctuation properties of waves and particles to
great advantage. In particular he noted that in Planck's problem, there were
particle-like fluctuations in photon number in addition to the wave-like
contribution, i.e.
\begin{equation}
\Delta n_{k}^{2}={\bar{n}_{k}^{2}}\text{(wave)}+{\bar{n}_{k}}\text{(particle)%
} ,  \label{waveparticle}
\end{equation}
and in this way he argued for a particle picture of light.

In his studies on Bose-Einstein condensation, he reversed the logic arguing
that the fluctuations in the ideal quantum gas also show both wave-like and
particle-like attributes, just as in the case of photons. It is interesting
that Einstein was led to the wave nature of matter by studying fluctuations.
We note that he knew of and credited de Broglie at this point well before
wave mechanics was developed.

Another important contribution to the problem of BEC fluctuations came
from Fritz London's observation~\cite{London} that the specific heat is
proportional to the variance of a Bose-Einstein condensate and showed a cusp,
which he calculated as being around 3.1K. It is noteworthy that the so-called
lambda point in liquid Helium, marking the transition from normal to
superfluid, takes place at around 2.19K.

However, Ziff, Uhlenbeck and Kac~\cite{Ziff} note several decades later that
there is a problem with the usual treatment of fluctuations. They say:
\begin{quote}
[When] the grand canonical properties for the ideal Bose gas are derived,
it turns out that some of them differ from the corresponding canonical
properties --- even in the bulk limit! \ldots The grand canonical ensemble
\ldots \emph{loses its validity} for the ideal Bose gas in the condensed
region.
\end{quote}

One of us (M.~H.) has noted elsewhere~\cite{HKK} that:
\begin{quote}
This grand canonical fluctuation catastrophe has been discussed by
generations of physicists \ldots
\end{quote}

Let us sharpen the preceding remarks. Large fluctuations are a feature of
the thermal behavior of systems of bosons. If $\bar{n}$ is the mean number
of non-interacting particles occupying a particular one particle state, then
the mean square occupation fluctuation in the grand canonical picture is $%
\bar{n}(\bar{n}+1)$. If, however, the system has a fixed total number of
particles $N$ confined in space by a trapping potential, then at low enough
temperature $T$ when a significant fraction of $N$ are in the ground state,
such large fluctuations are impossible. No matter how large $N$, the grand
canonical description cannot be even approximately true. This seems to be
one of the most important examples that different statistical ensembles give
agreement or disagreement in different regimes of temperatures. To avoid the
catastrophe, the acclaimed statistical physicist D. ter Haar \cite{DterHaar}
proposed that the fluctuations in the condensate particle number in the low
temperature regime (adapted to a harmonic trap) might go as
\begin{equation}
\Delta n_{0}\equiv \sqrt{\langle (n_{0}-\bar{n}_{0})^{2}\rangle }=N-\bar{n}%
_{0}=N\left( \frac{T}{T_{c}}\right) ^{3}.
\end{equation}
This had the correct zero limit as $T\rightarrow 0$, but is not right for
higher temperatures where the leading term actually goes as
$[N\left( \frac{T}{T_{c}}\right) ^{3}]^{1/2}$. The point is that fluctuations
are subtle; even the ideal Bose gas is full of interesting physics in this
regard.

In this paper, we resolve the grand canonical fluctuation catastrophe in
several ways. In particular, recent application of techniques developed in the quantum
theory of the laser~\cite{sl,ssl} and in quantum optics~\cite{Walls,sz}
allow us to formulate a consistent and physically appealing analytical
picture of the condensate fluctuations in the ideal and interacting Bose
gases. Our present understanding of the statistics of the BEC fluctuations
goes far beyond the results that were formulated before the 90's BEC boom, as
summarized by Ziff, Uhlenbeck, and Kac in their classical review~\cite{Ziff}.
Theoretical predictions for the BEC fluctuations, which are anomalously
large and non-Gaussian even in the thermodynamical limit, are derived and
explained on the basis of the simple analytical expressions \cite{KKS-PRL,KKS-PRA}. The 
results are in excellent agreement with the exact numerical simulations. The existence of the 
infrared singularities in the moments of fluctuations and the universal fact that these singularities are responsible for the anomalously large fluctuations in BEC, are among the recent conceptual discoveries. The quantum theory of laser threshold behavior constitutes another important advance in the physics of bosonic systems.

4. The laser made its appearance in the early 60's and provided us with a
new source of light with a new kind of photon statistics. Before the laser,
the statistics of radiation were either those of black-body photons
associated with Planck's radiation, which for a single mode of frequency $%
\nu $ takes the form
\begin{equation}
p_{n}=e^{-n\beta \hbar \nu }(1-e^{-\beta \hbar \nu })\text{,}
\end{equation}
or when one considers radiation from a coherent oscillating current such as
a radio transmitter or a microwave klystron the photon distribution becomes
Poissonian,
\begin{equation}
p_{n}=\frac{\bar{n}^{n}}{n!}e^{-\bar{n}}  \label{Poisson}
\end{equation}
where $\bar{n}$ is the average photon number.

However, laser photon statistics, as derived from the quantum theory of the
laser, goes from black-body statistics below threshold to Poissonian statistics far above threshold. In between, when we are in the threshold region (and even above threshold as in the case, for example, of the helium neon laser), we have a new distribution. We present a review of the laser photon statistical problem.

It has been said that the Bose-Einstein condensate is to atoms what the laser
is to photons; even the concept of an atom laser has emerged. In such a case,
one naturally asks, ``what is the statistical distribution of atoms in the
condensate?'' For example, let us first address the issue of an ideal gas of
$N$ atoms in contact with a reservoir at temperature $T$. The condensate occupation distribution in the harmonic trap under these conditions at low enough temperatures is given by the BEC master equation analysis as
\begin{equation}
p_{n_{0}}=\frac{1}{Z_{N}}\frac{[N(T/T_{c})^{3}]^{N-n_{0}}}{(N-n_{0})!}\text{.%
}  \label{pn0 BEC}
\end{equation}
The mean number and variance obtained from the condensate master equation are in excellent agreement with computer simulation (computer experiment) as shown
in Fig.~\ref{n0n2me}. We will discuss this aspect of the fluctuation problem
in some detail and indicate how the fluctuations change when we go to the
case of the interacting Bose gas.

5. The fascinating interface between superfluid He~II and BEC in a dilute gas
was mapped out by the experiments of Reppy and coworkers~\cite{rep};
and finite-size effects were studied theoretically by
M. Fisher and co-workers \cite{Fisher}. They carried out
experiments in which He~II was placed in a porous
glass medium which serves to keep the atoms well separated.
These experiments are characterized by a dilute gas BEC of $N$ atoms at
temperature $T$.

Of course, it was the successful experimental demonstration of Bose-Einstein
condensation in the ultracold atomic alkali-metal~\cite{bec,morebec,miesner},
hydrogen~\cite{kleppner} and helium gases~\cite{rep,Sant01,Robe01} that
stimulated the renaissance in the theory of BEC. In less than a decade, many
intriguing problems in the physics of BEC, that were not studied, or
understood before the 90's~\cite{LL,Huang,hm,fw,Griffin93,Griffin95,Griffin},
were formulated and resolved.

6. Finally, we turn on the interaction between atoms in the BEC and find
explicit expressions for the characteristic function and all cumulants of the
probability distribution of the number of atoms in the (bare) ground state of
a trap for the weakly interacting dilute Bose gas in equilibrium. The
surprising result is that the BEC statistics is not Gaussian, i.e., the ratio
of higher cumulants to an appropriate power of the variance does not vanish,
even in the thermodynamic limit. We calculate explicitly the effect of
Bogoliubov coupling between excited atoms on the suppression of the BEC
fluctuations in a box (``homogeneous gas'') at moderate temperatures and
their enhancement at very low temperatures. We find that there is a strong
pair-correlation effect in the occupation of the coupled atomic modes with
the opposite wavevectors ${\bf k}$ and ${\bf -k}$. This explains why the ground-state
occupation fluctuations remain anomalously large to the same extent as in the
non-interacting gas, except for a factor of 1/2 suppression. We find that,
roughly speaking, this is so because the atoms are strongly coupled in
correlated pairs such that the number of independent stochastic occupation
variables (``degrees of freedom'') contributing to the fluctuations of the
total number of excited atoms is only 1/2 the atom number $N$. This is a
particular feature of the well studied quantum optics phenomenon of two-mode
squeezing (see, e.g.,~\cite{squeezing} and \cite{Walls,sz}). The
squeezing is due to the quantum correlations that build up in the bare excited
modes via Bogoliubov coupling and is very similar to the noise squeezing in a
nondegenerate parametric amplifier.

Throughout the review, we will check main approximate analytical results
(such as in Eqs.~(\ref{II40}), (\ref{II54}), (\ref{III19}), (\ref{III71}),
(\ref{III72a})) against the ``exact'' numerics based on the recursion
relations~(\ref{III76}) and (\ref{III77}) which take into account exactly
all mesoscopic effects near the critical temperature $T_c$.
Unfortunately, the recursion relations are known only for the ideal Bose gas.
In the present review we discuss the BEC fluctuations mainly in the canonical
ensemble, which cures misleading predictions of the grand canonical ensemble
and, at the same time, does not have any essential differences with the
microcanonical ensemble for most physically interesting quantities and
situations. Moreover, as we discuss below, the canonical partition function
can be used for an accurate calculation of the microcanonical partition
function via the saddle-point method.

\section{History of the Bose-Einstein distribution}

In late 1923, a certain Satyendranath Bose, reader in physics at the
University of Dacca in East Bengal, submitted a paper on Planck's law of
blackbody radiation to the \emph{Philosophical Magazine\/}. Six months later
he was informed that the paper had received a negative referee report, and
consequently been rejected~\cite{Blanpied72}. While present authors may find
consolation in the thought that the rejection of a truly groundbreaking
paper after an irresponsibly long refereeing process is not an invention of
our times, few of their mistreated works will eventually meet with a
recognition comparable to Bose's. Not without a palpable amount of
self-confidence, Bose sent the rejected manuscript to Albert Einstein in
Berlin, together with a handwritten cover letter dated June 4, 1924,
beginning \cite{Quote}:

\begin{quote}
Respected Sir:

I have ventured to send you the accompanying article for your perusal and
opinion. I am anxious to know what you think of it. You will see that I have
tried to deduce the coefficient $8\pi\nu^2/c^3$ in Planck's Law independent
of the classical electrodynamics, only assuming that the ultimate elementary
regions in the phase-space has the content $h^3$. I do not know sufficient
German to translate the paper. If you think the paper worth publication I
shall be grateful if you arrange its publication in \emph{Zeitschrift f\"ur
Physik\/}. Though a complete stranger to you, I do not hesitate in making
such a request. Because we are all your pupils though profiting only from
your teachings through your writings \ldots
\end{quote}

In hindsight, it appears curious that Bose drew Einstein's attention only to
his derivation of the prefactor in Planck's law. Wasn't he aware of the fact
that his truly singular achievement, an insight not even spelled out
explicitly in Einstein's translation of his paper as it was received by the
\emph{Zeitschrift f\"ur Physik\/} on July 2, 1924 \cite{Bose,TheimerRam76},
but contained implicitly in the mathematics, lay elsewhere?

\subsection{What Bose did}

In the opening paragraph of his paper \cite{Bose,TheimerRam76}, Bose pounces
on an issue which he considers unsatisfactory: When calculating the energy
distribution of blackbody radiation according to
\begin{equation}
\varrho_\nu \mathrm{d} \nu = \frac{8\pi\nu^2 \mathrm{d} \nu}{c^3} E_\nu \; ,
\label{eq:Planck}
\end{equation}
that is,
\begin{eqnarray}
& & \mbox{energy per volume of blackbody radiation with frequency between}
\; \nu \; \mbox{and} \; \nu + \mathrm{d} \nu  \nonumber \\
& = &
\mbox{number of modes contained in that frequency interval of the
    radiation field per volume}  \nonumber \\
& \times & \mbox{thermal energy} \; E_\nu \;
\mbox{of a radiation mode with
    frequency} \; \nu \; ,  \nonumber
\end{eqnarray}
the number of modes had previously been derived only with reference to
classical physics. In his opinion, the logical foundation of such a recourse
was not sufficiently secure, and he proposed an alternative derivation,
based on the hypothesis of light quanta.

Considering radiation inside some cavity with volume~$V$, he observed that
the squared momentum of such a light quantum is related to its frequency
through
\begin{equation}
p^2 = \frac{h^2\nu^2}{c^2} \; ,
\end{equation}
where $h$ denotes Planck's constant, and $c$ is the velocity of light.
Dividing the frequency axis into intervals of length $\mathrm{d} \nu^s$,
such that the entire axis is covered when the label $s$ varies from $s=0$ to
$s=\infty$, the phase space volume associated with frequencies between $\nu$
and $\nu + \mathrm{d}\nu^s$ therefore is
\begin{eqnarray}
\int \! \mathrm{d} x \, \mathrm{d} y \, \mathrm{d} z \, \mathrm{d} p_x
\mathrm{d} p_y \mathrm{d} p_z & = & V \, 4\pi \left(\frac{h\nu}{c}\right)^2
\frac{h \mathrm{d} \nu^s}{c}  \nonumber \\
& = & V \, 4\pi \frac{h^3 \nu^2}{c^3} \mathrm{d} \nu^s \; .
\end{eqnarray}
It does not seem to have bothered Bose that the concept of phase space again
brings classical mechanics into play. Relying on the assumption that a
single quantum state occupies a cell of volume $h^3$ in phase space, a
notion which, in the wake of the Bohr--Sommerfeld quantization rule, may
have appeared natural to a physicist in the early 1920s, and accounting for
the two states of polarization, the total number $A^s$ of quantum cells
belonging to frequencies between $\nu$ and $\nu + \mathrm{d}\nu^s$,
corresponding to the number of radiation modes in that frequency interval,
immediately follows:
\begin{equation}
A^s = V \, \frac{8\pi\nu^2}{c^3} \mathrm{d}\nu^s \; .  \label{eq:Nmodes}
\end{equation}
That's all, as far as the first factor on the r.h.s.\ of Eq.~(\ref{eq:Planck}%
) is concerned. This is what Bose announced in his letter to Einstein, but
this is, most emphatically, not his main contribution towards the
understanding of Planck's law. The few lines which granted him immortality
follow when he turns to the second factor. Back-translated from Einstein's
phrasing of his words \cite{Bose,TheimerRam76}:

\begin{quote}
Now it is a simple task to calculate the thermodynamic probability of a
(macroscopically defined) state. Let $N^{s}$ be the number of quanta
belonging to the frequency interval $\mathrm{d}\nu^s$. How many ways are
there to distribute them over the cells belonging to $\mathrm{d}\nu^s$? Let $%
p_0^s$ be the number of vacant cells, $p_1^s$ the number of those containing
one quantum, $p_2^s$ the number of cells which contain two quanta, and so
on. The number of possible distributions then is
\begin{equation}
\frac{A^s!}{p_0^s! \, p_1^s! \, \ldots} \; , \quad \mbox{where} \quad A^s = V%
\frac{8\pi\nu^2}{c^3}\mathrm{d} \nu^s \; ,  \label{eq:Bose}
\end{equation}
and where
\[
N^s = 0 \cdot p_0^s + 1 \cdot p_1^s + 2 \cdot p_2^s \, \ldots
\]
is the number of quanta belonging to $\mathrm{d}\nu^s$.
\end{quote}

What is happening here? Bose is resorting to a fundamental principle of
statistical mechanics, according to which the probability of observing a
state with certain macroscopic properties --- in short: a macrostate --- is
proportional to the number of its microscopic realizations --- microstates
--- compatible with the macroscopically given restrictions. Let us, for
example, consider a model phase space consisting of four cells only, and let
there be four quanta. Let us then specify the macrostate by requiring that
one cell remain empty, two cells contain one quantum each, and one cell be
doubly occupied, \emph{i.e.\/}, $p_0 = 1$, $p_1 = 2$, $p_2 = 1$, and $p_r =
0 $ for $r \ge 3$. How many microstates are compatible with this
specification? We may place two quanta in one out of four cells, and then
choose one out of the remaining three cells to be the empty one. After that
there is no further choice left, since each of the two other cells now has
to host one quantum. Hence, there are $4 \times 3 = 12$ possible
configurations, or microstates: $12 = 4! / (1!\,2!\,1!)$. In general, when
there are $A^s$ cells belonging to $\mathrm{d} \nu^s$, they can be arranged
in $A^s!$ ways. However, if a cell pattern is obtained from another one
merely by a rearrangement of those $p_0^s$ cells containing no quantum, the
configuration remains unchanged. Obviously, there are $p_0^s!$ such
``neutral'' rearrangements which all correspond to the same configuration.
The same argument then applies, for any $r \ge 1$, to those $p_r^s$ cells
containing $r$ quanta: Each of the $p_r^s!$ possibilities of arranging the
cells with~$r$ quanta leads to the same configuration. Thus, each
configuration is realized by $p_0^s! \, p_1^s! \ldots$ equivalent
arrangements of cells, and the number of different configurations, or
microstates, is given by the total number of arrangements divided by the
number of equivalent arrangements, that is, by Bose's expression~(\ref
{eq:Bose}).

There is one proposition tacitly made in this way of counting microstates
which might even appear self-evident, but which actually constitutes the
very core of Bose's breakthrough, and which deserves to be spelled out
explicitly: When considering equivalent arrangements as representatives of
merely one microstate, it is implied that the quanta are \emph{%
indistinguishable\/}. It does not matter ``which quantum occupies which
cell''; all that matters are the occupation numbers $p_r^s$. Even more, the
``which quantum''-question is rendered meaningless, since there is, as a
matter of principle, no way of attaching some sort of label to individual
quanta belonging to the same $\mathrm{d} \nu^s$, with the purpose of
distinguishing them. This ``indistinguishability in principle'' does not
occur in classical physics. Two classical particles may have the same mass,
and identical other properties, but it is nevertheless taken for granted
that one can tell one from the other. Not so, according to Bose, with light
quanta.

The rest of Bose's paper has become a standard exercise in statistical
physics. Taking into account all frequency intervals $\mathrm{d}\nu ^{s}$,
the total number of microstates corresponding to a pre-specified set $%
\{p_{r}^{s}\}$ of cell occupation numbers is
\begin{equation}
W[\{p_{r}^{s}\}]=\prod_{s}\frac{A^{s}!}{p_{0}^{s}!\,p_{1}^{s}!\,\ldots }\;.
\label{W Bose}
\end{equation}
The logarithm of this functional yields the entropy associated with the
considered set $\{p_{r}^{s}\}$. Since, according to the definition of $%
p_{r}^{s}$,
\begin{equation}
A^{s}=\sum_{r}p_{r}^{s}\qquad \mbox{for each}\;s\;,  \label{eq:Cells}
\end{equation}
and assuming the statistically relevant $p_{r}^{s}$ to be large, Stirling's
approximation $\ln n!\approx n\ln n-n$ gives
\begin{equation}
\ln W[\{p_{r}^{s}\}]=\sum_{s}A^{s}\ln A^{s}-\sum_{s}\sum_{r}p_{r}^{s}\ln
p_{r}^{s}\;.  \label{eq:Entropy}
\end{equation}
The most probable macrostate now is the one with the maximum number of
microstates, characterized by that set of occupation numbers which maximizes
this expression~(\ref{eq:Entropy}). Stipulating that the radiation field be
thermally isolated, so that its total energy
\begin{equation}
E=\sum_{s}N^{s}h\nu ^{s}\qquad \mbox{with}\qquad N^{s}=\sum_{r}rp_{r}^{s}
\label{eq:energy}
\end{equation}
is fixed, the maximum is found by variation of the $p_{r}^{s}$, subject to
this constraint~(\ref{eq:energy}). In addition, the constraints~(\ref
{eq:Cells}) have to be respected. Introducing Lagrangian multipliers $%
\lambda ^{s}$ for these ``number-of-cells'' constraints, and a further
Lagrangian multiplier $\beta $ for the energy constraint, the maximum is
singled out by the condition
\begin{equation}
\delta \left( \ln W[\{p_{r}^{s}\}]-\sum_{s}\lambda
^{s}\sum_{r}p_{r}^{s}-\beta \sum_{s}h\nu ^{s}\sum_{r}rp_{r}^{s}\right) =0\;,
\label{eq:VarB}
\end{equation}
giving
\begin{equation}
\sum_{r,s}\delta p_{r}^{s}\left( \ln p_{r}^{s}+1+\lambda ^{s}\right) +\beta
\sum_{s}h\nu ^{s}\sum_{r}r\delta p_{r}^{s}=0\;.
\end{equation}
Since the $\delta p_{r}^{s}$ can now be taken as independent, the maximizing
configuration $\{\widehat{p}_{r}^{s}\}$ obeys
\begin{equation}
\ln \widehat{p}_{r}^{s}+1+\lambda ^{s}+r\beta h\nu ^{s}=0\;,
\end{equation}
or
\begin{equation}
\widehat{p}_{r}^{s}=B^{s}\mathrm{e}^{-r\beta h\nu ^{s}}\;,
\end{equation}
with normalization constants $B^{s}$ to be determined from the constraints~(%
\ref{eq:Cells}):
\begin{eqnarray}
A^{s} &=&\sum_{r}\widehat{p}_{r}^{s}  \nonumber \\
&=&\frac{B^{s}}{1-\mathrm{e}^{-\beta h\nu ^{s}}}\;.
\end{eqnarray}
The total number of quanta for the maximizing configuration then is
\begin{eqnarray}
\widehat{N}^{s} &=&\sum_{r}r\widehat{p}_{r}^{s}  \nonumber \\
&=&A^{s}\left( 1-\mathrm{e}^{-\beta h\nu ^{s}}\right) \sum_{r}r\mathrm{e}%
^{-r\beta h\nu ^{s}}  \nonumber \\
&=&\frac{A^{s}}{\mathrm{e}^{\beta h\nu ^{s}}-1}\;.  \label{eq:Nquanta}
\end{eqnarray}
Still, the physical meaning of the Lagrangian multiplier $\beta $ has to be
established. This can be done with the help of the entropy functional, since
inserting the maximizing configuration yields the thermodynamical
equilibrium entropy:
\begin{eqnarray}
S &=&k_{\mathrm{B}}\ln W[\{\widehat{p}_{r}^{s}\}]  \nonumber \\
&=&k_{\mathrm{B}}\left[ \beta E-\sum_{s}A^{s}\ln \left( 1-\mathrm{e}^{-\beta
h\nu ^{s}}\right) \right] \;,
\end{eqnarray}
where $k_{\mathrm{B}}$ denotes Boltzmann's constant. From the identity $%
\partial S/\partial E=1/T$ one then finds $\beta =1/(k_{\mathrm{B}}T)$, the
inverse energy equivalent of the temperature~$T$. Hence, from Eqs.~(\ref
{eq:Nquanta}) and (\ref{eq:Nmodes}) Bose obtains the total energy of the
radiation contained in the volume~$V$ in the form
\begin{eqnarray}
E &=&\sum_{s}N^{s}h\nu ^{s}  \nonumber \\
&=&\sum_{s}\frac{8\pi h{\nu ^{s}}^{3}}{c^{3}}V\frac{1}{\exp (\frac{h\nu ^{s}%
}{k_{\mathrm{B}}T}-1)}\mathrm{d}\nu ^{s}\;,
\end{eqnarray}
which is equivalent to Planck's formula: With the indistinguishability of
quanta, \emph{i.e.\/}, Bose's enumeration~(\ref{eq:Bose}) of microstates as
key input, the principles of statistical mechanics immediately yield the
thermodynamic properties of radiation.

\subsection{What Einstein did}

Unlike that unfortunate referee of the \emph{Philosophical Magazine\/},
Einstein immediately realized the power of Bose's approach. Estimating that
it took the manuscript three weeks to travel from Dacca to Berlin, Einstein
may have received it around June~25 \cite{Delbruck}. Only one week later, on
July~2, his translation of the manuscript was officially received by the
\emph{Zeitschrift f\"ur Physik\/}. The author's name was lacking its
initials --- the byline of the published paper \cite{Bose} simply reads: By
Bose (Dacca University, India) --- but otherwise Einstein was doing Bose
fair justice: He even sent Bose a handwritten postcard stating that he
regarded his paper as a most important contribution; that postcard seems to
have impressed the German Consulate in Calcutta to the extent that Bose's
visa was issued without requiring payment of the customary fee \cite
{Blanpied72}.

Within just a few days, Einstein then took a further step towards exploring
the implications of the ``indistinguish\-ability in principle'' of quantum
mechanical entities. At the end of the printed, German version of Bose's
paper \cite{Bose}, there appears the parenthetical remark ``Translated by
A.~Einstein'', followed by an announcement:

\begin{quote}
Note added by the translator: Bose's derivation of Planck's formula
constitutes, in my opinion, an important step forward. The method used here
also yields the quantum theory of the ideal gas, as I will explain in detail
elsewhere.
\end{quote}

``Elsewhere'' in this case meant the Proceedings of the Prussian Academy of
Sciences. In the session of the Academy on July 10, Einstein delivered a
paper entitled ``Quantum theory of the monoatomic ideal gas'' \cite{Einstein}%
. In that paper, he considered nonrelativistic free particles of mass~$m$,
so that the energy-momentum relation simply reads
\begin{equation}
E = \frac{p^2}{2m} \; ,
\end{equation}
and the phase-space volume for a particle with an energy not exceeding~$E$
is
\begin{equation}
\Phi = V \frac{4\pi}{3}(2mE)^{3/2} \; .  \label{eq:Phase}
\end{equation}
Again relying on the notion that a single quantum state occupies a cell of
volume $h^3$ in phase space, the number of such cells belonging to the
energy interval from $E$ to $E + \Delta E$ is
\begin{equation}
\Delta s = \frac{2\pi V}{h^3} (2m)^{3/2} E^{1/2} \Delta E \; .
\end{equation}
Thus, for particles with nonzero rest mass $\Delta s$ is the analog of
Bose's $A^s$ introduced in Eq.~(\ref{eq:Nmodes}). Einstein then specified
the cell occupation numbers by requiring that, out of these $\Delta s$
cells, $p_r^s \Delta s$ cells contain $r$~particles, so that $p_r^s$ is the
probability of finding $r$ particles in any one of these cells,
\begin{equation}
\sum_r p_r^s = 1 \; .  \label{eq:ConP}
\end{equation}
Now comes the decisive step. Without attempt of justification or even
comment, Einstein adopts Bose's way~(\ref{eq:Bose}) of counting the number
of corresponding microstates. This is a far-reaching hypothesis, which
implies that, unlike classical particles, atoms of the same species with
energies in the same range $\Delta E$ are indistinguishable: Interchanging
two such atoms does not yield a new microstate; as with photons, it does not
matter ``which atom occupies which cell''. Consequently, the number of
microstates associated with a pre-specified set of occupation probabilities $%
\{p_r^s\}$ for the above $\Delta s$ cells is
\begin{equation}
W^s = \frac{\Delta s !}{\prod_{r=0}^\infty (p_r^s \Delta s)!} \; ,
\label{W Einstein}
\end{equation}
giving, with the help of Stirling's formula,
\begin{equation}
\ln W^s = -\Delta s \sum_r p_r^s \ln p_r^s \; .
\end{equation}
Einstein then casts this result into a more attractive form. Stipulating
that the index~$s$ does no longer refer jointly to the cells within a
certain energy interval, but rather labels individual cells, the above
expression naturally generalizes to
\begin{equation}
\ln W[\{ p_r^s \}] = - \sum_s \sum_r p_r^s \ln p_r^s \; ,  \label{eq:Shannon}
\end{equation}
where the cell index~$s$ now runs over all cells, so that $p_r^s$ here is
the probability of finding $r$ particles in the $s$-th cell. It is
interesting to observe that this functional~(\ref{eq:Shannon}) has precisely
the same form as the Shannon entropy introduced in 1948 in an
information-theoretical context \cite{Shannon48}.

Since $\sum_r r p_r^s$ gives the expectation value of the number of
particles occupying the cell labelled~$s$, the total number of particles is
\begin{equation}
N = \sum_s \sum_r r p_r^s \; ,  \label{eq:ConN}
\end{equation}
while the total energy of the gas reads
\begin{equation}
E = \sum_s E^s \sum_r r p_r^s \; ,  \label{eq:ConE}
\end{equation}
where $E^s$ is the energy of a particle in the $s$-th cell. Since, according
to Eq.~(\ref{eq:Phase}), a cell's number~$s$ is related to the energy~$E^s$
through
\begin{eqnarray}
s & = & \frac{\Phi^s}{h^3}  \nonumber \\
& = & \frac{V}{h^3} \frac{4\pi}{3}(2 m E^s)^{3/2} \; ,
\end{eqnarray}
one has
\begin{equation}
E^s = c s^{2/3}
\end{equation}
with
\begin{equation}
c = \frac{h^2}{2m} \left( \frac{4\pi V}{3} \right)^{-2/3} \; .
\end{equation}
Considering an isolated system, with given, fixed particle number~$N$ and
fixed energy~$E$, the macrostate realized in nature is characterized by that
set $\{ \widehat{p}_r^s \}$ which maximizes the entropy functional~(\ref
{eq:Shannon}), subject to the constraints~(\ref{eq:ConN}) and~(\ref{eq:ConE}%
), together with the constraints~(\ref{eq:ConP}) expressing normalization of
the cell occupation probabilities. Hence,
\begin{equation}
\delta \left( \ln W[\{ p_r^s \}] -\sum_s \lambda^s \sum_r p_r^s - \alpha
\sum_s \sum_r r p_r^s - \beta \sum_s E^s \sum_r r p_r^s \right) = 0 \; ,
\label{eq:VarE}
\end{equation}
so that, seen from the conceptual viewpoint, the only difference between
Bose's variational problem~(\ref{eq:VarB}) and Einstein's variational
problem~(\ref{eq:VarE}) is the appearance of an additional Lagrangian
multiplier $\alpha$ in the latter: In the case of radiation, the total
number of light quanta adjusts itself in thermal equilibrium, instead of
being fixed beforehand; in the case of a gas of particles with nonzero rest
mass, the total number of particles is conserved, requiring the introduction
of the entailing multiplier~$\alpha$. One then finds
\begin{equation}
\ln \widehat{p}_r^s + 1 + \lambda^s + \alpha r + \beta r E^s = 0
\end{equation}
or
\begin{equation}
\widehat{p}_r^s = B^s \mathrm{e}^{-r(\alpha + \beta E^s)} \; ,
\label{eq:Pmax}
\end{equation}
with normalization constants to be determined from the constraints~(\ref
{eq:ConP}):
\begin{equation}
B^s = 1 - \mathrm{e}^{-(\alpha + \beta E^s)} \; .
\end{equation}
Here we deviate from the notation in Einstein's paper \cite{Einstein}, in
order to be compatible with modern conventions. The expectation value for
the occupation number of the cell with energy $E^s$ then follows from an
elementary calculation similar to Bose's reasoning~(\ref{eq:Nquanta}):
\begin{equation}
\sum_r r p_r^s = \frac{1}{\mathrm{e}^{\alpha + \beta E^s} - 1} \; .
\label{eq:ExpE}
\end{equation}
Therefore, the total number of particles and the total energy of the gas can
be expressed as
\begin{eqnarray}
N & = & \sum_s \frac{1}{\mathrm{e}^{\alpha + \beta E^s} - 1} \; ,
\label{eq:Nsum} \\
E & = & \sum_s \frac{E^s}{\mathrm{e}^{\alpha + \beta E^s} - 1} \; .
\label{eq:Esum}
\end{eqnarray}
Inserting the maximizing set~(\ref{eq:Pmax}) into the functional~(\ref
{eq:Shannon}) yields, after a brief calculation, the equilibrium entropy of
the gas in the form
\begin{eqnarray}
S & = & k_{\mathrm{B}} \ln W[\{ \widehat{p}_r^s \}]  \nonumber \\
& = & k_{\mathrm{B}} \left[ \alpha N + \beta E - \sum_s \ln \left( 1 -
\mathrm{e}^{-(\alpha + \beta E^s)} \right) \right] \; .
\end{eqnarray}
In order to identify the Lagrangian multiplier~$\beta$, Einstein considered
an infinitesimal heating of the system, assuming its volume and, hence, the
cell energies $E^s$ to remain fixed. This gives
\begin{eqnarray}
\mathrm{d} E & = & T \mathrm{d} S  \nonumber \\
& = & k_{\mathrm{B}} T \left[ N \mathrm{d} \alpha + \beta \mathrm{d} E + E
\mathrm{d} \beta - \sum_s \frac{\mathrm{d} (\alpha + \beta E^s)}{\mathrm{e}%
^{\alpha + \beta E^s} - 1} \right]  \nonumber \\
& = & k_{\mathrm{B}} T \beta \, \mathrm{d} E \; ,
\end{eqnarray}
requiring
\begin{equation}
\beta = \frac{1}{k_{\mathrm{B}} T} \; .
\end{equation}
As in Bose's case, the Lagrangian multiplier $\beta$ accounting for the
energy constraint is the inverse energy equivalent of the temperature~$T$.
The other multiplier $\alpha$, guaranteeing particle number conservation,
then is determined from the identity~(\ref{eq:Nsum}).

In the following two sections of his paper \cite{Einstein}, Einstein shows
how the thermodynamics of the classical ideal gas is recovered if one
neglects unity against $\mathrm{e}^{\alpha + \beta E^s}$, and derives the
virial expansion of the equation of state for the quantum gas obeying Eqs.~(%
\ref{eq:Nsum}) and~(\ref{eq:Esum}).

\subsection{Was Bose--Einstein statistics arrived at by serendipity?}

The title of this subsection is a literal quote from the title of a paper by
M.~Delbr\"uck \cite{Delbruck}, who contents that Bose made an elementary
mistake in statistics in that he \emph{should\/} have bothered ``which
quantum occupies which cell'', which would have been the natural approach,
and that Einstein first copied that mistake without paying much attention to
it. Indeed, such a suspicion does not seem to be entirely unfounded. In his
letter to Einstein, Bose announces only his comparatively straightforward
derivation of the number~(\ref{eq:Nmodes}) of radiation modes falling into
the frequency range from $\nu$ to $\nu + \mathrm{d} \nu^s$, apparently being
unaware that his revolutionary deed was the implicit exploitation of the
``indistinguishability in principle'' of quanta --- a concept so far unheard
of. In Einstein's translation of his paper \cite{Bose} this notion of
indistinguishability does not appear in words, although it is what underlies
the breakthrough. Even more, it does not appear in the first
paper \cite{Einstein} on the ideal Bose gas --- until the very last
paragraph, where Einstein ponders over

\begin{quote}
...a paradox which I have been unable to resolve. There is no difficulty in
treating also the case of a mixture of two different gases by the method
explained here. In this case, each molecular species has its own ``cells''.
From this follows the additivity of the entropies of the mixture's
components. Therefore, with respect to molecular energy, pressure, and
statistical distribution each component behaves as if it were the only one
present. A mixture containing $n_{1}$ and $n_{2}$ molecules, with the
molecules of the first kind being distinguishable (in particular with
respect to the molecular masses $m_{1}$, $m_{2}$) only by an arbitrarily
small amount from that of the second, therefore yields, at a given
temperature, a pressure and a distribution of states which differs from that
of a uniform gas with $n_{1}+n_{2}$ molecules with practically the same
molecular mass, occupying the same volume. However, this appears to be as
good as impossible.
\end{quote}

Interestingly, Einstein here considers ``distinguishability to some variable
degree'', which can be continuously reduced to indistinguishability. But
this notion is flawed: Either the molecules have some feature which allows
us to tell one species from the other, in which case the different species
can be distinguished, or they have none at all, in which case they are
indistinguishable in principle. Thus, at this point, about two weeks after
the receipt of Bose's manuscript and one week after sending its translation
to the \emph{Zeitschrift f\"ur Physik\/}, even Einstein may not yet have
fully grasped the implications of Bose's way~(\ref{eq:Bose}) of counting
microstates.

But there was more to come. In December 1924, Einstein submitted a second
manuscript on the quantum theory of the ideal Bose gas \cite
{Einstein25,Einstein25b}, formally written as a continuation of the first
one. He began that second paper by pointing out a curiosity implied by his
equation of state of the ideal quantum gas: Given a certain number of
particles~$N$ and a temperature~$T$, and considering a compression of the
volume~$V$, there is a certain volume below which a segregation sets in.
With decreasing volume, an increasing number of particles has to occupy the
first quantum cell, \emph{i.e.\/}, the state without kinetic energy, while the
rest is distributed over the other cells according to Eq.~(\ref{eq:ExpE}),
with $\mathrm{e}^{\alpha} = 1$. Thus, Bose--Einstein condensation was
unveiled! But this discovery merely appears as a small addendum to the
previous paper~\cite{Einstein}, for Einstein then takes up a different, more
fundamental scent. He mentioned that Ehrenfest and other colleagues of his had
criticized that in Bose's and his own theory the quanta or particles had not
been treated as statistically independent entities, a fact which had not
been properly emphasized. Einstein agrees, and then he sets out to put
things straight. He abandons his previous ``single-cell'' approach and again
considers the collection of quantum cells with energies between $E_\nu $ and
$E_\nu + \Delta E_\nu$, the number of which is
\begin{equation}
z_\nu = \frac{2\pi V}{h^3} (2m)^{3/2} E_\nu^{1/2} \Delta E_\nu \; .
\end{equation}
Then he juxtaposes in detail Bose's way of counting microstates to what is
done in classical statistics. Assuming that there are $n_\nu$ quantum
particles falling into $\Delta E_\nu$, Bose's approach~(\ref{eq:Bose})
implies that there are
\begin{equation}
W_\nu = \frac{(n_\nu + z_\nu - 1)!}{n_\nu! \, (z_\nu - 1)!}  \label{eq:Binom}
\end{equation}
possibilities of distributing the particles over the cells. This expression
can easily be visualized: Drawing the $n_\nu$ particles as a sequence of $%
n_\nu$ ``dots'' in a row, they can be organized into a microstate with
specific occupation numbers for $z_\nu$ cells --- again assuming that it
does not matter which particle occupies which cell --- by inserting $z_\nu -
1$ separating ``lines'' between them. Thus, there are $n_\nu + z_\nu -1$
positions carrying a symbol, $n_\nu$ of which are dots. The total number of
microstates then equals the total number of possibilities to select the $%
n_\nu$ positions carrying a ``dot'' out of these $n_\nu + z_\nu -1$
positions, which is just the binomial coefficient stated in Eq.~(\ref
{eq:Binom}). To give an example: Assuming that there are $n_\nu = 4$
particles and $z_\nu = 4$ cells, Eq.~(\ref{eq:Binom}) states that there are
altogether
\[
\frac{(4+4-1)!}{4! \, 3!} = \frac{7!}{4! \, 3!} = \frac{7 \cdot 6 \cdot 5}{2
\cdot 3} = 35
\]
microstates. On the other hand, there are several sets of occupation numbers
which allow one to distribute the particles over the cells:

\begin{center}
\begin{tabular}{||l|c||}
\hline
occupation numbers & number of microstates \\ \hline
$p_4 = 1$, $p_0 = 3$ & $\;\frac{4!}{1! \, 3!} = 4$ \\ \hline
$p_3 = 1$, $p_1 = 1$, $p_0 = 2$ & $\frac{4!}{1! \, 1! \, 2!} = 12$ \\ \hline
$p_2 = 2$, $p_0 = 2$ & $\;\frac{4!}{2! \, 2!} = 6$ \\ \hline
$p_2 = 1$, $p_1 = 2$, $p_0 = 1$ & $\frac{4!}{1! \, 2! \, 1!} = 12$ \\ \hline
$p_1 = 4$ & $\;\;\;\frac{4!}{4!} = 1$ \\ \hline
\end{tabular}
\end{center}

The right column of this table gives, for each set, the number of
microstates according to Bose's formula~(\ref{eq:Bose}); obviously, these
numbers add up to the total number 35 anticipated above. Thus, the binomial
coefficient~(\ref{eq:Binom}) conveniently accounts for \emph{all\/} possible
microstates, without the need to specify the occupation numbers, according
to the combinatorial identity
\begin{equation}
\sum \frac{Z!}{p_0! \, \ldots \, p_N!} = {\binom{N + Z -1}{N}} \; ,
\label{combinatory}
\end{equation}
where the sum is restricted to those sets $\{p_0, p_1, \ldots, p_N \}$ which
comply with the two conditions $\sum_r p_r = Z$ and $\sum_r r p_r = N$, as
in the example above. In Appendix \ref{ap:proof} we provide a proof of this
identity. With this background, let us return to Einstein's reasoning: When
taking into account all energy intervals $\Delta E_{\nu }$, the
total number of microstates is given by the product $W=\prod_{\nu}W_{\nu }$,
providing the entropy functional
\begin{eqnarray}
\ln W[\{n_{\nu }\}] &=&\sum_{\nu }\Big[(n_{\nu }+z_{\nu })\ln (n_{\nu
}+z_{\nu })-n_{\nu }\ln n_{\nu }-z_{\nu }\ln z_{\nu }\Big]  \nonumber \\
&=&\sum_{\nu }\left[ n_{\nu }\ln \!\left( 1+\frac{z_{\nu }}{n_{\nu }}\right)
+z_{\nu }\ln \!\left( \frac{n_{\nu }}{z_{\nu }}+1\right) \right] \;.
\end{eqnarray}
The maximizing set $\{\widehat{n}_{\nu }\}$ now has to obey the two
constraints
\begin{eqnarray}
\sum_{\nu }n_{\nu } &=&N \\
\sum_{\nu }n_{\nu }E_{\nu } &=&E\;,
\end{eqnarray}
but there is no more need for the multipliers~$\lambda ^{s}$ appearing in
the previous Eqs.~(\ref{eq:VarB}) and (\ref{eq:VarE}), since the
constraints~(\ref{eq:Cells}) or (\ref{eq:ConP}) are automatically respected
when starting from the convenient expression~(\ref{eq:Binom}). Hence, one
has
\begin{equation}
\delta \left( \ln W[\{n_{\nu }\}]-\alpha \sum_{\nu }n_{\nu }-\beta \sum_{\nu
}n_{\nu }E_{\nu }\right) =0
\end{equation}
or
\begin{equation}
\sum_{\nu }\left[ \ln \!\left( 1+\frac{z_{\nu }}{n_{\nu }}\right) -\alpha
-\beta E_{\nu }\right] \delta n_{\nu }=0\;,  \label{eq:VarF}
\end{equation}
leading immediately to
\begin{equation}
\widehat{n}_{\nu }=\frac{z_{\nu }}{\mathrm{e}^{\alpha +\beta E_{\nu }}-1}\;,
\label{eq:DistE}
\end{equation}
in agreement with the previous result~(\ref{eq:ExpE}).

But what, Einstein asks, would have resulted had one not adopted Bose's
prescription~(\ref{eq:Bose}) and thus counted equivalent arrangements with
equal population numbers only once, but rather had treated the particles as
classical, statistically independent entities? Then there obviously are
\begin{equation}
W_\nu = (z_\nu)^{n_\nu}
\end{equation}
possibilities of distributing the $n_\nu$ particles belonging to $\Delta
E_\nu$ over the $z_\nu$ cells: Each particle simply is placed in one of the $%
z_\nu$ cells, regardless of the others. Now, when considering all intervals $%
\Delta E_\nu$, with distinguishable particles it does matter how those $%
n_\nu $ particles going into the respective $\Delta E_\nu$ are selected from
all $N $~particles; for this selection, there are $N!/ \prod_\nu n_\nu!$
possibilities. Thus, taking classical statistics seriously, there are
\begin{equation}
W = N! \prod_\nu \frac{(z_\nu)^{n_\nu}}{n_\nu !}  \label{eq:Class}
\end{equation}
possible microstates, yielding
\begin{eqnarray}
\ln W[\{n_\nu\}] & = & N \ln N - N + \sum_\nu \Big[
n_\nu \ln z_\nu - n_\nu \ln n_\nu + n_\nu \Big]  \nonumber \\
& = & N \ln N + \sum_\nu \left[ n_\nu \ln\!\left(\frac{z_\nu}{n_\nu}\right)
+ n_\nu \right] \; .  \label{eq:Cfun}
\end{eqnarray}
This is a truly vexating expression, since it gives a thermodynamical
entropy which is not proportional to the total number of particles, \emph{%
i.e.\/}, no extensive quantity, because of the first term on the r.h.s.
Hence, already in the days before Bose and Einstein one had got used to
ignoring the leading factor~$N!$ in Eq.~(\ref{eq:Class}), with the
half-hearted concession that microstates which result from each other by a
mere permutation of the $N$ particles should not be counted as different. Of
course, this is an intrinsic inconsistency of the classical theory: Instead
of accepting that, shouldn't one abandon Eq.~(\ref{eq:Class}) straight away
and accept the more systematic quantum approach, despite the apparently
strange consequence of losing the particles' independence? And Einstein
gives a further, strong argument in favor of the quantum theory: At zero
temperature, all particles occupy the lowest cell, giving $n_1 = N$ and $%
n_\nu = 0$ for $\nu > 1$. With $z_1 = 1$, the quantum way of counting based
on Eq.~(\ref{eq:Binom}) gives just one single microstate, which means zero
entropy in agreement with Nernst's theorem, whereas the classical
expression~(\ref{eq:Class}) yields an incorrect entropy even if one ignores
the disturbing~$N!\,$ Finally, the variational calculation based on the
classical functional~(\ref{eq:Cfun}) proceeds via
\begin{equation}
\sum_\nu \left[ \ln\!\left(\frac{z_\nu}{n_\nu} \right) - \alpha - \beta
E_\nu \right] \delta n_\nu = 0 \; ,
\end{equation}
furnishing the Boltzmann-like distribution
\begin{equation}
\widehat{n}_\nu = z_\nu \mathrm{e}^{-\alpha - \beta E_\nu}
\end{equation}
for the maximizing set $\{ \widehat{n}_\nu \}$. In short, the quantum ideal
gas of nonzero mass particles with its distribution~(\ref{eq:DistE})
deviates from the classical ideal gas in the same manner as does Planck's
law of radiation from Wien's law. This observation convinced Einstein, even
in the lack of any clear experimental evidence, that Bose's way of counting
microstates had to be taken seriously, since, as he remarks in the
introduction to his second paper \cite{Einstein25}, ``if it is justified to
consider radiation as a quantum gas, the analogy between the quantum gas and
the particle gas has to be a complete one''. This belief also enabled him to
accept the sacrifice of the statistical independence of quantum particles
implied by the formula~(\ref{eq:Binom}), which, by the end of 1924, he had
clearly realized:

\begin{quote}
The formula therefore indirectly expresses a certain hypothesis about a
mutual influence of the molecules on each other which is of an entirely
mysterious kind ...
\end{quote}

But what might be the physics behind that mysterious influence which
non-interacting particles appear to exert on each other? In a further section
of his paper \cite{Einstein25}, Einstein's reasoning takes an amazing
direction: He considers the density fluctuations of the ideal quantum gas,
and from this deduces the necessity to invoke wave mechanics! Whereas he had
previously employed what is nowadays known as the microcanonical ensemble,
formally embodied through the constraints that the total number of particles
and the total energy be fixed, he now resorts to a grand canonical framework
and considers a gas within some finite volume $V$ which communicates with a
gas of the same species contained in an infinitely large volume. He then
stipulates that both volumes be separated from each other by some kind of
membrane which can be penetrated only by particles with an energy in a
certain infinitesimal range $\Delta E_\nu$, and quantifies the ensuing
fluctuation $\Delta n_\nu$ of the number of particles in~$V$, not admitting
energy exchange between particles in different energy intervals. Writing $%
n_\nu = \widehat{n}_\nu + \Delta n_\nu$, the entropy of the gas within~$V$
is expanded in the form
\begin{equation}
S_{\mathrm{gas}}(\Delta n_\nu) = \widehat{S}_{\mathrm{gas}} + \frac{\partial
\widehat{S}_{\mathrm{gas}}} {\partial \Delta n_\nu} \, \Delta n_\nu + \frac{1%
}{2} \, \frac{\partial^2 \widehat{S}_{\mathrm{gas}}} {\partial (\Delta
n_\nu)^2} \, (\Delta n_\nu)^2 \; ,
\end{equation}
whereas the entropy of the reservoir changes with the transferred particles
according to
\begin{equation}
S_0(\Delta n_\nu) = \widehat{S}_0 - \frac{\partial \widehat{S}_0}{\partial
\Delta n_\nu} \, \Delta n_\nu \; .
\end{equation}
In view of the assumed infinite size of the reservoir, the quadratic term is
negligible here. Since the equilibrium state is characterized by the
requirement that the total entropy $S = S_{\mathrm{gas}} + S_0$ be maximum,
one has
\begin{equation}
\frac{\partial \widehat{S}}{\partial \Delta n_\nu} = 0 \; ,
\end{equation}
so that
\begin{equation}
S(\Delta n_\nu) = \widehat{S} + \frac{1}{2} \, \frac{\partial^2 \widehat{S}_{%
\mathrm{gas}}} {\partial (\Delta n_\nu)^2} \, (\Delta n_\nu)^2 \; .
\end{equation}
Hence, the probability distribution for finding a certain fluctuation $%
\Delta n_\nu$ is Gaussian,
\begin{eqnarray}
P(\Delta n_\nu) & = & \mathrm{const} \cdot \mathrm{e}^{S(\Delta n_\nu)/k_{%
\mathrm{B}}}  \nonumber \\
& = & \mathrm{const} \cdot \exp\left(\frac{1}{2k_{\mathrm{B}}} \, \frac{%
\partial^2 \widehat{S}_{\mathrm{gas}}} {\partial (\Delta n_\nu)^2} \,
(\Delta n_\nu)^2 \right) \; ,
\end{eqnarray}
from which one reads off the mean square of the fluctuations,
\begin{equation}
\langle (\Delta n_\nu )^2 \rangle = \frac{k_{\mathrm{B}}}{- \frac{\partial^2
\widehat{S}_{\mathrm{gas}}} {\partial (\Delta n_\nu)^2}} \; .
\end{equation}
Since, according to the previous Eq.~(\ref{eq:VarF}), one has
\begin{equation}
\frac{1}{k_{\mathrm{B}}} \frac{\partial \widehat{S}_{\mathrm{gas}}}{\partial
\Delta n_\nu} = \ln\!\left(1 + \frac{z_\nu}{\widehat{n}_\nu}\right) \; ,
\end{equation}
one deduces
\begin{equation}
\frac{1}{k_{\mathrm{B}}} \frac{\partial^2 \widehat{S}_{\mathrm{gas}}}{%
\partial (\Delta n_\nu)^2} = \frac{- z_\nu}{\widehat{n}_\nu^2 + z_\nu%
\widehat{n}_\nu} \; ,
\end{equation}
resulting in
\begin{equation}
\langle (\Delta n_\nu )^2 \rangle = \widehat{n}_\nu + \frac{\widehat{n}_\nu^2%
}{z_\nu}  \label{eq:Delta}
\end{equation}
or
\begin{equation}
\langle (\Delta n_\nu / n_\nu)^2 \rangle = \frac{1}{\widehat{n}_\nu} + \frac{%
1}{z_\nu} \; .  \label{eq:Fluc}
\end{equation}
With $z_\nu = 1$, this gives the familiar grand canonical expression for the
fluctuation of the occupation number of a single quantum state. With a
stroke of genius, Einstein now interprets this formula: Whereas the first
term on the r.h.s.\ of Eq.~(\ref{eq:Fluc}) would also be present if the
particles were statistically independent, the second term reminds him of
interference fluctuations of a radiation field \cite{Einstein25}:

\begin{quote}
One can interpret it even in the case of a gas in a corresponding manner, by
associating with the gas a radiation process in a suitable manner, and
calculating its interference fluctuations. I will explain this in more
detail, since I believe that this is more than a formal analogy.
\end{quote}

He then refers to de Broglie's idea of associating a wavelike process with
single material particles and argues that, if one associates a scalar wave
field with a gas of quantum particles, the term $1/z_\nu$ in Eq.~(\ref
{eq:Fluc}) describes the corresponding mean square fluctuation of the wave
field. What an imagination --- on the basis of the fluctuation formula~(\ref
{eq:Fluc}) Einstein anticipates many-body matter waves, long before wave
mechanics was officially enthroned! Indeed, it was this paper of his which
led to a decisive turn of events: From these speculations on the relevance
of matter waves Schr\"odinger learned about de Broglie's thesis, acquired a
copy of it, and then formulated his wave mechanics.

Having recapitulated this history, let us once again turn to the title of
this subsection: Was Bose--Einstein statistics arrived at by serendipity?
Delbr\"{u}ck's opinion that it arose out of an elementary mistake in
statistics that Bose made almost certainly is too harsh. On the other hand,
the important relation~(\ref{combinatory}), see also the Appendix \ref
{ap:proof}, does not seem to have figured in Bose's thinking. When writing
down the crucial expression~(\ref{eq:Bose}), Bose definitely must have been
aware that he was counting the number of microstates by determining the
number of different distributions of quanta over the available quantum
cells, regardless of ``which quantum occupies which cell''. He may well have
been fully aware that his way of counting implied the indistinguishability
of quanta occupying the same energy range, but he did not reflect on this
curious issue. On the other hand, he didn't have to, since his way of
counting directly led to one of the most important formulas in physics, and
therefore simply had to be correct.

Many years later, Bose recalled \cite{Pais}:

\begin{quote}
I had no idea that what I had done was really novel \ldots. I was not a
statistician to the extent of really knowing that I was doing something
which was really different from what Boltzmann would have done, from
Boltzmann statistics. Instead of thinking of the light-quantum just as a
particle, I talked about these states.
\end{quote}

By counting the number of ways to fill a number of photonic states
(cells) Bose
obtained Eq. (\ref{W Bose}) which is exactly the same form as
Boltzmann's Eq. (\ref{eq:Class}) for $z_\nu=1$, but with new
meanings attached to the new symbols: $n_{s}$ replaced by
$p_{r}^{s}$ and $N$ replaced by $A^{s}$. Bose's formula leads to
an entirely different new statistics - the quantum statistics for
indistinguishable particles, in contrast to Boltzmann's
distinguishable particles. It took a while for this to sink in
even with Einstein, but that is the nature of research.


Contrary to Bose, Einstein had no experimental motivation when adapting
Bose's work to particles with nonzero rest mass. He seems to have been
guided by a deep-rooted belief in the essential simplicity of physics, so
that he was quite ready to accept a complete analogy between the gas of
light quanta and the ideal gas of quantum particles, although he may not yet
have seen the revolutionary implications of this concept when he submitted
his first paper~\cite{Einstein} on this matter. But the arrival at a deep
truth on the basis of a well-reflected conviction can hardly be called
serendipity. His second paper \cite{Einstein25} is, by all means, a singular
intellectual achievement, combining daring intuition with almost prophetical
insight. And who would blame Einstein for trying to apply, in another
section of that second paper, his quantum theory of the ideal Bose gas to the
electron gas in metals?

In view of the outstanding importance which Einstein's fluctuation formula~(%
\ref{eq:Fluc}) has had for the becoming of wave mechanics, it appears
remarkable that a puzzling question has long remained unanswered: What
happens if one faces, unlike Einstein in his derivation of this relation~(%
\ref{eq:Fluc}), a closed system of Bose particles which does not communicate
with some sort of particle reservoir? When the temperature~$T$ approaches
zero, all $N$~particles are forced into the system's ground state, so that
the mean square $\langle (\Delta n_0)^2 \rangle$ of the fluctuation of the
ground-state occupation number has to vanish for $T \to 0$ --- but with $z_0
= 1$, the grand canonical Eq.~(\ref{eq:Delta}) gives $\langle (\Delta n_0)^2
\rangle \to N(N+1)$, clearly indicating that with respect to these
fluctuations the different statistical ensembles are no longer equivalent.
What, then, would be the correct expression for the fluctuation of the
ground-state occupation number within the \emph{canonical\/} ensemble, which
excludes any exchange of particles with the environment, but still allows for
the exchange of energy? To what extent does the \emph{microcanonical}
ensemble, which applies when even the energy is kept constant, differ from
the canonical one? Various aspects of this riddle have appeared in the
literature over the years \cite{Ziff,DterHaar,Fierz}, mainly inspired
by academic curiosity, before it resurfaced in 1996 \cite{gh96,pol,gr,ww},
this time triggered by the experimental realization of mesoscopic
Bose--Einstein condensates in isolated microtraps. Since then, much insight
into this surprisingly rich problem has been gained, and some of the answers
to the above questions have been given by now. In the following sections of
this article, these new developments will be reviewed in detail.

\subsection{Comparison between Bose's and Einstein's counting of the number
of microstates $W$}

\begin{figure}[tbp]
\center \epsfxsize=9cm\epsffile{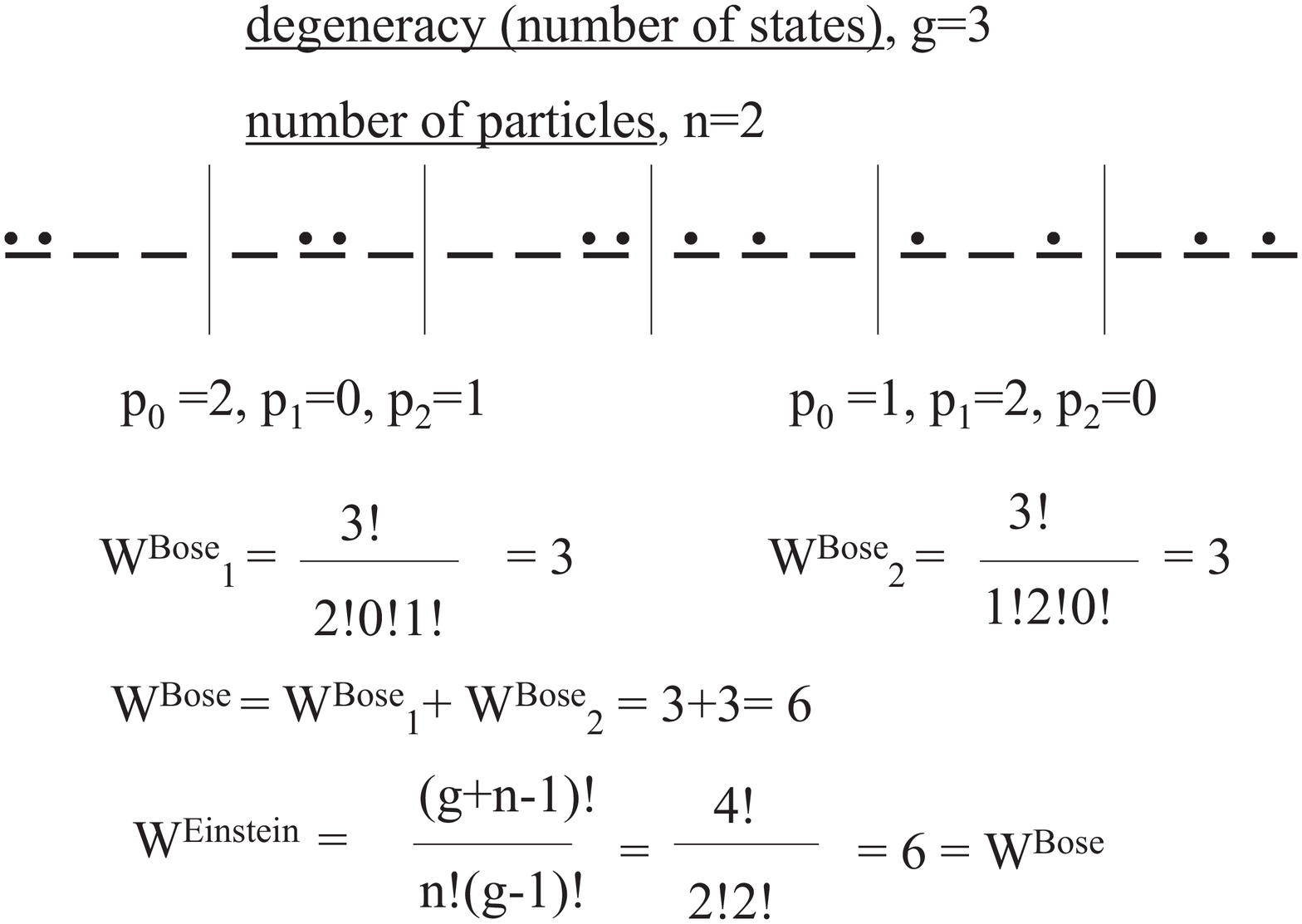}
\caption{A simple example showing Bose's and Einstein's (textbook) methods
of counting the number of possible ways to put $n=2$ particles in a level
having $A=g=3$ states.}
\label{Bosecount}
\end{figure}

In Bose's original counting~(\ref{eq:Bose}), he considered the numbers
$p_{r}^{s}$ of cells occupied with $r$ photons, so that the total number
of cells is given by $A^{s}=\sum\limits_{r}p_{r}^{s}$. While Einstein
still had adopted this way of counting in his first paper~\cite{Einstein}
on the ideal Bose gas, as witnessed by his Eq.~(\ref{W Einstein}), he used,
without comment, more economical means in his second paper~\cite{Einstein25},
relying on the binomial coefficient~(\ref{eq:Binom}).
Figure~\ref{Bosecount} shows an example in which $N^s=2$ particles are
distributed over $A^s=3$ cells, and visualizes that Bose's formula for
counting the number of possible arrangements (or the number of microstates
which give the same macrostate) gives the same result as Einstein's formula
only after one has summed over all possible configurations, i.e., over those
sets $p_{0}^{s},p_{1}^{s},...$ which simultaneously obey the two conditions
$A^{s}=\sum_{r}p_{r}^{s}$ and $N^s =\sum_{r}r p_{r}^{s}$. In this example,
there are only two such sets; a general proof is given in
Appendix~\ref{ap:proof}. Thus, it is actually possible to state the number
of microstates without specifying the individual arrangements, by summing
over all of them:
\begin{equation}
\left(
\sum\limits_{p_{0}^{s}=0}^{A^{s}}...\sum\limits_{p_{N^{s}}^{s}=0}^{A^{s}}%
\frac{A^{s}!}{p_{0}^{s}!\,p_{1}^{s}!\,\ldots \,p_{N^{s}}^{s}}, \quad
A^{s}=\sum_{r=0}^{N^{s}}p_{r}^{s}
\;\text{ and }\;N^{s}=\sum_{r=0}^{N^{s}}rp_{r}^{s}%
\right) = \frac{(N^s+A^s-1)!}{N^{s}!(A^s-1)!}.
\end{equation}

\section{Grand canonical versus canonical statistics of BEC fluctuations}

The problem of BEC fluctuations in a Bose gas is well
known~\cite{Ziff,DterHaar}.
However, as with many other problems which are well-posed physically and
mathematically, it is highly nontrivial and deep, especially for the
interacting Bose gas.

\subsection{Relations between statistics of BEC fluctuations in the grand
canonical, canonical, and microcanonical ensembles}

To set the stage, we briefly review here some basic notions and facts from
the statistical physics of BEC involving relations between statistics of BEC
fluctuations in the grand canonical, canonical, and microcanonical ensembles.

Specific experimental conditions determine which statistics should be
applied to describe a particular system. In view of the present experimental
status the canonical and microcanonical descriptions of the BEC are of
primary importance. Recent BEC experiments on harmonically trapped atoms of
dilute gases deal with a finite and well defined number of particles. This
number, even if it is not known exactly, certainly does not fluctuate once
the cooling process is over. Magnetic or optical confinement suggests that
the system is also thermally isolated and, hence, the theory of the trapped
condensate based on a microcanonical ensemble is needed. In this
microcanonical ensemble the total particle number and the total energy are
both exactly conserved, i.e., the corresponding operators are constrained to
be the $c$-number constants, $\hat N = const$ and $\hat H = const$. On the
other hand, in experiments with two (or many) component BECs, Bose-Fermi
mixtures, and additional gas components, e.g., for sympathetic cooling,
there is an energy exchange between the components. As a result, each of the
components can be described by the canonical ensemble that applies to
systems with conserved particle number while exchanging energy with a heat
bath of a given temperature. Such a description is also appropriate for
dilute $^{4}$He in a porous medium \cite{rep}. In the canonical ensemble
only the total number of particles is constrained to be an exact,
non-fluctuating constant,
\begin{equation}
N = \hat n_0 + \sum_{\mathbf{k} \neq 0} \hat n_{\mathbf{k}} ,
\label{CEconstraint}
\end{equation}
but the energy $\hat H$ has non-zero fluctuations, $\langle (\hat H - \bar H
)^2 ) \rangle \neq 0$, and only its average value is a constant, $\langle
\hat H \rangle = \bar H = E = const$, determined by a fixed temperature $T$
of the system.

However, the microcanonical and canonical descriptions of a many-body system
are difficult because of the operator constraints imposed on the total energy
and particle number. As a result, the standard textbook formulation of the
BEC problem assumes either a grand canonical ensemble (describing a system
which is allowed to exchange both energy and particles with a reservoir at a
given temperature $T$ and chemical potential $\mu$, which fix only the
average total number of particles, $\langle \hat N \rangle = const$, and the
average total energy $\langle \hat H \rangle = const$) or some restricted
ensemble that selects states so as to ensure a condensate wave function with
an almost fixed phase and amplitude \cite{b60,Anderson,hm,LL}. These
standard formulations focus on and provide effective tools for the study of
the thermodynamic and hydrodynamic properties of the many-body Bose system
at the expense of an artificial modification of the condensate statistics
and dynamics of BEC formation. While the textbook grand canonical prediction
of the condensate mean occupation agrees, in some sense, with the
Bose-Einstein condensation of trapped atomic gases, this is not even
approximately true as concerns the grand canonical counting statistics
\begin{equation}
\rho _{\nu}^{GC}(n_{\nu})=\frac{1}{1+\bar{n}_{\nu}}\left( \frac{\bar{n}_{\nu}%
}{1+\bar{n}_{\nu}}\right)^{n_{\nu}},  \label{GC0}
\end{equation}
which gives the probability to find $n_{\nu}$ particles in a given
single-particle state $\nu$, where the mean occupation is $\bar{n}_{\nu}$.
Below the Bose-Einstein condensation temperature, where the ground state
mean-occupation number is macroscopic, $\bar{n}_{0}\sim N$, the distribution
$\rho_{0}^{GC}(n_0)$ becomes extremely broad with the squared variance
$\langle (n_{0}-\bar{n}_{0})^{2}\rangle \approx \bar{n}_{0}^{2}\propto N^{2}$
even at $T \to 0$~\cite{Ziff,bb}. This prediction is surely at odds with the
isolated Bose gas, where for sufficiently low temperature all particles are
expected to occupy the ground state with no fluctuations left. It was argued
by Ziff et al.~\cite{Ziff} that this unphysical behavior of the variance is
just a mathematical artefact of the standard grand canonical ensemble, which
becomes unphysical below the condensation point. Thus, the grand canonical
ensemble is irrelevant to experiment if not revised properly. Another
extremity, namely, a complete fixation of the amplitude and phase of the
condensate wave function, is unable to address the condensate formation and
the fluctuation problems at all.

It was first realized by Fierz~\cite{Fierz} that the canonical ensemble with
an exactly fixed total number of particles removes the pathologies of the
grand canonical ensemble. He exploited the fact that the description of BEC
in a homogeneous ideal Bose gas is exactly equivalent to the spherical model
of statistical physics, and that the condensate serves as a particle reservoir
for the non-condensed phase. Recently BEC fluctuations were studied by a
number of authors in different statistical ensembles, both in the ideal and
interacting Bose gases. The microcanonical treatment of the ground-state
fluctuations in a one-dimensional harmonic trap is closely related to the
combinatorics of partitioning an integer, opening up an interesting link
to number theory~\cite{gh96,wh0203,Lewenstein05}.

It is worthwhile to compare the counting statistics (\ref{GC0}) with the
predictions of other statistical ensembles \cite{ww}. For high
temperatures, $T>T_c$, all three ensembles predict the same counting
statistics. However, this is not the case for low temperatures, $T<T_c$.
Here the broad distribution of the grand canonical statistics differs most
dramatically from $\rho _{0}(n_0)$ in the canonical and microcanonical
statistics, which show a narrow single-peaked distribution around the
condensate mean occupation number. In particular, the master equation
approach \cite{MOS99,KSZZ} (Section IV) yields a finite negative binomial
distribution for the probability distribution of the ground-state occupation
in the ideal Bose gas in the canonical ensemble (see Fig.~\ref{FigII6} and
Eqs.~(\ref{II35}), (\ref{II40}), and (\ref{II41})). The width of the peak
decreases with decreasing temperature. In fact, it is this sharply-peaked
statistical distribution which one would naively expect for a Bose condensate.

Each statistical ensemble is described by a different partition function.
The microcanonical partition function $\Omega (E,N)$ is equal to the number
of $N$-particle microstates corresponding to a given total energy $E$.
Interestingly, at low temperatures canonical and microcanonical fluctuations
have been found to agree in the large-$N$ (thermodynamic) limit for
one-dimensional harmonic trapping potentials, but to differ in the case of
three-dimensional isotropic harmonic traps~\cite{Navez}. More precisely, for
the $d$-dimensional power-law traps characterized by an exponent~$\sigma$,
as considered later in Section~V.C, microcanonical and canonical
fluctuations agree in the large-$N$ limit when $d/\sigma <2$, but the
microcanonical fluctuations remain smaller than the canonical ones when
$d/\sigma > 2$~\cite{gh97b}. Thus, in $d=3$ dimensions the homogeneous Bose
gas falls into the first category, but the harmonically trapped one into
the second. The difference between the fluctuations in these two ensembles is
expressed by a formula which is similar in spirit to the one expressing the
familiar difference between the heat capacities at constant pressure and at
constant volume~\cite{Navez,gh97b}.

The direct numerical computation of the microcanonical partition function
becomes very time consuming or not possible for $N>10^{5}$. For $N\gg 1$,
and for large numbers of occupied excited energy levels, one can invoke,
e.g., the approximate technique based on the saddle-point method, which is
widely used in statistical physics~\cite{Huang}. When employing this method,
one starts from the known grand canonical partition function and utilizes
the saddle-point approximation for extracting its required canonical and
microcanonical counterparts, which then yield all desired quantities by
taking suitable derivatives. Recently, still another statistical ensemble,
the so called Maxwell's demon ensemble, has been introduced~\cite{Navez}.
Here, the system is divided into the condensate and the particles occupying
excited states, so that only particle transfer (without energy exchange)
between these two subsystems is allowed, an idea that had also been
exploited by Fierz~\cite{Fierz} and by Politzer~\cite{pol}. This ensemble has
been used to obtain an approximate analytical expression for the ground-state
BEC fluctuations both in the canonical and in the microcanonical ensemble.
The Maxwell's demon approximation can be understood on the basis of the
canonical-ensemble quasiparticle approach \cite{KKS-PRL,KKS-PRA}, which is discussed in Section~V, and which is readily generalizable to the case of the interacting Bose gas (see Section~VI). It also directly proves that the higher statistical moments for a homogeneous Bose gas depend on the particular boundary conditions imposed, even in the thermodynamic limit~\cite{KKS-PRL,KKS-PRA,hks02}.

The canonical partition function $Z_{N}(T)$ is defined as
\begin{equation}
Z_{N}(T)=\sum_{E}^{\infty }e^{-E/k_B T}\Omega (E,N),  \label{II42}
\end{equation}
where the sum runs over all allowed energies, and $k_B$ is the Boltzmann
constant. Eq. (\ref{II42}) can be used to calculate also the microcanonical
partition function $\Omega (E,N)$ by means of the inversion of this
definition. Likewise, the canonical partition function $Z_{N}(T)$ can be
obtained by the inversion of the definition of the grand canonical partition
function $\Xi (\mu ,T)$:
\begin{equation}
\Xi (\mu ,T)=\sum_{N=0}^{\infty }e^{\mu N/k_B T}Z_{N}(T) .  \label{gc1}
\end{equation}
Inserting Eq.~(\ref{II42}) into Eq.~(\ref{gc1}), we obtain the following
relation between $\Xi (\mu ,T)$ and $\Omega (E,N)$:
\begin{equation}
\Xi (\mu ,T)=\sum_{N=0}^{\infty }e^{\mu N/k_B T}\sum_{E}^{\infty }e^{-E/k_B
T}\Omega (E,N) .  \label{gc2}
\end{equation}
As an example, let us consider an isotropic 3-dimensional harmonic trap with
eigenfrequency $\omega$. In this case, a relatively compact expression for
the grand canonical partition function,
\begin{equation}
\Xi (\mu
,T)=\prod\limits_{E}^{\infty }\left( \frac{1}{1-e^{(\mu -E)/k_B T}} \right)
^{(E/\hbar \omega +1) (E/\hbar \omega +2)/2} \; ,
\end{equation}
allows us to find $\Omega(E,N) $ from Eq.~(\ref{gc2}) by an application of
the saddle-point approximation to the contour integral
\begin{equation}
\Omega (E,N)=\frac{1}{(2\pi i)^{2}}\oint_{\gamma _{z}}dz\oint_{\gamma
_{x}}dx \frac{\Xi (z,x)}{z^{N+1}x^{E/\hbar \omega +1}}\quad ,\qquad
x=e^{-\hbar \omega /k_B T},\qquad z=e^{\mu /k_B T},  \label{II43}
\end{equation}
where the contours of integration $\gamma _{z}$ and $\gamma _{x}$ include
$z=0$ and $x=0$, respectively. It is convenient to rewrite the function
under the integral in Eq.~(\ref{II43}) as $\exp [\varphi (z,x)]$, where
$\varphi (z,x)=\ln \Xi (z,x)-(N+1)\ln z-(E/\hbar \omega +1)\ln x$. Taking
the contours through the extrema (saddle points) $z_{0}$ and $x_{0}$
of $\varphi (z,x)$, and employing the usual Gaussian approximation, we get
for $N\rightarrow \infty $ and $E/\hbar \omega \rightarrow \infty$ the
following asymptotic formula: $\Omega (E,N)=[2\pi
D(z_{0},x_{0})]^{-1/2}\Xi (z_{0},x_{0})/ z_{0}^{N+1}x_{0}^{E/\hbar \omega
+1} $, where $D(z_{0},x_{0})$ is the determinant of the second derivatives
of the function $\varphi (z,x)$, evaluated at the saddle points~\cite{gr}.
For $N\rightarrow \infty$ and $E/\hbar \omega \rightarrow \infty $ the
function $\exp [\varphi (z,x)]$ is sharply peaked at $z_{0}$ and $x_{0}$,
which ensures good accuracy of the Gaussian approximation. However,
the standard result becomes incorrect for $E < N\varepsilon_1$, where
$\varepsilon_1 = \hbar \omega_{1}$ is the energy of the first excited
state (see \cite{gr,gh}); in this case, a more refinded version of
the saddle-point method is required~\cite{Schu46,Dinglebook,Holt99}.
We discuss this improved version of the saddle-point method in
Appendix \ref{ap:saddle}.

An accurate knowledge of the canonical partition function is helpful for
the calculation of the microcanonical condensate fluctuations by the
saddle point method, as has been demonstrated~\cite{gh} by a numerical
comparison with exact microcanonical simulations. In principle, the
knowledge of the canonical partition function allows us to calculate
thermodynamic and statistical equilibrium properties of the system in the
standard way (see, e.g., \cite{Ziff,la}). An important fact is that the
usual thermodynamic quantities (average energy, work, pressure, heat
capacities, etc.) and the average number of condensed atoms do not have
any infrared-singular contributions and do not depend on a choice of the
statistical ensemble in the thermodynamic (bulk) limit. However, the
variance and higher moments of the BEC fluctuations do have the infrared
anomalies and do depend on the statistical ensemble, so that their
calculation is much more involved and subtle.

\subsection{Exact recursion relation for the statistics of the number of
condensed atoms in an ideal Bose gas}

It is worth noting that there is one useful reference result in the theory
of BEC fluctuations, namely, an exact recursion relation for the statistics
of the number of condensed atoms in an ideal Bose gas. Although it does not
give any simple analytical answer or physical insight into the problem, it
can be used for ``exact" numerical simulations for traps containing a finite
number of atoms. This is very useful as a reference to be compared against
different approximate analytical formulas. This exact recursion relation for
the ideal Bose gas had been known for a long time~\cite{la}, and re-derived
independently by several authors~\cite{ww,bo,br,recursion}. In the canonical
ensemble, the probability to find $n_{0}$ particles occupying the
single-particle ground state is given by
\begin{equation}
\rho _{0}(n_{0})=\frac{Z_{N-n_{0}}(T)-Z_{N-n_{0}-1}(T)}{Z_{N}(T)};\, \quad
Z_{-1}\equiv 0\;.  \label{III76}
\end{equation}
The recurrence relation for the ideal Bose gas then states~\cite{la,bo}
\begin{equation}
Z_{N}(T)=\frac{1}{N}\sum_{k=1}^{N}Z_{1}(T/k)Z_{N-k}(T),\qquad
Z_{1}(T)=\sum_{\nu =0}^{\infty }e^{-\varepsilon _{\nu }/k_B T},\qquad
Z_{0}(T)=1,  \label{III77}
\end{equation}
which enables one to numerically compute the entire counting statistics
(\ref{III76}). Here $\nu$ stands for a set of quantum numbers which label a
given single-particle state, $\varepsilon _{\nu }$ is the associated energy,
and the ground-state energy is taken as $\varepsilon _{0}=0$ by convention.
For an isotropic, three-dimensional harmonic trap one has
\[
Z_1(T)=\sum_{n=0}^{\infty}\frac{1}{2} (n+2)(n+1)e^{-n\hbar\omega/k_BT}=
\frac{1}{\left( 1-e^{-\hbar\omega/k_BT}\right)^3},
\]
where $\frac{1}{2} (n+2)(n+1)$ is the degeneracy of the level
$\varepsilon _{n}=n\hbar\omega$.

In the microcanonical ensemble the ground-state occupation probability is
given by a similar formula
\begin{equation}
\rho _{0}^{MC}(n_{0})=\frac{\Omega (E,N-n_{0})-\Omega (E,N-n_{0}-1)}{\Omega
(E,N)};\,\quad \Omega (E,-1)\equiv 0\;,  \label{MC1}
\end{equation}
where the microcanonical partition function obeys the recurrence relation
\begin{equation}
\Omega (E,N)=\frac{1}{N}\sum_{k=1}^{N}\sum_{\nu =0}^{\infty }\Omega
(E-k\varepsilon _{\nu },N-k),\qquad \Omega (0,N)=1,\qquad \Omega (E>0,0)=0.
\label{MC2}
\end{equation}
For finite $E$ the sum over $\nu $ is finite because of $\Omega (E<0,N)=0$.

\subsection{Grand canonical approach}

Here we discuss the grand canonical ensemble, and show that it loses its
validity for the ideal Bose gas in the condensed region. Nevertheless,
reasonable approximate results can be obtained if the canonical-ensemble
constraint is properly incorporated in the grand canonical approach,
especially if we are not too close to $T_{c}$. In principle, the statistical
properties of BECs can be probed with light~\cite{Idzi00}. In particular, the
variance of the number of scattered photons may distinguish between the
Poisson and microcanonical statistics.

\subsubsection{Mean number of condensed particles in an ideal Bose gas}

The Bose-Einstein distribution can be easily derived from the density matrix
approach. Consider a collection of particles with the Hamiltonian $%
\hat{H}=\sum\limits_{k}\hat{a}_{k}^{\dagger }\hat{a}_{k}(\varepsilon_{k}-\mu
)$, where $\mu $ is the chemical potential. The equilibrium state of the
system is described by

\begin{equation}
\hat{\rho}=\frac{1}{Z}\exp (-\beta \hat{H})\quad
\end{equation}
where $Z=$Tr$\{\exp (-\beta \hat{H})\}=\prod_{k}(1-e^{-\beta (
\varepsilon_{k}-\mu )})^{-1}$. Then the mean number of particles with energy
$\varepsilon_{k}$ is
\begin{eqnarray}
\langle n_{k}\rangle &=&\text{Tr}\{\hat{\rho}\hat{a}_{k}^{\dagger }\hat{a}%
_{k}\}=(1-e^{-\beta (\varepsilon_{k}-\mu )})\sum_{n_{k}}\langle n_{k}|\hat{a}%
_{k}^{\dagger }\hat{a}_{k}e^{-\beta \hat{a}_{k}^{\dagger }\hat{a}%
_{k}(\varepsilon_{k}-\mu )}|n_{k}\rangle  \nonumber \\
&=&(1-e^{-\beta (\varepsilon_{k}-\mu )})\frac{d(1-e^{-\beta (
\varepsilon_{k}-\mu )})^{-1}}{d(-\beta (\varepsilon_{k}-\mu ))}=\frac{1}{%
\exp [\beta ( \varepsilon_{k}-\mu )]-1}.\allowbreak  \label{nk simple}
\end{eqnarray}

In the grand canonical ensemble the average condensate particle number $\bar{%
n}_{0}$ is determined from the equation for the total number of particles in
the trap,
\begin{equation}
N=\sum_{k=0}^{\infty }\bar{n}_{\mathbf{k}}=\sum_{\mathbf{k}=0}^{\infty }%
\frac{1}{\exp \beta (\varepsilon_{\mathbf{k}}-\mu )-1},  \label{c1x}
\end{equation}
where $\varepsilon_{\mathbf{k}}$ is the energy spectrum of the trap. In
particular, for the three dimensional (3D) isotropic harmonic trap we have $%
\varepsilon_{\mathbf{k}}=\hbar \Omega (k_{x}+k_{y}+k_{z})$. For simplicity,
we set $\varepsilon_0 =0$.

For 3D and 1D traps with non-interacting atoms, Eq.~(\ref{c1x}) was
studied by Ketterle and van Druten~\cite{kd}, and by Grossmann and
Holthaus~\cite{gh96,gh95}. They calculated the fraction of ground-state
atoms versus temperature for various $N$ and found that BEC also exists in
1D traps, where the condensation phenomenon looks very similar to the 3D
case. Later Herzog and Olshanii~\cite{Herzog} used the known analytical
formula for the canonical partition function of bosons trapped in a
1D harmonic potential~\cite{Auluck,Toda} and showed that the discrepancy
between the grand canonical and the canonical predictions for the 1D
condensate fraction becomes less than a few per cent for $N>10^{4}$. The
deviation decreases according to a $1/\ln N$ scaling law for fixed $T/T_{c}$.
In 3D the discrepancy is even less than in the 1D system~\cite{recursion}.
The ground state occupation number and other thermodynamic properties were
studied by Chase, Mekjian and Zamick~\cite{recursion} in the grand canonical,
canonical and microcanonical ensembles by applying combinatorial techniques
developed earlier in statistical nuclear fragmentation models. In such models
there are also constraints, namely the conservation of proton and neutron
number. The specific heat and the occupation of the ground state were found
substantially in agreement in all three ensembles. This confirms the essential
validity of the use of the different ensembles even for small groups of
particles as long as the usual thermodynamic quantities, which do not have
any infrared singular contributions, are calculated.

Let us demonstrate how to calculate the mean number of condensed particles for a particular example of a 3D isotropic harmonic trap. Following
Eq.~(\ref{c1x}), we can relate the chemical potential $\mu$ to the mean
number of condensed particles
$\bar{n}_{0}$ as $1+1/\bar{n}_{0}=\exp (-\beta \mu )$. Thus, we have
\begin{equation}
N=\sum_{\mathbf{k}=0}^{\infty }\langle n_{\mathbf{k}}\rangle =\sum_{\mathbf{k%
}=0}^{\infty }\frac{1}{(1+1/\bar{n}_{0})\exp \beta \varepsilon_{\mathbf{k}}-1%
}.  \label{N relates n0}
\end{equation}
The standard approach is to consider the case $N\gg 1$ and separate off the ground state so that Eq. (\ref{N relates n0}) approximately yields
\begin{equation}
N- \bar{n}_{0}=\sum_{k>0}\frac{1}{\exp (\beta \varepsilon _{k})-1} \; . \label{sum1}
\end{equation}
For an isotropic harmonic trap with frequency $\Omega$,
\begin{equation}
\sum_{k>0}\frac{1}{\exp (\beta \varepsilon _{k})-1}=\frac{1}{2}%
\sum_{n=1}^{\infty } \frac{(n+2)(n+1)}{\exp (\beta n\hslash \Omega )-1}%
\approx \frac{1 }{2}\int_{1}^{\infty }\frac{(x+2)(x+1)}{\exp (x\beta \hslash
\Omega )-1}dx.  \label{sum2}
\end{equation}
In the limit $k_{B}T\gg \hslash \Omega $ we obtain
\begin{equation}
\sum_{k>0}\frac{1}{\exp (\beta \varepsilon _{k})-1}\approx \frac{1}{2}%
\int_{0}^{\infty }\frac{x^{2}}{\exp (x\beta \hslash \Omega )-1}dx=\left(
\frac{k_{B}T}{\hslash \Omega }\right) ^{3}\zeta (3).  \label{sum3}
\end{equation}
Furthermore, when $T=T_{c}$ we take $\bar{n}_{0}=0$. Then Eqs.~(\ref{sum1}) and
(\ref{sum3}) yield
\begin{equation}
k_{B}T_{c}=\hslash \Omega \left( \frac{N}{\zeta (3)}\right)^{1/3}
\label{critt}
\end{equation}
and the temperature dependence of the mean condensate occupation with a cusp at $T=T_c$ in the thermodynamic limit
\begin{equation}
\bar{n}_{0} (T)=N\left( 1-\left( \frac{T}{T_{c}}\right)^{3}\right) .
\label{crit}
\end{equation}

\begin{figure}[tbp]
\bigskip
\centerline{\epsfxsize=0.65\textwidth\epsfysize=0.5\textwidth
\epsfbox{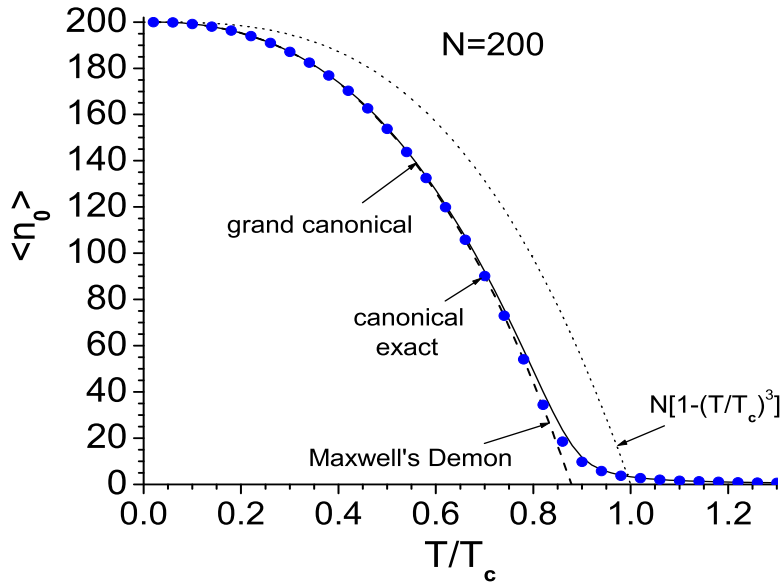}}
\par
\caption{Mean number of condensed particles as a function of temperature for
$N=200$. The solid line is the plot of the numerical solution of Eq.~(\ref{N
relates n0}). The exact numerical result for the canonical ensemble
(Eqs.~(\ref{III76}) and (\ref{III77})) is plotted as large dots.
The dashed line is obtained using Maxwell's demon ensemble approach, which
yields $\bar{n}%
_{0}=N-\sum_{k>0}^{\infty} 1/[\exp(\protect\beta\protect\varepsilon_k)-1]$. }
\label{n0}
\end{figure}

Figure \ref{n0} compares the numerical
solution of Eq.~(\ref{N relates n0})
(solid line) for $N=200$ with the numerical calculation of $\bar{n}_{0}(T)$
from the exact recursion relations in Eqs.~(\ref{III76}) and (\ref{III77})
in the canonical ensemble (large dots). Small dots show the plot of the solution (\ref{crit}),
which is valid only for a large number of particles, $N$.
Obviously, more accurate solution of the equation for the mean number of condensed particles (\ref{N relates n0})
in a trap with a finite number of particles does not show the cusp.
In Appendix \ref{ap:mean} we derive an
analytical solution of Eq. (\ref{N relates n0}) for $\bar{n}_{0}(T)$ valid for $\bar{n}_{0}(T)\gg 1$.
One can see that for the average particle number both ensembles yield very close answers. However, this
is not the case for the BEC fluctuations.

\subsubsection{Condensate fluctuations in an ideal Bose gas}

Condensate fluctuations are characterized by the central moments $\langle
(n_{0}-\langle n_{0}\rangle )^{s}\rangle =\sum_{r}\binom{s}{r}\langle
n_{0}^{r}\rangle \langle n_{0}\rangle ^{s-r}$. The first few of them are
\begin{equation}
\langle (n_{0}-\bar{n}_{0})^{2}\rangle =\langle n_{0}^{2}\rangle -\langle
n_{0}\rangle ^{2} ,
\end{equation}
\begin{equation}
\langle (n_{0}-\bar{n}_{0})^{3}\rangle =\langle n_{0}^{3}\rangle -3\langle
n_{0}^{2}\rangle \langle n_{0}\rangle +2\langle n_{0}\rangle ^{3} ,
\end{equation}
\begin{equation}
\langle (n_{0}-\bar{n}_{0})^{4}\rangle =\langle n_{0}^{4}\rangle \allowbreak
-4\langle n_{0}^{3}\rangle \langle n_{0}\rangle +6\langle n_{0}^{2}\rangle
\langle n_{0}\rangle ^{2}-3\langle n_{0}\rangle ^{4}.
\end{equation}

At arbitrary temperatures, BEC fluctuations in the canonical ensemble can be
described via a stochastic variable $n_{0}=N-\sum_{k\neq 0}n_{k}$ that
depends on and is complementary to the sum of the independently fluctuating
numbers~$n_{k}$, $k\neq 0$, of the excited atoms. In essence, the canonical
ensemble constraint in Eq.~(\ref{CEconstraint}) eliminates one degree of
freedom (namely, the ground-state one) from the set of all independent
degrees of freedom of the original grand canonical ensemble, so that only
transitions between the ground and excited states remain independently
fluctuating quantities. They describe the canonical-ensemble quasiparticles,
or excitations, via the creation and annihilation operators $\hat{\beta}^{+}$
and $\hat{\beta}$ (see Sections V and VI below).

At temperatures higher than $T_{c}$ the condensate fraction is small and one
can approximately treat the condensate as being in contact with a reservoir
of non-condensate particles. The condensate exchanges particles with the big
reservoir. Hence, the description of particle fluctuations in the grand
canonical picture, assuming that the number of atoms in the ground state
fluctuates independently from the numbers of excited atoms, is accurate in
this temperature range.

At temperatures close to or below $T_{c}$ the situation becomes completely
different. One can say that at low temperatures, $T \ll T_{c}$, the picture is
opposite to the picture of the Bose gas above the BEC phase transition at
$T>T_{c}$. The canonical-ensemble quasiparticle approach, suggested in
Refs.~\cite{KKS-PRL,KKS-PRA} and valid both for the ideal and interacting
Bose gases, states that at low temperatures the non-condensate particles can
be treated as being in contact with the big reservoir of the $n_{0}$
condensate particles. This idea had previously been spelled out by
Fierz~\cite{Fierz} and been used by Politzer~\cite{pol}, and was then
employed for the construction of the ``Maxwell's demon ensemble''~\cite{Navez}, named so since a permanent selection of the excited (moving) atoms from the ground state (static) atoms is a problem resembling the famous Maxwell's Demon problem in statistical physics.

Note that although this novel statistical concept can be studied approximately
with the help of the Bose-Einstein expression~(\ref{nk simple}) for the mean
number of excited (only excited, $k\neq 0$!) states with some new chemical
potential~$\mu$, it describes the canonical-ensemble fluctuations and is
essentially different from the standard grand canonical description of
fluctuations in the Bose gas. Moreover, it is approximately valid only if we
are not too close to the critical temperature $T_{c}$, since otherwise the
``particle reservoir'' is emptied. Also, the fluctuations obtained from the
outlined ``grand'' canonical approach (see Eqs.~(\ref{d3})--(\ref{2D1D})
and (\ref{mom2})--(\ref{mom4})), an approach complementary to the grand
canonical one, provide an accurate description in this temperature range in
the thermodynamic limit, but do not take into account all mesoscopic effects,
especially near $T_c$ (for more details, see Sections IV and V). Thus,
although the mean number of condensed atoms $\bar{n}_{0}$ can be found from
the grand canonical expression used in Eq.~(\ref{c1x}), we still need to
invoke the conservation of the total particle number $N=n_{0}+n$ in order to
find the higher moments of condensate fluctuations, $\langle n_{0}^{r}\rangle$.
Namely, we can use the following relation between the central moments of the
$m$th order of the number of condensed atoms, and that of the non-condensed
ones: $\langle (n_{0}-\bar{n}_{0})^{m}\rangle =(-1)^{m}\langle (n-%
\bar{n} )^{m}\rangle $. As a result, at low enough temperatures one can
approximately write the central moments in the well-known canonical form via
the cumulants in the ideal Bose gas (see Eqs.~(\ref{III1}), (\ref{III2}),
(\ref{III16}), (\ref{III19}) and Section V below for more details):
\begin{equation}
\langle (n_{0}-\bar{n}_{0})^{2}\rangle =\kappa _{2}=\tilde{\kappa}_{2}+%
\tilde{\kappa}_{1}\approx \sum_{k>0}(\bar{n}_{k}^{2}+\bar{n}_{k}),
\label{d3}
\end{equation}
\begin{equation}
\langle (n_{0}-\bar{n}_{0})^{3}\rangle =-\kappa _{3}=-(\tilde{\kappa}_{3}-3%
\tilde{\kappa}_{2}-\tilde{\kappa}_{1})\approx -\sum_{k>0}(2\bar{n}_{k}^{3}+3%
\bar{n}_{k}^{2}+\bar{n}_{k}),  \label{a16}
\end{equation}
\begin{equation}
\langle (n_{0}-\bar{n}_{0})^{4}\rangle =\kappa _{4}+3\kappa _{2}^{2}=\tilde{%
\kappa}_{4}+6\tilde{\kappa}_{3}+7\tilde{\kappa}_{2}+\tilde{\kappa}_{1}+3(%
\tilde{\kappa}_{2}+\tilde{\kappa}_{1})^{2}\approx \sum_{k>0}(6\bar{n}%
_{k}^{4}+12\bar{n}_{k}^{3}+7\bar{n}_{k}^{2}+\bar{n}_{k})+3\left[ \sum_{k>0}(\bar{n}_{k}^{2}+\bar{n}_{k})\right]^2
.  \label{a5}
\end{equation}
These same equations can also be derived by means of the straightforward
calculation explained in Appendix~\ref{ap:fluct}.

\begin{figure}[tbp]
\center \epsfxsize=13cm\epsffile{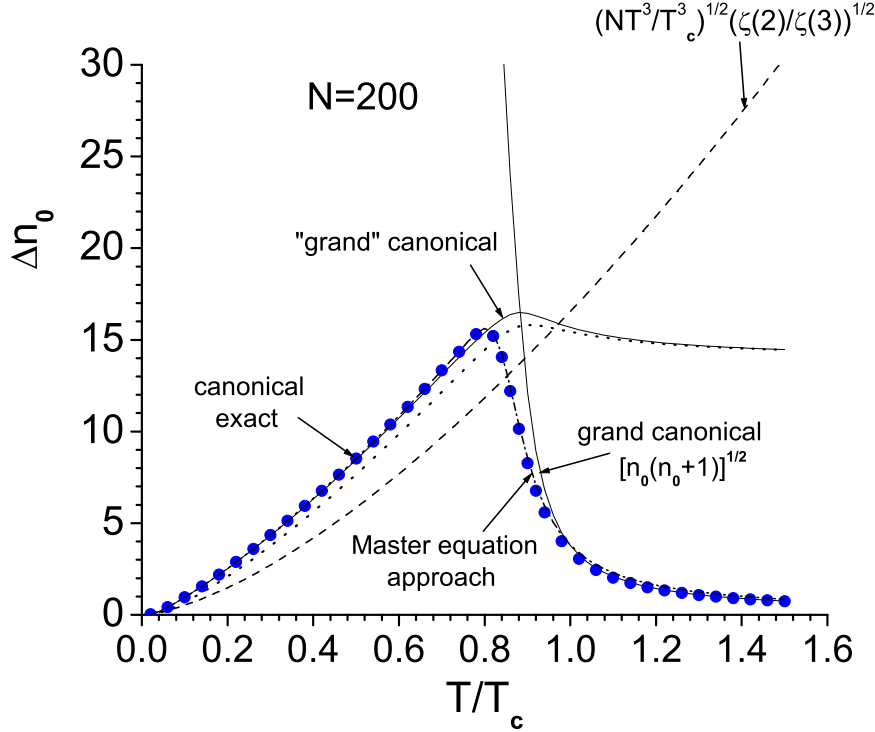}
\caption{Variance $\langle (n_{0}-\langle n_{0}\rangle )^{2}\rangle ^{1/2}$
of the condensate particle number for $N = 200$. The solid line is the
``grand'' canonical result obtained from Eq.~(\ref{a1}) and the numerical
solution of $\langle n_{0}\rangle $ from Eq.~(\ref{N relates n0}). Exact
numerical data for the canonical ensemble (Eqs.~(\ref{III76}) and
(\ref{III77})) are shown as dots. The asymptotic Politzer
approximation~\cite{pol}, given by Eq.~(\ref{a6}), is shown by the dashed
line, while small dots result from Eq. (\ref{Dn02 analytic}). The dash-dotted
line is obtained from the master equation approach (see Eq.~(\ref{II54})
below).}
\label{n2v1}
\end{figure}

Combining the hallmarks of the grand canonical approach, namely, the value
of the chemical potential $\mu =-\beta ^{-1}\ln (1+1/\bar{n}_{0})$ and the
mean non-condensate occupation
$\bar{n}_{k}=\{\exp [\beta (E_{k}-\mu )]-1\}^{-1}$, with the Eq.~(\ref{d3})
describing the fluctuations in the canonical ensemble, we find the BEC
variance
\begin{equation}
\Delta n_{0}^{2}\equiv \langle (n_{0}-\bar{n}_{0})^{2}\rangle
=\sum_{k>0}\left\{ \frac{1}{\left[ \exp (\beta E_{k})\left( 1+\frac{1}{\bar{n%
}_{0}}\right) -1\right] ^{2}}+\frac{1}{\exp (\beta E_{k})\left( 1+\frac{1}{%
\bar{n}_{0}}\right) -1}\right\} .  \label{a1}
\end{equation}
In the case $k_{B}T\gg \hbar \Omega $, this Eq.~(\ref{a1}) can be evaluated
analytically, as is shown in Appendix \ref{ap:variance}. The variance up
to second order in $1/{\bar{n}}_{0}$ from Eq.~(\ref{Dn02 analytic}) is
\begin{equation}
\Delta n_{0}\approx \sqrt{\frac{N}{\zeta (3)}\left( \frac{T}{T_{c}}\right)
^{3}}\left[ \frac{\pi ^{2}}{6}-\frac{1}{\bar{n}_{0}}(1+\ln \bar{n}_{0})+%
\frac{1}{\bar{n}_{0}^{2}}\left( \frac{1}{2}\ln \bar{n}_{0}-\frac{1}{4}%
\right) \right] ^{1/2}.  \label{a6x}
\end{equation}
The leading term in this expression yields Politzer's result~\cite{pol},
\begin{equation}
\Delta n_{0}\approx \sqrt{\frac{\zeta (2)N}{\zeta (3)}\left( \frac{T}{T_{c}}%
\right) ^{3}}\approx 1.17\sqrt{N\left( \frac{T}{T_{c}}\right) ^{3}} \; ,
\label{a6}
\end{equation}
plotted as a dashed line in Fig.~\ref{n2v1}, where $\zeta (2)=\frac{\pi ^{2}%
}{6}\approx \allowbreak 1.\,\allowbreak 644\,9$ and $\zeta (3)\approx 1.202\,1$
(compare this with D.~ter Haar's~\cite{DterHaar} result $\Delta n_{0}\approx
N\left( \frac{T}{T_{c}}\right)^{3}$, which is missing the square root). The
same formula was obtained later by Navez et al.\ using the Maxwell's demon
ensemble~\cite{Navez}. For the microcanonical ensemble the Maxwell's
demon approach yields \cite{Navez}
\begin{equation}
\Delta n_{0}\approx \sqrt{N\left( \frac{\zeta (2)}{\zeta (3)}-\frac{3\zeta
(3)}{4\zeta (4)}\right) \left( \frac{T}{T_{c}}\right) ^{3}}\approx 0.73\sqrt{%
N\left( \frac{T}{T_{c}}\right) ^{3}};  \label{mca6}
\end{equation}
higher order terms have been derived in Ref.~\cite{Holt99}.
The microcanonical fluctuations are smaller than the canonical ones due to
the additional energy constraint. For 2D and 1D harmonic traps the Maxwell's
demon approach leads to \cite{gh97b,ww}
\begin{equation}
\Delta n_{0}\sim \sqrt{N}\frac{T}{T_{c}}\sqrt{\ln \frac{T_{c}}{T}}
\quad \mbox{and} \quad
\Delta n_{0}\sim \frac{T}{T_{c}}N \; , \label{2D1D}
\end{equation}
respectively.

\begin{figure}[tbp]
\center \epsfxsize=13cm\epsffile{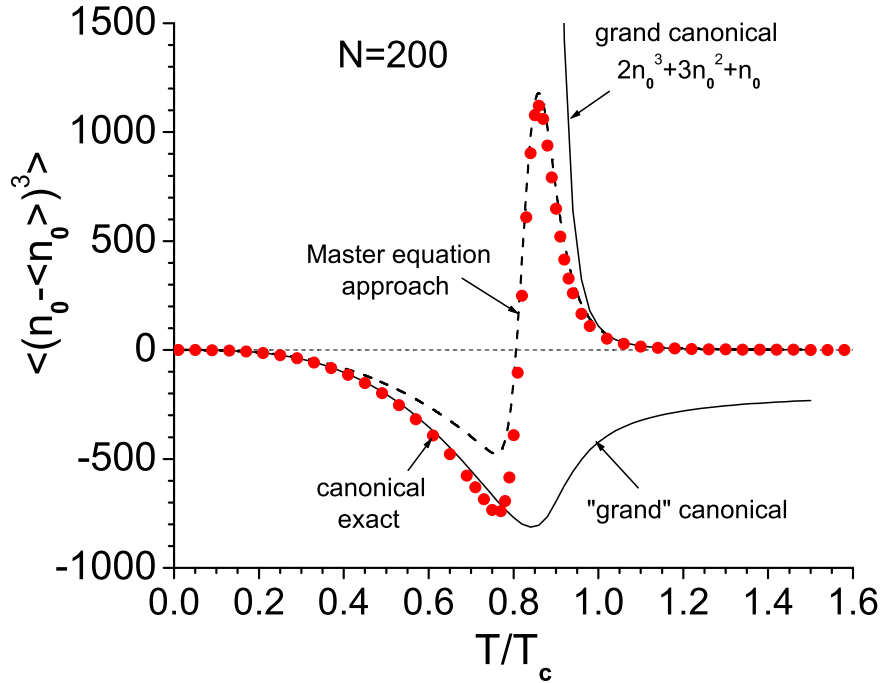}
\caption{The third central moment $\langle (n_{0}-\bar{n}_{0})^{3}\rangle$
for $N=200$, plotted using Eqs.~(\ref{a16}) and (\ref{N relates n0}). Exact
numerical data for the canonical ensemble (Eqs.~(\ref{III76})
and (\ref{III77})) are shown as dots. The dashed line is the result of
master equation approach (see Eq.~(\ref{third mom}) below).}
\label{n3}
\end{figure}

Figure \ref{n2v1} shows the BEC variance $\Delta n_{0}(T)$ as a function
of temperature for a 3D trap with the total particle number $N=200$. The
``grand'' canonical curve refers to Eq.~(\ref{a1}) and shows good agreement
for $T<T_{c}$ with the numerical result for $\Delta n_{0}(T)$ (large dots),
obtained within the exact recursion relations (\ref{III76}) and (\ref{III77})
for the canonical ensemble. The plot of Politzer's asymptotic
formula~(\ref{a6}) (dashed line) does not give good agreement, since the
particle number considered here is fairly low. We also plot
$\sqrt{\bar{n}_{0}(\bar{n}_{0}+1)}$, which is the expression for the
condensate number fluctuations $\Delta n_{0}$ in the grand canonical ensemble;
it works well above $T_{c}$. Figure~\ref{n3} shows the third central moment
$\langle (n_{0}-\bar{n}_{0})^{3}\rangle$ as a function of temperature for
the total particle number $N=200$, plotted using Eqs.~(\ref{a16}) and
(\ref{N relates n0}). Dots are the exact numerical result obtained within
the canonical ensemble. We also plot the standard grand canonical formula
$2\bar{n}_{0}^{3} + 3\bar{n}_{0}^{2}+\bar{n}_{0}$, which again works well
only above $T_{c}$.

At high temperatures the main point, of course, is the validity of the
standard grand canonical approach where the average occupation of the ground
state alone gives a correct description of the condensate fluctuations, since
the excited particles constitute a valid ``particle reservoir''. Condensate
fluctuations obtained from the ``grand'' canonical approach for the canonical
ensemble quasiparticles and from the standard grand canonical ensemble provide
an accurate description of the canonical-ensemble fluctuations at temperatures
not too close to the narrow crossover region between low (Eq.~(\ref{a1})) and
high ($\sqrt{\bar{n}_{0}(\bar{n}_{0}+1)}$) temperature regimes; i.e., in the
region not too near $T_c$. In this crossover range both approximations fail,
since the condensate and the non-condensate fractions have comparable numbers
of particles, and there is no any valid particle reservoir. Note, however,
that a better description, that includes mesoscopic effects and works in the
whole temperature range, can be obtained using the condensate master equation,
as shown in Section IV; see, e.g., Fig.~\ref{VVKIIm1234}. Another
(semi-analytic) technique, the saddle-point method, is discussed in
Appendix~\ref{ap:saddle}.

\section{Dynamical master equation approach and laser phase-transition
analogy}

One approach to the canonical statistics of ideal Bose gases, presented in
\cite{MOS99} and developed further in \cite{KSZZ}, consists in setting up a
master equation for the condensate and finding its equilibrium solution.
This approach has the merit of being technically lucid and physically
illuminating. Furthermore, it reveals important parallels to the quantum
theory of the laser. In deriving that master equation, the approximation of
detailed balance in the excited states is used, in addition to the
assumption that given an arbitrary number $n_{0}$ of atoms in the
condensate, the remaining $N-n_{0}$ excited atoms are in an equilibrium
state at the prescribed temperature $T$.

In Section IV.B we summarize the master equation approach against the
results provided by independent techniques. In Section IV.A we motivate our
approach by sketching the quantum theory of the laser with special emphasis
on the points relevant to BEC.

\subsection{Quantum theory of the laser}

In order to set the stage for the derivation of the BEC master equation, let
us remind ourselves of the structure of the master equations for a few basic
systems that have some connection with $N$ particles undergoing Bose-Einstein
condensation while exchanging energy with a thermal reservoir.

\subsubsection{Single mode thermal field}

The dissipative dynamics of a small system $(s)$ coupled to a large
reservoir $(r)$ is described by the reduced density matrix equation up to
second order in the interaction Hamiltonian $\hat{V}_{sr}$%
\begin{equation}
\frac{\partial }{\partial t}\hat{\rho}_{s}(t)=-\frac{1}{\hbar ^{2}}%
Tr_{r}\int_{0}^{t}[\hat{V}_{sr}(t),[\hat{V}_{sr}(t^{\prime }),\hat{\rho}_s
(t)\otimes \hat{\rho}_{r}^{th}]]dt^{\prime }  \label{reduced master eqt}
\end{equation}
where $\hat{\rho}_{r}^{th}$ is the density matrix of the thermal reservoir,
and we take the Markovian approximation.

We consider the system as a single radiation mode cavity field $(f)$ of
frequency $\nu $ coupled to a reservoir $(r)$ of two-level thermal atoms,
and show how the radiation field thermalizes. The multiatom Hamiltonian in
the interaction picture is
\begin{equation}
\hat{V}_{fr}=\hbar \sum\limits_{j}g_{j}(\hat{\sigma}_{j}\hat{a}^{\dagger
}e^{i(\nu -\omega _{j})t}+adj.)  \label{Vfr}
\end{equation}
where $\hat{\sigma}_{j}=|b_{j}\rangle \langle a_{j}|$ is the atomic (spin)
operator of the $j$-th particle corresponding to the downward transition, $%
\hat{a}^{\dagger }$ is the creation operator for the single mode field and $%
g_{j}$ is the coupling constant. The reduced density matrix equation for the
field can be obtained from Eqs. (\ref{reduced master eqt}) and (\ref{Vfr}).
By using the density matrix for the thermal ensemble of atoms
\begin{equation}
\hat{\rho}_{r}^{th}=\prod\limits_{j}(|a_{j}\rangle \langle a_{j}|e^{-\beta
E_{a,j}}+|b_{j}\rangle \langle b_{j}|e^{-\beta E_{b,j}})/Z_{j}  \label{pth}
\end{equation}
where $Z_{j}=e^{-\beta E_{a,j}}+e^{-\beta E_{b,j}}$, we obtain
\begin{equation}
\frac{\partial }{\partial t}\hat{\rho}_{f}=-\frac{1}{2}\sum\limits_{j}\kappa
_{j}\{P_{a_{j}}(\hat{a}\hat{a}^{\dagger }\hat{\rho}_{f}-2\hat{a}^{\dagger }%
\hat{\rho}_{f}\hat{a}+\hat{\rho}_{f}\hat{a}\hat{a}^{\dagger })+P_{b_{j}}(%
\hat{a}^{\dagger }\hat{a}\hat{\rho}_{f}-2\hat{a}\hat{\rho}_{f}\hat{a}%
^{\dagger }+\hat{\rho}_{f}\hat{a}^{\dagger }\hat{a})\}  \label{dpf/dt}
\end{equation}
where $P_{x_{j}}=e^{-\beta E_{x_{j}}}/Z_{j}$ with $x=a,b$ and the
dissipative constant is $\frac{1}{2}\kappa_j =$Re\{$g_{j}^{2}%
\int_{0}^{t}e^{i(\nu -\omega _{j})(t-t^{\prime })}dt^{\prime }$\}.
Note that the same structure of the master equation is obtained
for a phonon bath modelled as a collection of harmonic
oscillators, as shown in Appendix \ref{ap:reservoir}.

Taking the diagonal matrix elements $\rho_{n,n}(t)=\langle n|\hat{\rho}%
_{f}|n\rangle $ of Eq. (\ref{dpf/dt}), we have
\begin{equation}
\frac{\partial }{\partial t}\rho_{n,n}(t)=-\kappa
P_{a}\{(n+1)\rho_{n,n}(t)-n\rho_{n-1,n-1}(t)\}-\kappa
P_{b}\{n\rho_{n,n}(t)-(n+1)\rho_{n+1,n+1}(t)\} ,
\label{master eqt nn thermal}
\end{equation}
where $\kappa$ is $\kappa_j$ times a density of states factor and $%
\rho_{n,n} $.

The steady state equation gives
\begin{equation}
\rho_{n,n} \equiv p_{n}=e^{-n\beta \hbar \omega }p_{0} .  \label{xxx0}
\end{equation}

From $\sum\limits_{n=0}^{\infty }p_{n}=1$ we obtain the thermal photon
number distribution
\begin{equation}
p_{n}=e^{-n\beta \hbar \omega }(1-e^{-\beta \hbar \omega })
\end{equation}
which is clearly an exponentially decaying photon number distribution.

\begin{figure}[tbp]
\center \epsfxsize=9cm\epsffile{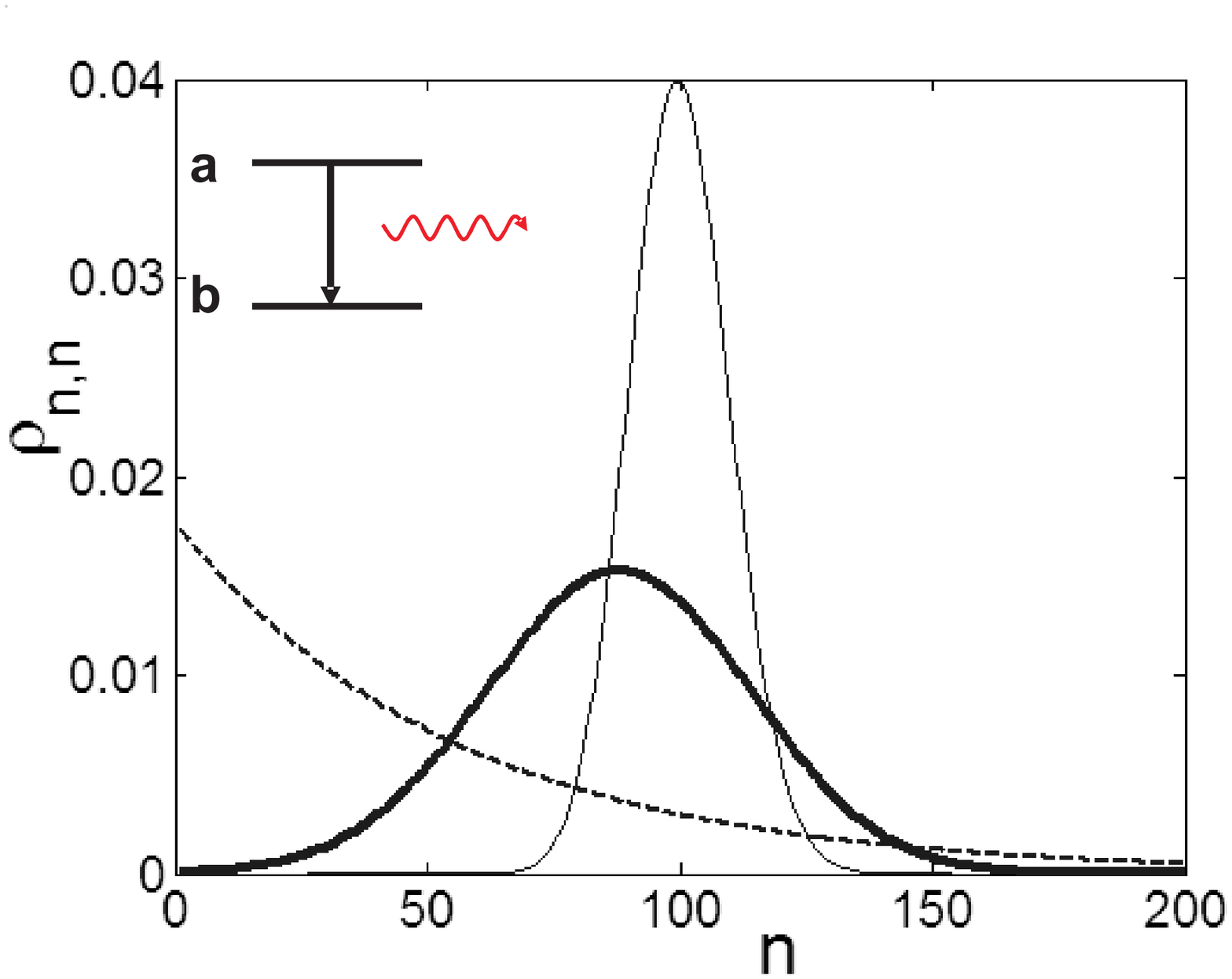}
\caption{Photon number distributions for a) thermal photons plotted from Eq.
(\ref{xxx0}) (dashed line), b) coherent state (Poissonian) (thin solid
line), and c) He-Ne laser plotted using Eq. (\ref{pnlaser}) (thick solid
line). Insert shows an atom making a radiation transition.}
\label{thermallaser}
\end{figure}

\subsubsection{Coherent state}

Consider the interaction of a single mode field with a classical current $%
\mathbf{J}$ described by
\begin{eqnarray}
\hat{V}_{\text{coh}}(t) &=&\int_{V}\mathbf{J(\mathbf{r},}t\mathbf{)\cdot
\hat{A}}(\mathbf{r},t)d^{3}r  \nonumber \\
&=&\hbar (j(t)\hat{a}+j^{\ast }(t)\hat{a}^{\dagger })  \label{Vcoh}
\end{eqnarray}
where the complex time dependent coefficient is $j(t)=\frac{A_{0}}{\hbar }%
\int_{V}\mathbf{J(\mathbf{r},}t\mathbf{)\cdot }\hat{x}e^{i(\mathbf{k}\cdot
\mathbf{r-\nu }t)}d^{3}r$ and $A_{o}$ is the amplitude vector potential $%
\mathbf{\hat{A}(\mathbf{r},}t\mathbf{)}$ of the single mode field, assumed
to be polarized along $x$ axis. An example of such interaction is a
klystron. Clearly, the unitary time evolution of $\hat{V}_{\text{coh}}$ is
in the form of a displacement operator $\exp (\alpha ^{\ast}(t)\hat{a}%
-\alpha (t)\hat{a}^{\dagger })$ associated with a coherent state when
dissipation is neglected.

\begin{figure}[tbp]
\center \epsfxsize=11cm\epsffile{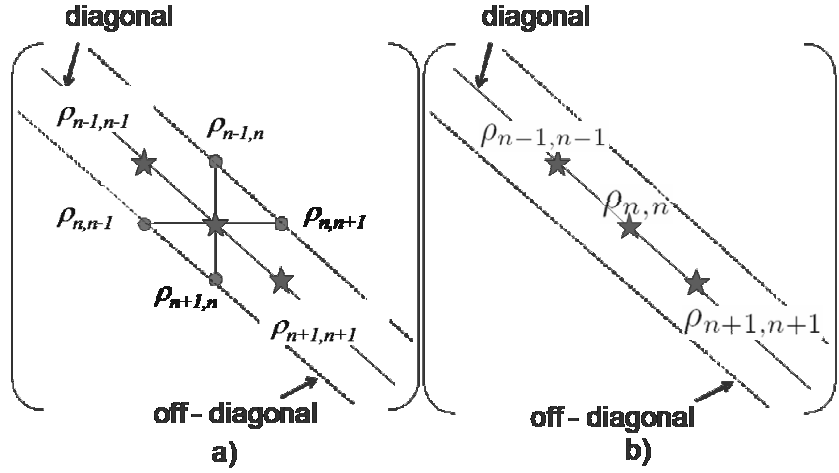}
\caption{Diagonal (star) and off-diagonal (circle) density matrix elements
that govern temporal dynamics in a) klystron and b) thermal field.}
\label{densop}
\end{figure}

Thus, the density matrix equation for a klystron including coupling with a
thermal bath is
\begin{equation}
\frac{\partial \hat{\rho}_{f}(t)}{\partial t}=\frac{1}{i\hbar }[\hat{V}_{%
\text{coh}}(t),\hat{\rho}_{f}(t)]-\frac{1}{\hbar ^{2}}Tr_{r}\int_{0}^{t}[%
\hat{V}_{fr}(t),[\hat{V}_{fr}(t^{\prime }),\hat{\rho}_{f}(t)\otimes \hat{\rho%
}_{r}^{th}]]dt^{\prime }  \label{master eqt coh}
\end{equation}
where $\hat{V}_{fr}$ is given by Eq. (\ref{Vsr}). The second term of Eq. (%
\ref{master eqt coh}) describes the damping of the single mode field given
by Eq. (\ref{Louivillean}). By taking the matrix elements of Eq. (\ref
{master eqt coh}), after a bit of analysis, we have
\[
\frac{\partial \rho_{n,n^{\prime }}(t)}{\partial t}=-i(j(t)\sqrt{n+1}%
\rho_{n+1,n^{\prime }}+j^{\ast }(t)\sqrt{n}\rho_{n-1,n^{\prime }}-j(t)\sqrt{%
n^{\prime }}\rho_{n,n^{\prime }-1}-j^{\ast }(t)\sqrt{n^{\prime }+1}%
\rho_{n,n^{\prime }+1})
\]
\begin{equation}
-\frac{1}{2}\mathcal{C}[(n+n^{\prime })\rho_{n,n^{\prime }}-2\sqrt{%
(n+1)(n^{\prime }+1)}\rho_{n+1,n^{\prime }+1}]-\frac{1}{2}\mathcal{D}%
[(n+1+n^{\prime }+1)\rho_{n,n^{\prime }}-2\sqrt{nn^{\prime }}\hat{\rho}%
_{n-1,n^{\prime }-1}]  \label{master eqt nn' coh}
\end{equation}
Clearly, the first line of Eq. (\ref{master eqt nn' coh}) shows that the
change in the photon number is effected by the off-diagonal field density
matrix or the coherence between two states of different photon number. On
the other hand, the damping mechanism only causes a change in the photon
number through the diagonal matrix element or the population of the number
state. This is depicted in Fig. \ref{densop}.

It can be shown that the solution of Eq.~(\ref{master eqt nn' coh}) for $%
\mathcal{D}=0$ is the matrix element of a coherent state $|\beta \rangle
=e^{|\beta |^{2}/2}\sum\limits_{n}\frac{\beta ^{n}}{\sqrt{n!}}|n\rangle $,
i.e.
\begin{equation}
\rho_{n,n^{\prime }}(t)=\langle n|\{|\beta \rangle \langle \beta
|\}|n^{\prime }\rangle =\frac{\beta (t)^{n}\beta ^{\ast }(t)^{n^{\prime }}}{%
\sqrt{n!}\sqrt{n^{\prime }!}}e^{-|\beta (t)|^{2}}  \label{pnn' solution}
\end{equation}
where $\beta (t)=\alpha (t)-\frac{1}{2}\mathcal{C}\int_{0}^{t}\alpha
(t^{\prime })dt^{\prime }$. This can be verified if we differentiate Eq. (%
\ref{pnn' solution})
\[
\frac{d\rho_{n,n^{\prime }}}{dt}=\{\frac{d\alpha (t)}{dt}-\frac{1}{2}%
\mathcal{C}\alpha (t)\}\sqrt{n}\rho_{n-1,n^{\prime }}+\sqrt{n^{\prime }}\{%
\frac{d\alpha ^{\ast }(t)}{dt}-\frac{1}{2}\mathcal{C}\alpha ^{\ast
}(t)\}\rho_{n,n^{\prime }-1}
\]
\begin{equation}
-\{\frac{d\alpha (t)}{dt}-\frac{1}{2}\mathcal{C}\alpha (t)\}\sqrt{n^{\prime
}+1}\rho_{n,n^{\prime }+1}-\{\frac{d\alpha ^{\ast }(t)}{dt}-\frac{1}{2}%
\mathcal{C}\alpha ^{\ast }(t)\}\sqrt{n+1}\rho_{n+1,n^{\prime }}  \label{diff}
\end{equation}
and using
\begin{eqnarray}
\sqrt{n}\rho_{n,n^{\prime }} &=&\rho_{n-1,n^{\prime }}\text{, \ }\sqrt{%
n^{\prime }}\rho_{n,n^{\prime }}=\rho_{n,n^{\prime }-1},  \nonumber \\
\sqrt{n+1}\rho_{n+1,n^{\prime }+1} &=&\rho_{n,n^{\prime }+1}\text{, \ }\sqrt{%
n^{\prime }+1}\rho_{n+1,n^{\prime }+1}=\rho_{n+1,n^{\prime }}
\label{relations}
\end{eqnarray}
where $\frac{d\alpha (t)}{dt}=-ij^{\ast }(t)$ is found by comparing Eq. (\ref
{diff}) with Eq. (\ref{master eqt nn' coh}).

\subsubsection{Laser master equation}

The photon number equation (\ref{master eqt nn thermal}) for thermal field
is linear in photon number, $n$, and it describes only the thermal damping
and pumping due to the presence of a thermal reservoir. Now, we introduce a
laser pumping scheme to drive the single mode field and show how the
atom-field nonlinearity comes into the laser master equation.

\begin{figure}[tbp]
\center \epsfxsize=10cm\epsffile{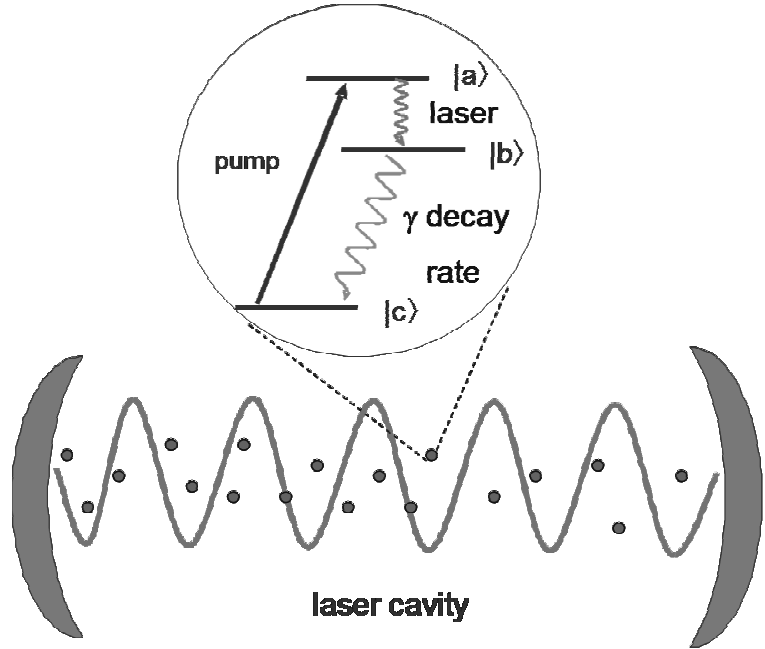}
\caption{Typical setup of a laser showing an ensemble of atoms driving a
single mode field. A competition between lasing and dissipation through
cavity walls leads to a detailed balance.}
\label{lasing}
\end{figure}

We consider a simple three level system where the cavity field couples level
$a$ and level $b$ of lasing atoms in a molecular beam injected into a cavity
in the excited state at rate $r$ (see Fig. \ref{lasing}). The atoms undergo
decay from level $b$ to level $c$. The pumping mechanism from level $c$ up
to level $a$ can thus produce gain in the single mode field. As shown in \cite
{sz}, we find
\begin{equation}
\left( \frac{\partial \rho_{n,n}(t)}{\partial t}\right) _{\text{gain}}=-\{%
\mathcal{A}(n+1)-\mathcal{B}(n+1)^{2}\}\rho_{n,n}+\{\mathcal{A}n-\mathcal{B}%
n^{2}\}\rho_{n-1,n-1} ,  \label{dpnn/dt gain}
\end{equation}
where $\mathcal{A}=\frac{2rg^{2}}{\gamma ^{2}}$ is the linear gain
coefficient and $\mathcal{B}=\frac{4g^{2}}{\gamma ^{2}}\mathcal{A}$ is the
self-saturation coefficient. Here $g$ is the atom-field coupling constant
and $\gamma$ is the $b \to c$ decay rate. We take the damping of the field
to be
\begin{equation}
\left( \frac{\partial \rho_{n,n}(t)}{\partial t}\right) _{\text{loss}}=-
\mathcal{C}n \rho_{n,n} + \mathcal{C}(n+1)\rho_{n+1,n+1} .
\label{dpnn/dt loss}
\end{equation}

Thus, the overall master equation for the laser is
\begin{equation}
\frac{\partial \rho_{n,n}(t)}{\partial t} =-\{\mathcal{A}(n+1)-\mathcal{B}
(n+1)^{2}\}\rho_{n,n}+\{\mathcal{A}n-\mathcal{B}n^{2}\}\rho_{n-1,n-1} -%
\mathcal{C}n\rho_{n,n}+\mathcal{C}(n+1)\rho_{n+1,n+1} ,
\label{master eqt nn laser}
\end{equation}
which is valid for small $\mathcal{B}/\mathcal{A} \ll 1$.

We emphasize that the nonlinear process associated with $\mathcal{B}$ is a
key physical process in the laser (but not in a thermal field) because the
laser field is so large.

We proceed with detailed balance equation between level $n-1$ and $n$
\begin{eqnarray}
-\{ \mathcal{A}(n+1)-\mathcal{B}(n+1)^{2}\}p_{n}+\mathcal{C}(n+1)p_{n+1}
&=&0 , \\
\{ \mathcal{A} n-\mathcal{B}n^{2}\}p_{n-1}-\mathcal{C} np_{n} &=&0 .
\end{eqnarray}

By iteration of $p_{k}=\frac{(\mathcal{A}-\mathcal{B}k}{\mathcal{C}}p_{k-1}$%
, we have
\begin{equation}
p_{n}=p_{0}\prod\limits_{k=1}^{n}\frac{\mathcal{A}-\mathcal{B}k}{\mathcal{C}}
\label{pnlaser}
\end{equation}
where $p_{0}=1/(1+\sum\limits_{n=1}^{\infty }\prod\limits_{k=1}^{n} \frac{%
\mathcal{A}-\mathcal{B}k}{\mathcal{C}})$ follows from $\sum\limits_{n=0}^{%
\infty }p_{n}=1.$ Eq. (\ref{pnlaser}) is plotted in Fig. \ref{thermallaser}.
There we clearly see that the photon statistics of, e.g. a He-Ne, laser is
not Poissonian $p_{n}=\langle n\rangle ^{n}e^{-\langle n\rangle }/n!$, as
would be expected for a coherent state.

\subsection{Laser phase-transition analogy}

Bose-Einstein condensation of atoms in a trap has intriguing similarities
with the threshold behavior of a laser which also can be viewed as a kind of
a phase transition \cite{ds,ph-tra}. Spontaneous formation of a long range
coherent order parameter, i.e., macroscopic wave function, in the course of
BEC second order phase transition is similar to spontaneous generation of a
macroscopic coherent field in the laser cavity in the course of lasing. In
both systems stimulated processes are responsible for the appearance of the
macroscopic order parameter. The main difference is that for the Bose gas in
a trap there is also interaction between the atoms which is responsible for
some processes, including 
stimulated effects in BEC. Whereas for the laser there are two subsystems,
namely the laser field and the active atomic medium. The crucial point for
lasing is the interaction between the field and the atomic medium which is
relatively small and can be treated perturbatively. Thus, the effects of
different interactions in the laser system are easy to trace and relate to
the observable characteristics of the system. This is not the case in BEC
and it is more difficult to separate different effects.

As is outlined in the previous subsection, in the quantum theory of laser,
the dynamics of laser light is conveniently described by a master equation
obtained by treating the atomic (gain) media and cavity dissipation (loss)
as reservoirs which when ``traced over" yield the coarse grained equation of
motion for the reduced density matrix for laser radiation. In this way we
arrive at the equation of motion for the probability of having $n$ photons
in the cavity given by Eq.~(\ref{master eqt nn laser}). From Eq.~(\ref
{pnlaser}) we have the important result that partially coherent laser light
has a sharp photon distribution (with width several times Poissonian for a
typical He-Ne laser) due to the presence of the saturation nonlinearity, $B$%
, in the laser master equation. Thus, we see that the saturation
nonlinearity in the radiation-matter interaction is essential for laser
coherence.

One naturally asks: is the corresponding nonlinearity in BEC due to
atom-atom scattering? Or is there a nonlinearity present even in an ideal
Bose gas? The master equation presented in this Section proves that the
latter is the case.

More generally we pose the question: Is there a similar nonequilibrium
approach for BEC in a dilute atomic gas that helps us in understanding the
underlying physical mechanisms for the condensation, the critical behavior,
and the associated nonlinearities? The answer to this question is ``yes"
\cite{MOS99,KSZZ}.

\subsection{Derivation of the condensate master equation}

We consider the usual model of a dilute gas of Bose atoms wherein
interatomic scattering is neglected. This ideal Bose gas is confined inside
a trap, so that the number of atoms, $N$, is fixed but the total energy, $E$%
, of the gas is not fixed. Instead, the Bose atoms exchange energy with a
reservoir which has a fixed temperature $T$. As we shall see, this canonical
ensemble approach is a useful tool in studying the current laser cooled
dilute gas BEC experiments \cite{bec,morebec,miesner,kleppner,at-la}. It is
also directly relevant to the He-in-vycor BEC experiments \cite{rep}.

This ``ideal gas + reservoir'' model allows us to demonstrate most clearly
the master equation approach to the analysis of dynamics and statistics of
BEC, and in particular, the advantages and mathematical tools of the method.
Its extension for the case of an interacting gas which includes usual
many-body effects due to interatomic scattering will be discussed elsewhere.

Thus, we are following the so-called canonical-ensemble approach. It
describes, of course, an intermediate situation as compared with the
microcanonical ensemble and the grand canonical ensemble. In the
microcanonical ensemble, the gas is completely isolated, $E=const,N=const$,
so that there is no exchange of energy or atoms with a reservoir. In the
grand canonical ensemble, only the average energy per atom, i.e., the
temperature $T$ and the average number of atoms $\langle N\rangle $ are
fixed. In such a case there is an exchange of both energy and atoms with the
reservoir.

The ``ideal gas + thermal reservoir'' model provides the simplest
description of many qualitative and, in some cases, quantitative
characteristics of the experimental BEC. In particular, it explains many
features of the condensate dynamics and fluctuations and allows us to
obtain, for the first time, the atomic statistics of the BEC as discussed in
the introduction and in the following.

\subsubsection{The ``ideal gas + thermal reservoir" model}

\begin{figure}[tbp]
\center\epsfxsize=10cm\epsffile{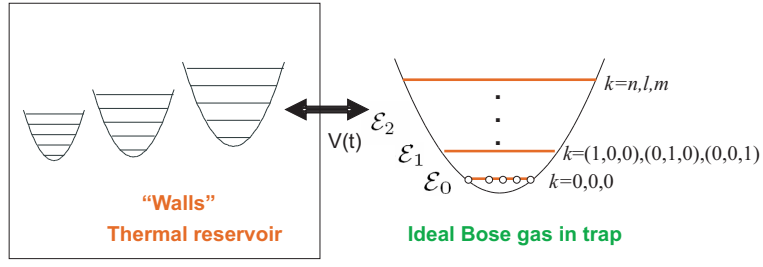} 
\caption{Simple harmonic oscillators as a thermal reservoir for the ideal
Bose gas in a trap.}
\label{reservoir}
\end{figure}

For many problems a concrete realization of the reservoir system is not very
important if its energy spectrum is dense and flat enough. For example, one
expects (and we find) that the equilibrium (steady state) properties of the
BEC are largely independent of the details of the reservoir. For the sake of
simplicity, we assume that the reservoir is an ensemble of simple harmonic
oscillators whose spectrum is dense and smooth, see Fig. \ref{reservoir} .
The interaction between the gas and the reservoir is described by the
interaction picture Hamiltonian
\begin{equation}
V= \sum_j\sum_{k>l} g_{j, kl} b_j^{\dagger} a_k a_l^{\dagger} e^{-i(\omega
_j-\nu_k+\nu_l)t} + H.c. ,  \label{II9}
\end{equation}
where $b_j^{\dagger}$ is the creation operator for the reservoir $j$
oscillator (``phonon"), and $a_k^{\dagger}$ and $a_k$ ($k\neq 0$) are the
creation and annihilation operators for the Bose gas atoms in the $k\mathrm{%
th}$ level. Here $\hbar \nu _k$ is the energy of the $k\mathrm{th}$ level of
the trap, and $g_{j, kl}$ is the coupling strength.

\subsubsection{Bose gas master equation}

The motion of the total ``gas + reservoir" system is governed by the
equation for the total density matrix in the interaction representation, $%
\dot{\rho}_{total}(t)=-i[V(t),\rho_{total}(t)]/\hbar$. Integrating the above
equation for $\rho_{total}$, inserting it back into the commutator in Eq. (%
\ref{II11}), and tracing over the reservoir, we obtain the exact equation of
motion for the density matrix of the Bose-gas subsystem
\begin{equation}
\dot{\rho}(t)=-\frac 1{\hbar ^2}\int_0^tdt^{\prime } \mathrm{Tr}_{\mathrm{res%
}}[V(t),[V(t^{\prime }),\rho _{total}(t^{\prime })]],  \label{II11}
\end{equation}
where $\mathrm{{Tr}_{res}}$ stands for the trace over the reservoir degrees
of freedom.

We assume that the reservoir is large and remains unchanged during the
interaction with the dynamical subsystem (Bose gas). As discussed in \cite
{KSZZ}, the density operator for the total system ``gas+reservoir" can then
be factored, i.e., $\rho_{\mathit{total}} (t^{\prime}) \approx \rho
(t^{\prime}) \otimes \rho_{res}$, where $\rho_{res}$ is the equilibrium
density matrix of the reservoir. If the spectrum is smooth, we are justified
in making the Markov approximation, viz.\ $\rho(t^{\prime}) \to \rho(t)$. We
then obtain the following equation for the reduced density operator of the
Bose-gas subsystem,
\begin{eqnarray}
\dot{\rho} &=&-\frac \kappa 2\sum_{k>l}(\eta _{kl}+1)[a_k^{\dagger
}a_la_l^{\dagger }a_k\rho -2a_l^{\dagger }a_k\rho a_k^{\dagger }a_l+\rho
a_k^{\dagger }a_la_l^{\dagger }a_k]  \nonumber \\
&&-\frac \kappa 2\sum_{k>l}\eta _{kl}[a_ka_l^{\dagger }a_la_k^{\dagger }\rho
-2a_la_k^{\dagger }\rho a_ka_l^{\dagger }+\rho a_ka_l^{\dagger
}a_la_k^{\dagger }].  \label{II12}
\end{eqnarray}
In deriving Eq. (\ref{II12}), we replaced the summation over reservoir modes
by an integration (with the density of reservoir modes $D ( \omega_{kl} )$)
and neglected the frequency dependence of the coefficient $\kappa =2\pi Dg^2
/ \hbar^2$. Here
\begin{equation}
\eta _{kl}=\eta (\omega _{kl}) = Tr_{res} b^{\dagger}(\omega _{kl}) b(\omega
_{kl}) = [ \exp ( \hbar \omega_{kl} / T ) - 1]^{-1}  \label{II13}
\end{equation}
is the average occupation number of the heat bath oscillators at frequency $%
\omega_{kl} \equiv \nu_k - \nu_l$. Equation (\ref{II12}) is then the
equation of motion for an $N$ atom Bose gas driven by a heat bath at
temperature $T$.

\subsubsection{Condensate master equation}

What we are most interested in is the probability distribution $%
p_{n_{0}}=\sum_{\{n_{k}\}_{n_{0}}}p_{n_{0},\{n_{k}\}_{n_{0}}}$ for the
number of condensed atoms $n_{0}$, i.e., the number of atoms in the ground
level of the trap. Let us introduce $p_{n_{0},\{n_{k}\}_{n_{0}}}=\langle
n_{0},\{n_{k}\}_{n_{0}}|\rho |n_{0},\{n_{k}\}_{n_{0}}\rangle $ as a diagonal
element of the density matrix in the canonical ensemble where $%
n_{0}+\sum_{k>0}n_{k}=N$ and $|n_{0},\{n_{k}\}_{n_{0}}\rangle $ is an
arbitrary state of $N$ atoms with occupation numbers of the trap's energy
levels, $n_{k}$, subject to the condition that there are $n_{0}$ atoms in
the ground state of the trap.

In order to get an equation of motion for the condensate probability
distribution $p_{n_{0}}$, we need to perform the summation over all possible
occupations $\{n_{k}\}_{n_{0}}$ of the excited levels in the trap. The
resulting equation of motion for $p_{n_{0}}$, from Eq. (\ref{II12}), is
\begin{eqnarray}
\frac{dp_{n_{0}}}{dt} &=&-\kappa \sum_{\{n_{k}\}_{n_{0}}}\sum_{k>l>0}\{(\eta
_{kl}+1)[(n_{l}+1)n_{k}p_{n_{0},\{n_{k}\}_{n_{0}}}  \nonumber \\
&&-n_{l}(n_{k}+1)p_{n_{0},\{\dots ,n_{l}-1,\dots ,n_{k}+1,\dots \}_{n_{0}}}]
\nonumber \\
&&+\eta _{kl}[n_{l}(n_{k}+1)p_{n_{0},\{n_{k}\}_{n_{0}}}  \nonumber \\
&&-(n_{l}+1)n_{k}p_{n_{0},\{\dots ,n_{l}+1,\dots ,n_{k}-1,\dots
\}_{n_{0}}}]\}  \nonumber \\
&&-\kappa \sum_{\{n_{k}\}_{n_{0}}}\sum_{k^{\prime }>0}[(\eta _{k^{\prime
}}+1)(n_{0}+1)n_{k^{\prime }}p_{n_{0},\{n_{k}\}_{n_{0}}}  \nonumber \\
&&-(\eta _{k^{\prime }}+1)n_{0}(n_{k^{\prime }}+1)p_{n_{0}-1,\{n_{k}+\delta
_{k,k^{\prime }}\}_{n_{0}-1}}  \nonumber \\
&&+\eta _{k^{\prime }}n_{0}(n_{k^{\prime }}+1)p_{n_{0},\{n_{k}\}_{n_{0}}}
\nonumber \\
&&-\eta _{k^{\prime }}(n_{0}+1)n_{k^{\prime }}p_{n_{0}+1,\{n_{k}-\delta
_{k,k^{\prime }}\}_{n_{0}+1}}],  \label{II16}
\end{eqnarray}
where $\eta _{k^{\prime }}=\eta (\nu _{k^{\prime }})$ is the mean number of
thermal phonons of mode $k^{\prime }$ and the sum $\sum_{k^{\prime }}$ runs
over all excited levels.

\begin{figure}[tbp]
\center\epsfxsize=10cm\epsffile{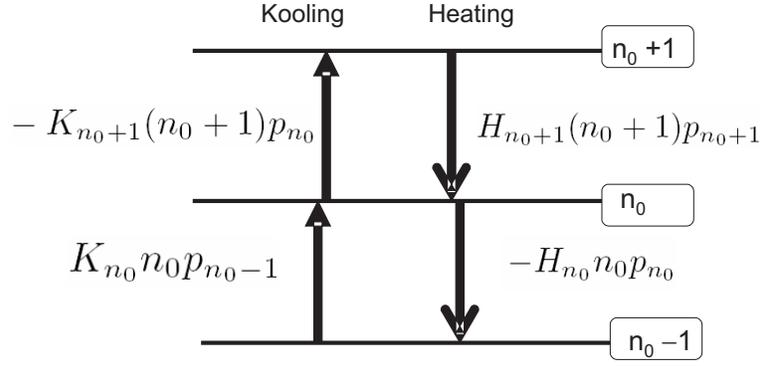} 
\caption{Detailed balance and the corresponding probability flow diagram. We
call $K_{n_0}$ cooling (kooling) rate since the laser loss rate is denote by
``C".}
\label{V2}
\end{figure}

To simplify Eq. (\ref{II16}) we assume that the atoms in the excited levels
with a given number of condensed atoms $n_0$ are in an equilibrium state at
the temperature $T$, i.e.,
\begin{equation}
p_{ n_0 , \{ n_k \}_{n_0} } = p_{ n_0 } \frac{\exp ( - \frac{\hbar}{T}
\sum_{k>0} \nu_k n_k )}{\sum_{\{ n^{\prime}_k \}_{n_0}} \exp ( - \frac{\hbar%
}{T} \sum_{k>0} \nu_k n^{\prime}_k )} ,  \label{II17}
\end{equation}
where $\sum_{k>0} n_k = N - n_0$, and we assume that the sum $\sum_{k>0}$
runs over all energy states of the trap, including degenerate states whose
occupations $n_k$ are treated as different stochastic variables. Equation (%
\ref{II17}) implies that the sum $\sum_{k>l>0}$ in Eq. (\ref{II16}) is equal
to zero, since as depicted in Fig. \ref{V2}

\begin{eqnarray}
(\eta _{kl}+1)p_{n_{0},\{n_{k}\}_{n_{0}}} &=&\eta _{kl}p_{n_{0},\{\dots
,n_{l}+1,\dots ,n_{k}-1,\dots \}_{n_{0}}},  \nonumber \\
(\eta _{kl}+1)p_{n_{0},\{\dots ,n_{l}-1,\dots ,n_{k}+1,\dots \}_{n_{0}}}
&=&\eta _{kl}p_{n_{0},\{n_{k}\}_{n_{0}}}.  \label{II18}
\end{eqnarray}
Equation (\ref{II18}) is precisely the detailed balance condition. The
average number of atoms in an excited level, subject to the condition that
there are $n_{0}$ atoms in the ground state, from Eq. (\ref{II17}), is
\begin{equation}
\langle n_{k^{\prime }}\rangle _{n_{0}}=\sum_{\{n_{k}\}_{n_{0}}}n_{k^{\prime
}}\frac{p_{n_{0},\{n_{k}\}_{n_{0}}}}{p_{n_{0}}}.  \label{II19}
\end{equation}
Therefore, the equation of motion for $p_{n_{0}}$ can be rewritten in the
symmetrical and transparent form
\begin{eqnarray}
\frac{d}{dt}p_{n_{0}} &=&-\kappa
\{K_{n_{0}}(n_{0}+1)p_{n_{0}}-K_{n_{0}-1}n_{0}p_{n_{0}-1}  \nonumber \\
&&+H_{n_{0}}n_{0}p_{n_{0}}-H_{n_{0}+1}(n_{0}+1)p_{n_{0}+1}\},  \label{II20}
\end{eqnarray}
where
\begin{equation}
K_{n_{0}}=\sum_{k^{\prime }>0}(\eta _{k^{\prime }}+1)\langle n_{k^{\prime
}}\rangle _{n_{0}},\qquad H_{n_{0}}=\sum_{k^{\prime }>0}\eta _{k^{\prime
}}(\langle n_{k^{\prime }}\rangle _{n_{0}}+1).  \label{II21}
\end{equation}

\begin{figure}[tbp]
\center\epsfxsize=15cm\epsffile{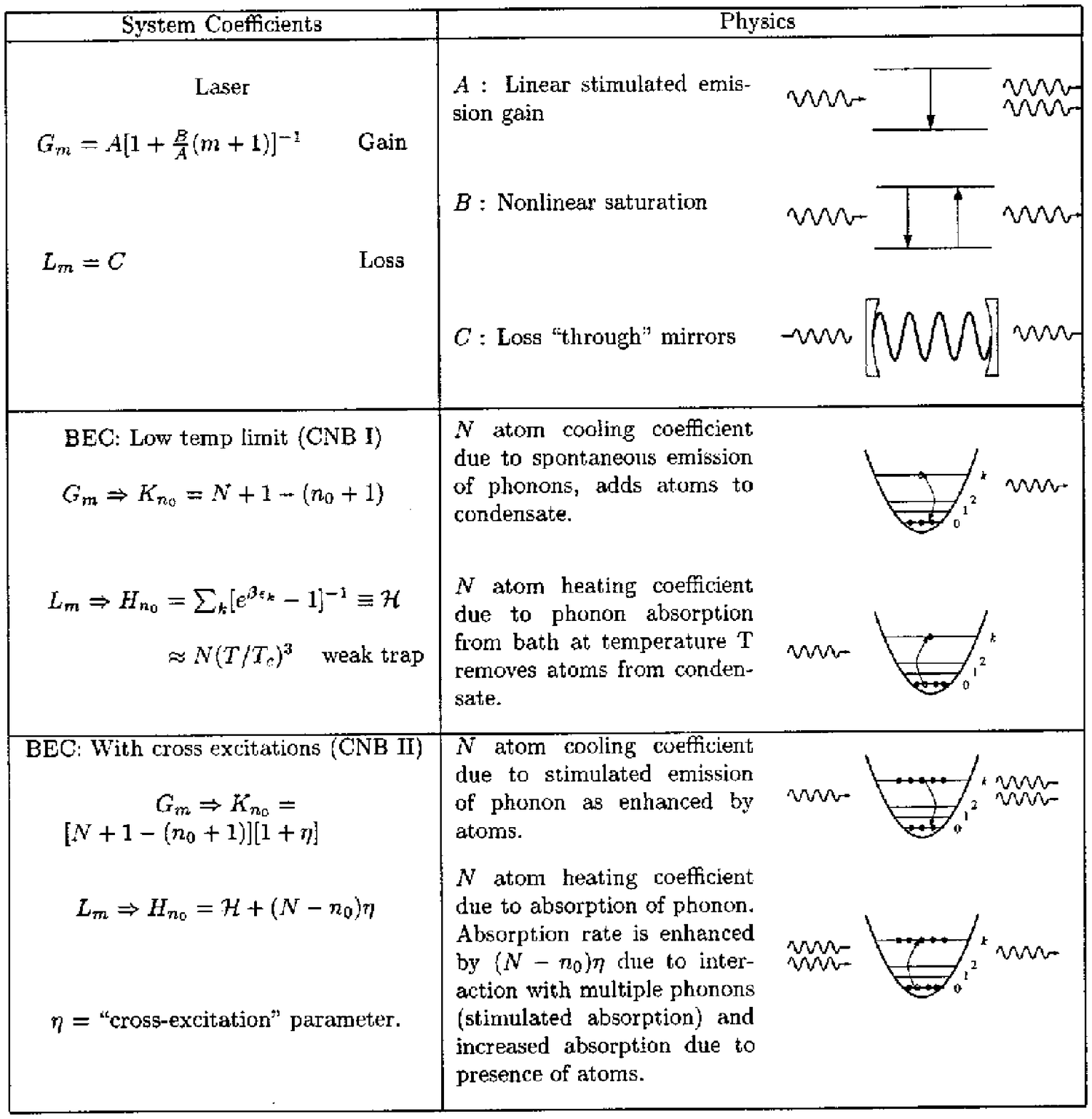} 
\caption{Physical interpretation of various coefficients in the master
equations}
\label{V1}
\end{figure}

We can obtain the steady state distribution of the number of atoms condensed
in the ground level of the trap from Eq.~(\ref{II20}). The mean value and
the variance of the number of condensed atoms can then be determined. It is
clear from Eq. (\ref{II20}) that there are two processes, cooling and
heating. The cooling process is represented by the first two terms with the
cooling coefficient $K_{n_{0}}$ while the heating by the third and fourth
terms with the heating coefficient $H_{n_{0}}$. The detailed balance
condition yields the following expression for the number distribution of the
condensed atoms
\begin{equation}
p_{n_{0}}=p_{0}\prod_{i=1}^{n_{0}}\frac{K_{i-1}}{H_{i}},  \label{II23}
\end{equation}
where the partition function
\begin{equation}
Z_{N}=\frac{1}{p_{N}}=\sum_{n_{0}=0}^{N}\prod_{i=n_{0}+1}^{N}\frac{H_{i}}{%
K_{i-1}}  \label{II24}
\end{equation}
is determined from the normalization condition $%
\sum_{n_{0}=0}^{N}p_{n_{0}}=1 $. The functions $H_{i}$ and $K_{i}$ as given
by Eq. (\ref{II21}), involve, along with $\eta _{k^{\prime }}$ (Eq. (\ref
{II13})), the function $\langle n_{k^{\prime }}\rangle _{n_{0}}$ (Eq. (\ref
{II19})). In the following sections, we shall derive closed form expressions
for these quantities under various approximations. The master equation (\ref
{II20}) for the distribution function for the condensed atoms is one of our
main results. It yields explicit expressions for the statistics of the
condensed atoms and the canonical partition function. Physical
interpretation of various coefficients in the master equations is summarized
in Fig. \ref{V1}.

Under the above assumption of a thermal equilibrium for non-condensed atoms,
we have
\begin{equation}
\langle n_{k^{\prime }}\rangle _{n_{0}}=\frac{\sum_{\{n_{k}\}_{n_{0}}}n_{k^{%
\prime }}\exp (-\frac{\hbar }{T}\sum_{k>0}\nu _{k}n_{k})}{%
\sum_{\{n_{k}^{^{\prime \prime }}\}_{n_{0}}}\exp (-\frac{\hbar }{T}%
\sum_{k>0}\nu _{k}n_{k}^{^{\prime \prime }})}.  \label{II25}
\end{equation}
In the next two sections we present different approximations that clarify
the general result (\ref{II23}).

Short summary of this subsection is as follows. We introduce the probability
of having $n_{0}$ atoms in the ground level and $n_{k}$ atoms in the $k$-th
level $P_{n_{0},n_{1},...,n_{k},...}$. We assume that the atoms in the
excited levels with a given number of condensed atoms $n_{0}$ are in
equilibrium state at the temperature $T$, then
\begin{equation}  \label{m1}
P_{n_{0},n_{1},...,n_{k},...}=\frac{1}{Z_{N}}\exp [-\beta
(E_{0}n_{0}+E_{1}n_{1}+...+E_{k}n_{k}+)].
\end{equation}
This equation yields
\begin{equation}  \label{m2}
P_{N}\equiv P_{n_{0}=N,n_{1}=0,...,n_{k}=0,...}=\frac{1}{Z_{N}}\exp [-\beta
E_{0}N].
\end{equation}
Assuming $E_{0}=0$ we obtain the following expression for the partition
function
\begin{equation}  \label{m3}
Z_{N}=\frac{1}{P_{N}}.
\end{equation}

We assume that Bose gas can exchange heat (but not particles) with a
harmonic-oscillator thermal reservoir. The reservoir has a dense and smooth
spectrum. The average occupation number of the heat bath oscillator at a
frequency $\omega _{q}=cq$ is
\begin{equation}
\eta _{q}=\frac{1}{\exp (\hbar \omega _{q}/k_{B}T)-1}.  \label{m4}
\end{equation}
The master equation for the distribution function of the condensed bosons $%
p_{n_{0}}\equiv \rho _{n_{0},n_{0}}$ takes the form
\[
\dot{p}_{n_{0}}=-\sum_{k,q}\kappa _{kq}\langle n_{k}\rangle _{n_{0}}(\eta
_{q}+1)[(n_{0}+1)p_{n_{0}}-n_{0}p_{n_{0}-1}]_{\text{Kool}}-
\]
\begin{equation}
-\sum_{k,q}\kappa _{kq}\langle n_{k}+1\rangle _{n_{0}}\eta
_{q}[n_{0}p_{n_{0}}-(n_{0}+1)p_{n_{0}+1}]_{\text{Heat}}.  \label{m5}
\end{equation}
The factors $\kappa _{kq}$ embody the spectral density of the bath and the
coupling strength of the bath oscillators to the gas particles, and
determine the rate of the condensate evolution since there is no direct
interaction between the particles of an ideal Bose gas. Since $\kappa
_{kq}=\kappa \cdot \delta (\hbar \Omega k-\hbar cq)$ the sum $\sum_{k,q}$
reduces to $\sum_{k}$.
\[
\frac{1}{\kappa }\dot{p}_{n_{0}}=-\sum_{k}\langle n_{k}\rangle _{n_{0}}(\eta
_{k}+1)[(n_{0}+1)p_{n_{0}}-n_{0}p_{n_{0}-1}]_{\text{Kool}}-
\]
\begin{equation}
-\sum_{k}\langle n_{k}+1\rangle _{n_{0}}\eta
_{k}[n_{0}p_{n_{0}}-(n_{0}+1)p_{n_{0}+1}]_{\text{Heat}}.  \label{m5a}
\end{equation}
Particle number constraint comes in a simple way: $\sum_{k}\langle
n_{k}\rangle _{n_{0}}=N-\bar{n}_{0}$.

\subsection{Low temperature approximation}

At low enough temperatures, the average occupations in the reservoir are
small and $\eta _{k}+1\simeq 1$ in Eq. (\ref{II21}). This suggests the
simplest approximation for the cooling coefficient
\begin{equation}
K_{n_{0}}\simeq \sum_{k}\langle n_{k}\rangle _{n_{0}}=N-\bar{n}_{0}.
\label{II26}
\end{equation}
In addition, at very low temperatures the number of non-condensed atoms is
also very small, we can therefore approximate $\langle n_{k^{\prime
}}\rangle _{n_{0}}+1$ by $1$ in Eq. (\ref{II21}). Then the heating
coefficient is a constant equal to the total average number of thermal
excitations in the reservoir at all energies corresponding to the energy
levels of the trap,
\begin{equation}
H_{n_{0}}\simeq \mathcal{H},\qquad \mathcal{H}\equiv \sum_{k>0}\eta
_{k}=\sum_{k>0}\bigl(e^{\hbar \nu _{k}/T}-1\bigr)^{-1}.  \label{II27}
\end{equation}

Under these approximations, the condensate master equation (\ref{II20})
simplifies considerably and contains only one non-trivial parameter $%
\mathcal{H}$. We obtain
\begin{eqnarray}
\frac{d}{dt}p_{n_{0}}=-\kappa
&&\{(N-n_{0})(n_{0}+1)p_{n_{0}}-(N-n_{0}+1)n_{0}p_{n_{0}-1}  \nonumber \\
&&+\mathcal{H}[n_{0}p_{n_{0}}-(n_{0}+1)p_{n_{0}+1}]\}.  \label{II28}
\end{eqnarray}
It may be noted that Eq. (\ref{II28}) has the same form as the equation~(\ref
{master eqt nn thermal}) of motion for the photon distribution function in a
laser operating not too far above threshold. The identification is complete
if we define the gain, saturation, and loss parameters in laser master
equation by $\kappa (N+1)$, $\kappa $, and $\kappa \mathcal{H}$,
respectively. The mechanism for gain, saturation, and loss are however
different in the present case.

A laser phase transition analogy exists via the P-representation \cite
{ds,ph-tra}. The steady-state solution of the Fokker-Planck equation for
laser near threshold is \cite{sz}
\begin{equation}
P(\alpha ,\alpha ^{\ast })=\frac{1}{\mathcal{N}}\exp [(\frac{\mathcal{A}-%
\mathcal{C}}{\mathcal{A}})|\alpha |^{2}-\frac{\mathcal{B}}{2\mathcal{A}}%
|\alpha |^{4}]  \label{P rep laser}
\end{equation}
which clearly indicates a formal similarity between
\begin{equation}
\ln P(\alpha ,\alpha ^{\ast })=-\ln \mathcal{N}+(1-\mathcal{H}%
/(N+1))n_{0}-(1/2(N+1))n_{0}^{2}  \label{lnP}
\end{equation}
for the laser equation and the Ginzburg-Landau type free energy \cite
{sz,ds,ph-tra}
\begin{equation}
G(n_{0})=\ln p_{n_{0}}\approx const+a(T)n_{0}+b(T)n_{0}^{2},  \label{II29}
\end{equation}
where $|\alpha |^{2}=n_{0},a(T)=-(N-\mathcal{H})/N$ and $b(T)=1/(2N)$ for
large $N$ near $T_{c}$.

The resulting steady state distribution for the number of condensed atoms is
given by
\begin{equation}
p_{n_0}=\frac {1}{Z_N}\frac{\mathcal{H}^{N-n_0}}{(N-n_0)!},  \label{II30}
\end{equation}
where $Z_N=1/p_N$ is the partition function. It follows from the
normalization condition $\sum_{n_0}p_{n_0}=1$ that
\begin{equation}
Z_N={e^{\mathcal{H}}}\Gamma(N+1,\mathcal{H})/N!,  \label{II31}
\end{equation}
where $\Gamma (\alpha,x) = \int^{\infty}_{x} t^{\alpha -1} e^{-t} dt$ is an
incomplete gamma-function.

The distribution (\ref{II30}) can be presented as a probability distribution
for the total number of non-condensed atoms, $n=N-n_{0}$,
\begin{equation}
P_{n}\equiv p_{N-n}=\frac{e^{-\mathcal{H}}N!}{\Gamma (N+1,\mathcal{H})}\frac{%
\mathcal{H}^{n}}{n!}.  \label{II32}
\end{equation}
It looks somewhat like a Poisson distribution, however, due to the
additional normalization factor, $N!/\Gamma (N+1,\mathcal{H})\neq 1$, and a
finite number of admissible values of $n=0,1,\dots ,N$, it is not
Poissonian. The mean value and the variance can be calculated from the
distribution (\ref{II30}) for an arbitrary finite number of atoms in the
Bose gas,
\begin{equation}
\langle n_{0}\rangle =N-\mathcal{H}+\mathcal{H}^{N+1}/Z_{N}N!,  \label{II33}
\end{equation}
\begin{equation}
\Delta n_{0}^{2}\equiv \langle n_{0}^{2}\rangle -\langle n_{0}\rangle ^{2}=%
\mathcal{H}\left( 1-(\langle n_{0}\rangle +1)\mathcal{H}^{N}/Z_{N}N!\right) .
\label{II34}
\end{equation}

As we shall see from the extended treatment in the next section, the
approximations (\ref{II26}), (\ref{II27}) and, therefore, the results (\ref
{II33}), (\ref{II34}) are clearly valid at low temperatures, i.e., in the
weak trap limit, $T\ll \varepsilon _{1}$, where $\varepsilon _{1}$ is an
energy gap between the first excited and the ground levels of a
single-particle spectrum in the trap. However, in the case of a harmonic
trap the results (\ref{II33}), (\ref{II34}) show qualitatively correct
behavior for all temperatures, including $T \gg \varepsilon _{1}$ and $T\sim
T_{c}$ \cite{MOS99}.

In particular, for a harmonic trap we have from Eq. (\ref{II27}) that
the heating rate is
\begin{equation}
\mathcal{H}=\sum_{k}\langle \eta _{k}\rangle =\sum_{l,m,n}\frac{1}{\exp
[\beta \hbar \Omega (l+m+n)-1]}\approx \left( \frac{k_{B}T}{\hbar \Omega }%
\right) ^{3}\zeta (3)=N\left( \frac{T}{T_{c}}\right) ^{3}.
\label{Hn0 harmonic}
\end{equation}
Thus, in the low temperature region the master equation~(\ref{II28}) for
the condensate in the harmonic trap becomes
\begin{eqnarray}
\frac{1}{\kappa }\dot{p}_{n_{0}}=- &&\left[ (N+1)(n_{0}+1)-(n_{0}+1)^{2}%
\right] p_{n_{0}}+[(N+1)n_{0}-n_{0}^{2}]p_{n_{0}-1}  \nonumber \\
&&-N\left( \frac{T}{T_{c}}\right) ^{3}[n_{0}p_{n_{0}}-(n_{0}+1)p_{n_{0}+1}].
\end{eqnarray}

\subsection{Quasithermal approximation for non-condensate occupations}

At arbitrary temperatures, a very reasonable approximation for the average
non-condensate occupation numbers in the cooling and heating coefficients in
Eq. (\ref{II21}) is suggested by Eq. (\ref{II25}) in a quasithermal form,
\begin{equation}
\langle n_{k}\rangle _{n_{0}}=\eta _{k}\sum_{k>0}\langle n_{k}\rangle
_{n_{0}}/\sum_{k^{\prime }}\eta _{k^{\prime }}=\frac{(N-\bar{n}_{0})}{%
(e^{\varepsilon_{k}/T}-1){\mathcal{H}}}  \; , \label{II35}
\end{equation}
where $\varepsilon _{k}=\hbar \nu _{k}$, $\eta _{k}$ is given by Eq. (\ref
{II13}) and $\mathcal{H}$ by Eq. (\ref{II27}). Equation (\ref{II35})
satisfies the canonical-ensemble constraint, $N=n_{0}+\sum_{k>0}n_{k}$,
independently of the resulting distribution $p_{n_{0}}$. This important
property is based on the fact that a quasithermal distribution (\ref{II35})
provides the same relative average occupations in excited levels of the trap
as in the thermal reservoir, Eq. (\ref{II13}).

To arrive at the quasithermal approximation in Eq. (\ref{II35}) one can go
along the following logic. In the low temperature limit we assumed $\eta
_{k}\ll 1$ and took
\[
\sum_{k}\langle n_{k}\rangle _{n_{0}}\langle \eta _{k}+1\rangle \approx
\sum_{k}\langle n_{k}\rangle _{n_{0}}=N-\bar{n}_{0}.
\]
To go further, still in the low temperature limit, we can write
\[
\langle n_{k}\rangle _{n_{0}}\approx (N-\bar{n}_{0})\left[ \frac{\exp
(-\beta E_{k})}{\sum_{k^{\prime }}\exp (-\beta E_{k^{\prime }})}\right] .
\]
This is physically motivated since the thermal factor in [...] is the
fraction of the excited atoms in the state~$k$, and $N-\bar{n}_{0}$ is the
total number of excited atoms. Note that
\[
\eta _{k}=\frac{1}{\exp (\beta E_{k})-1}\quad \Longrightarrow \quad \exp
(-\beta E_{k})=\frac{\eta _{k}}{1+\eta _{k}}.
\]
Since we are at low temperature we take $\exp (-\beta E_{k})\approx \eta
_{k} $ and therefore
\begin{equation}
\langle n_{k}\rangle _{n_{0}}\approx (N-\bar{n}_{0})\frac{\eta _{k}}{%
\sum_{k}\eta _{k}}=\frac{(N-\bar{n}_{0})}{[\exp (\beta E_{k})-1]\mathcal{H}},
\label{m9}
\end{equation}
where
\[
\mathcal{H}=\sum_{k}\eta _{k}.
\]
Now this ansatz is good for arbitrary temperatures. As a result,
\begin{equation}
\sum_{k}\langle n_{k}\rangle _{n_{0}}\langle \eta _{k}+1\rangle \approx
(N-n_{0})(1+\eta ),  \label{m10}
\end{equation}
where
\[
\eta =\frac{1}{(N-n_{0})}\sum_{k}\langle n_{k}\rangle _{n_{0}}\eta _{k}=%
\frac{1}{\mathcal{H}}\sum_{k>0}\eta _{k}^{2}=\frac{1}{\mathcal{H}}\sum_{k>0}%
\frac{1}{[\exp (\beta E_{k})-1]^{2}}.
\]

Another line of thought is the following:
\[
\langle n_{k}\rangle _{n_{0}}\approx (N-n_{0})\frac{\bar{n}_{k}}{\sum_{k}%
\bar{n}_{k}}.
\]
But by detailed balance in the steady state
\[
\kappa (\bar{n}_{0}+1)\bar{n}_{k}(\bar{\eta}_{k}+1)\approx \kappa \bar{n}%
_{0}(\bar{n}_{k}+1)\bar{\eta}_{k}
\]
and if the ground level is macroscopically occupied then $\bar{n}_{0}\approx
\bar{n}_{0}\pm 1$. Since even at $T=T_{c}$ one finds $\bar{n}_{0}\sim \sqrt{N}$,
one ``always'' has $\bar{n}_{0}\gg 1$. Therefore, $\bar{n}_{k}(\bar{\eta}%
_{k}+1)\approx (\bar{n}_{k}+1)\bar{\eta}_{k}$ and, hence, $\bar{n}%
_{k}\approx \bar{\eta}_{k}$. As a result we again obtain Eq. (\ref{m9}).

Calculation of the heating and cooling rates in this approximation is very
simple. For example, for the heating rate we have
\begin{equation}  \label{m11}
\sum_k\langle (n_k+1\rangle _{n_0}\langle \eta _k\rangle =\sum_k\langle \eta
_k\rangle +\sum_k\langle n_k\rangle _{n_0}\langle \eta _k\rangle \approx
\mathcal{H}+\eta (N-n_0).
\end{equation}

In summary, the cooling and heating coefficients (\ref{II21}) in the
quasithermal approximation of Eq. (\ref{II35}) are
\begin{equation}
K_{n_{0}}=(N-n_{0})(1+\eta ),\qquad H_{n_{0}}=\mathcal{H}+(N-n_{0})\eta .
\label{II36}
\end{equation}
Compared with the low temperature approximation (\ref{II26}) and (\ref{II27}%
), these coefficients acquire an additional contribution $(N-n_{0})\eta $
due to the cross-excitation parameter
\begin{equation}
\eta =\frac{1}{N-n_{0}}\sum_{k>0}\langle \eta _{k}\rangle \langle
n_{k}\rangle _{n_{0}}=\frac{1}{\mathcal{H}}\sum_{k>0}\frac{1}{%
(e^{\varepsilon _{k}/T}-1)^{2}}.  \label{II38}
\end{equation}

\subsection{Solution of the condensate master equation}

Now, at arbitrary temperatures, the condensate master equation (\ref{II20})
contains two non-trivial parameters, $\mathcal{H}$ and $\eta$,
\begin{eqnarray}
\frac{dp_{n_0}}{dt} = - \kappa \{(1+ \eta
)[(N-n_0)(n_0+1)p_{n_0}-(N-n_0+1)n_0p_{n_0-1}]  \nonumber \\
+ [ \mathcal{H}+ (N-n_0 ) \eta ]n_0p_{n_0}-[\mathcal{H}+(N-n_0-1) \eta
](n_0+1)p_{n_0+1}\} .  \label{II39}
\end{eqnarray}
It can be rewritten also in the equivalent form
\begin{eqnarray}
\frac{1}{\kappa }\frac{dp_{n_{0}}}{dt} =
-[(N+1)(n_{0}+1)-(n_{0}+1)^{2}]p_{n_{0}} +[(N+1)n_{0}-n_{0}{}^{2}]p_{n_{0}-1}
\nonumber \\
-\left(T/T_{c}\right)^{3}N[n_{0}p_{n_{0}}-(n_{0}+1)p_{n_{0}+1}].  \label{mas}
\end{eqnarray}

The steady-state solution of Eq. (\ref{II39}) is given by
\begin{equation}
p_{n_0} = \frac {1}{Z_N} \frac {(N-n_0+\mathcal{H}/\eta -1)!} {(\mathcal{H}%
/\eta -1)! (N-n_0)!} \Bigl( \frac{\eta}{1+\eta} \Bigr)^{N-n_0} = \frac{1}{Z_N%
}{\binom{N-n_0 +\frac{\mathcal{H}}{\eta} -1 }{N-n_0}} \Bigl( \frac{\eta}{%
1+\eta} \Bigr)^{N-n_0},  \label{II40}
\end{equation}
where the canonical partition function $Z_N=1/p_N$ is
\begin{equation}
Z_N = \sum^N_{n_0=0}{\binom{N-n_0+\mathcal{H}/\eta-1 }{N-n_0 }} \Bigl( \frac{%
\eta}{1+\eta} \Bigr)^{N-n_0} .  \label{II41}
\end{equation}
It is worth noting that the explicit formula (\ref{II40}) satisfies exactly
the general relation between the probability distribution of the number of
atoms in the ground state, $p_{n_0}$, and the canonical partition function
\cite{ww}, Eq. (\ref{III76}).

\begin{figure}[tbp]
\center \epsfxsize=9cm\epsffile{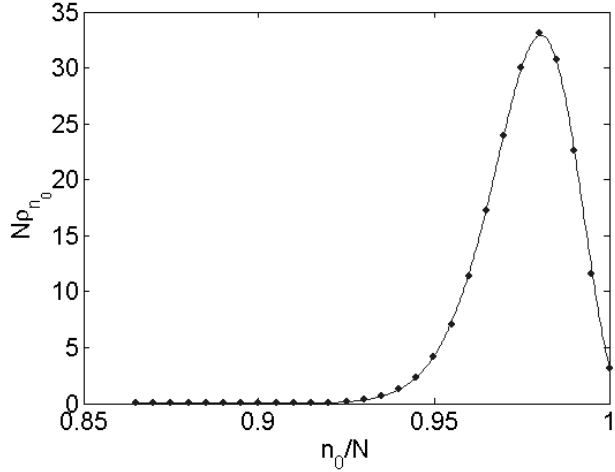}
\caption{Probability distribution of the ground-state occupation, $p_{n_0}$,
at the temperature $T=0.2 T_c$ in an isotropic harmonic trap with $N=200$
atoms as calculated from the solution of the condensate master equation (\ref
{II20}) in the quasithermal approximation, Eq. (\ref{II40}), (solid line)
and from the exact recursion relations in Eqs. (\ref{III76}) and (\ref{III77}%
) (dots).}
\label{FigII6}
\end{figure}

The master equation (\ref{II39}) for $p_{n_0}$, and the analytic approximate
expressions (\ref{II40}) and (\ref{II41}) for the condensate distribution
function $p_{n_0}$ and the partition function $Z_N$, respectively, are among
the main results of the condensate master equation approach. As we shall see
later, they provide a very accurate description of the Bose gas for a large
range of parameters and for different trap potentials. Now we are able to
present the key picture of the theory of BEC fluctuations, that is the
probability distribution $p_{n_0}$, Fig. \ref{FigII6}. Analogy with the evolution of
the photon number distribution in a laser mode (from thermal to coherent,
lasing) is obvious from a comparison of Fig. \ref{FigII6} and Fig. \ref
{thermallaser}. With an increase of the number of atoms in the trap, $N$,
the picture of the ground-state occupation distribution remains
qualitatively the same, just a relative width of all peaks becomes more
narrow.

The canonical partition function (\ref{II41}) allows us to calculate also
the microcanonical partition function $\Omega(E,N)$ by means of the
inversion of the definition in Eq. (\ref{II42}). Moreover, in principle, the
knowledge of the canonical partition function allows us to calculate all
thermodynamic and statistical equilibrium properties of the system in the
standard way (see, e.g., \cite{Ziff,la} and discussion in the Introduction).

Previously, a closed-form expression for the canonical partition
function~(\ref{II42}) was known only for one-dimensional harmonic
traps~\cite{Auluck,Toda}
\begin{equation}
Z_N (T) = \prod_{k=1}^N \frac{1}{1 - e^{-k \hbar \omega /T}} .  \label{II45}
\end{equation}
In the general case, there exists only the recursion relation (\ref{III77})
that is quite complicated, and difficult for analysis \cite{ww,la,bo,br}.

The distribution (\ref{II40}) can also be presented as a probability
distribution for the total number of non-condensed atoms, $n=N-n_0$,
\begin{equation}
P_{n} = p_{N-n} = \frac {1}{Z_N} {\binom{ n+\mathcal{H}/\eta-1 }{n }} \Bigl(
\frac{\eta}{1+\eta} \Bigr)^n .  \label{II47}
\end{equation}
The distribution (\ref{II47}) can be named as \textit{a finite negative
binomial distribution}, since it has the form of the well-known negative
binomial distribution~\cite{a},
\begin{equation}
P_n = {\binom{ n+M-1 }{n }} q^n (1-q)^M , \qquad n = 0, 1, 2, \dots , \infty
,  \label{II48}
\end{equation}
that was so named due to a coincidence of the probabilities $P_n$ with the
terms in the negative-power binomial formula
\begin{equation}
\frac{1}{(1-q)^M} = \sum^{\infty}_{n=0} {\binom{ n+M-1 }{n }} q^n .
\label{II49}
\end{equation}
It has a similar semantic origin as the well-known binomial distribution, $%
P_n = {\binom{ M }{n }} (1-q)^n q^{M-n}$, which was named after a Newton's
binomial formula $[q+(1-q)]^M = \sum^M_{n=0} {\binom{M }{n}} (1-q)^n q^{M-n}$%
. The finite negative binomial distribution (\ref{II47}) tends to the
well-known distribution (\ref{II48}) only in the limit $N \gg (1+\eta)
\mathcal{H}$.

The average number of atoms condensed in the ground state of the trap is
\begin{equation}
\langle n_{0}\rangle \equiv \sum_{n_{0}=0}^{N}n_{0}p_{n_{0}}.  \label{II50}
\end{equation}
It follows, on substituting for $p_{n_{0}}$ from Eq. (\ref{II40}), that
\begin{equation}
\langle n_{0}\rangle =N-\mathcal{H}+p_{0}\eta (N+\mathcal{H}/\eta )\quad .
\label{II51}
\end{equation}

The central moments of the $m$th order, $m>1$, of the
number-of-condensed-atom and number-of-non-condensed-atom fluctuations are
equal to each other for even orders and have opposite signs for odd orders,
\begin{equation}
\langle (n_{0}-\bar{n}_{0})^{m}\rangle =(-1)^{m}\langle (n-\bar{n}%
)^{m}\rangle .  \label{II52}
\end{equation}
The squared variance can be represented as
\begin{equation}
\Delta n_{0}^{2}=\langle n^{2}\rangle -\langle n\rangle
^{2}=\sum_{n=0}^{N}n(n-1)P_{n}+\langle n\rangle -\langle n\rangle ^{2}
\label{II53}
\end{equation}
and calculated analytically. We obtain
\begin{equation}
\Delta n_{0}^{2}=(1+\eta )\mathcal{H}-p_{0}(\eta N+\mathcal{H})\Bigl(N-%
\mathcal{H}+1+\eta \Bigr)-p_{0}^{2}(\eta N+\mathcal{H})^{2}.  \label{II54}
\end{equation}
where
\begin{equation}
p_{0}=\frac{1}{Z_{N}}\frac{(N+\mathcal{H}/\eta -1)!}{N!(\mathcal{H}/\eta -1)!%
}\left( \frac{\eta }{1+\eta }\right) ^{N}  \label{II55}
\end{equation}
is the probability that there are no atoms in the condensate.

\begin{figure}[tbp]
\center \epsfxsize=15cm\epsffile{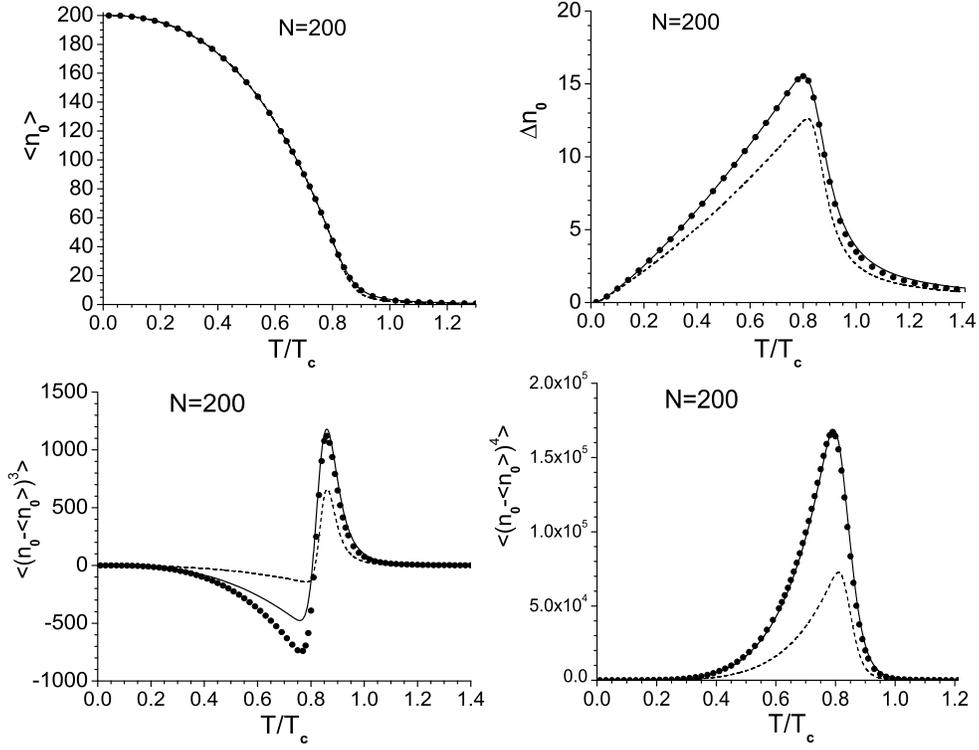}
\caption{The first four central moments for the ideal Bose gas in an
isotropic harmonic trap with $N=200$ atoms as calculated via the solution
of the condensate master equation
(solid lines - quasithermal approximation, Eq. (\ref{II40}); dashed lines -
low temperature approximation, Eq. (\ref{II30})) and via the exact recursion
relations in Eqs. (\ref{III76}) and (\ref{III77}) (dots).}
\label{VVKIIm1234}
\end{figure}

All higher central moments of the distribution Eq. (\ref{II40}) can be
calculated analytically using $\langle {n_{0}}^{s}\rangle
=\sum\limits_{n_{0}=0}^{N}n_{0}^{s}p_{n_{0}}$ and Eqs. (\ref{II40}), (\ref
{II41}). In particular, the third central moment is
\begin{eqnarray}
\langle (n_{0}-\langle n_{0}\rangle )^{3}\rangle &=&-(1+\eta )(1+2\eta )%
\mathcal{H}  \nonumber \\
&&+p_{0}(\mathcal{H}+\eta N)[1+(\mathcal{H}-N)^{2}+2(\eta ^{2}+N(1+\eta
))+3(\eta -\mathcal{H}(1+\eta ))]  \nonumber \\
&&+3p_{0}^{2}(\mathcal{H}+\eta N)^{2}(1+\eta -\mathcal{H}+N)+2p_{0}^{3}(%
\mathcal{H}+\eta N)^{3}.  \label{third mom}
\end{eqnarray}
The first four central moments for the Bose gas in a harmonic trap with $%
N=200$ atoms are presented in Fig. \ref{VVKIIm1234} as the functions of
temperature in different approximations.

For the ``condensed phase'' in the thermodynamic limit, the probability $%
p_{0}$ vanishes exponentially if the temperature is not very close to the
critical temperature. In this case only the first term in Eq. (\ref{II54})
remains, resulting in
\begin{equation}
\Delta n_{0}^{2}=(1+\eta )\mathcal{H}\equiv \sum_{k>0}(\langle n_{k}\rangle
^{2}+\langle n_{k}\rangle ).  \label{II56}
\end{equation}
This result was obtained earlier by standard statistical methods (see \cite
{Ziff} and references therein).

It is easy to see that the result (\ref{II47}) reduces to the simple
approximation (\ref{II32}) in the formal limit $\eta \rightarrow 0,\quad
\mathcal{H}/\eta \rightarrow \infty $, when
\begin{equation}
\frac{\Gamma (N-n_{0}+\mathcal{H}/\eta )}{\Gamma (N+\mathcal{H}/\eta )}%
\rightarrow \Bigl(\frac{\mathcal{H}}{\eta }\Bigr)^{-n_{0}}.  \label{II57}
\end{equation}
The limit applies to only very low temperatures, $T\ll\varepsilon _{1}$.
However, due to Eqs. (\ref{II27}) and (\ref{II38}), the parameter $\mathcal{H%
}/\eta $ tends to 1 as $T\rightarrow 0$, but never to infinity.
Nevertheless, the results (\ref{II51}) and (\ref{II54}) agree with the low
temperature approximation results (\ref{II33}) and (\ref{II34}) for $%
T\ll\varepsilon _{1}$. In this case the variance $\sqrt{\Delta n_{0}^{2}}$ is
determined mainly by a square root of the mean value $\langle n\rangle $
which is correctly approximated by Eq. (\ref{II33}) as $\langle n\rangle
\equiv N-\langle n_{0}\rangle \approx \mathcal{H}$.

\subsection{Results for BEC statistics in different traps}

As we have seen, the condensate fluctuations are governed mainly by two
parameters, the number of thermal excitations $\mathcal{H}$ and the
cross-excitation parameter $\eta$. They are determined by a single-particle
energy spectrum of the trap. We explicitly present them below for arbitrary
power-law trap. We discuss mainly the three-dimensional case. A
generalization to other dimensions is straightforward and is given in the
end of this subsection. First, we discuss briefly the case of the ideal Bose
gas in a harmonic trap. It is the simplest case since the quadratic energy
spectrum implies an absence of the infrared singularity in the variance of
the BEC fluctuations. However, because of the same reason it is not robust
relative to an introduction of a realistic weak interaction in the Bose gas
as is discussed in Section V.

\subsubsection{Harmonic trap}

The potential in the harmonic trap has, in general, an asymmetrical profile
in space, $V_{ext} (x,y,z) = \frac{m}{2} (x^2 \omega_x^2 + y^2 \omega_y^2 +
z^2 \omega_z^2 )$, with eigenfrequencies $\{ \omega_x , \omega_y , \omega_z
\} = \mathbf{\omega} , \quad \omega_x \geq \omega_y \geq \omega_x >0$. Here $%
m$ is the mass of the atom. The single-particle energy spectrum of the trap,
\begin{equation}
\varepsilon_{\mathbf{k}} = \hbar \mathbf{k} \mathbf{\omega} \equiv \hbar (
k_x \omega_x + k_y \omega_y + k_z \omega_z ) ,  \label{II59}
\end{equation}
can be enumerated by three non-negative integers $\{ k_x , k_y , k_z \} =
\mathbf{k} , \quad k_{x,y,z} \geq 0$. We have
\begin{equation}
\mathcal{H} = \sum_{\mathbf{k} >0} \frac{1}{e^{\hbar \mathbf{k}\mathbf{\omega%
}/T}-1} , \qquad \eta \mathcal{H} = \sum_{\mathbf{k} >0} \frac{1}{(e^{\hbar
\mathbf{k}\mathbf{\omega}/T}-1)^2} .  \label{II60}
\end{equation}
The energy gap between the ground state and the first excited state in the
trap is equal to $\varepsilon_1 = \hbar \omega_x$.

If the sums can be replaced by the integrals (continuum approximation),
i.e., if $\hbar \omega_x \ll T$, the parameters $\mathcal{H}$ and $\eta
\mathcal{H}$ are equal to
\begin{equation}
\mathcal{H} = \frac{T^3}{\hbar^3 \omega_x \omega_y \omega_z} \zeta (3) =
\Bigl( \frac{T}{T_c} \Bigr)^3 N ,  \label{II62}
\end{equation}
\begin{equation}
\eta \mathcal{H} = \frac{T^3}{\hbar^3 \omega_x \omega_y \omega_z} \Bigl(
\zeta (2) - \zeta (3) \Bigr) = \Bigl( \frac{T}{T_c} \Bigr)^3 N \frac{\zeta
(2) - \zeta (3)}{\zeta (3)} ,  \label{II63}
\end{equation}
where a standard critical temperature is introduced as
\begin{equation}
T_c = \hbar \Bigl( \frac{\omega_x \omega_y \omega_z N}{\zeta (3)}
\Bigr)^{1/3} ; \qquad \zeta (3) = 1.202\dots , \quad \zeta (2) = \frac{\pi^2%
}{6} .  \label{II64}
\end{equation}
Therefore, the cross-excitation parameter $\eta $ is a constant, independent
of the temperature and the number of atoms, $\eta = [\zeta (2) - \zeta
(3)]/\zeta (3) \approx 0.37$. The ratio $\mathcal{H}/\eta = N (T/T_c )^3
[\zeta (3)/(\zeta (2) - \zeta (3))]$ goes to infinity in the thermodynamic
limit proportionally to the number of atoms $N$.

In the opposite case of very low temperatures, $T \ll \hbar \omega_x$, we
have
\begin{equation}
\mathcal{H} \approx \exp (- \frac{\hbar \omega_x}{T}) + \exp (- \frac{\hbar
\omega_y}{T}) + \exp (- \frac{\hbar \omega_z}{T}) ,  \label{II67}
\end{equation}
\begin{equation}
\eta \mathcal{H} \approx \exp (- \frac{2\hbar \omega_x}{T}) + \exp (- \frac{%
2\hbar \omega_y}{T}) + \exp (- \frac{2\hbar \omega_z}{T})  \label{II68}
\end{equation}
with an exponentially good accuracy. Now the cross-excitation parameter $%
\eta $ depends exponentially on the temperature and, instead of the number
0.37, is exponentially small. The ratio
\begin{equation}
\frac{\mathcal{H}}{\eta} = \frac{ [ \exp (- \frac{\hbar \omega_x}{T}) + \exp
(- \frac{\hbar \omega_y}{T}) + \exp (- \frac{\hbar \omega_z}{T}) ]^2}{ \exp
(- \frac{2\hbar \omega_x}{T}) + \exp (- \frac{2\hbar \omega_y}{T}) + \exp (-
\frac{2\hbar \omega_z}{T})} \sim 1  \label{II69}
\end{equation}
becomes approximately a constant. The particular case of an isotropic
harmonic trap is described by the same equations if we substitute $\omega_x
= \omega_y = \omega_z = \omega$.

\subsubsection{Arbitrary power-law trap}

We now consider the general case of a $d$-dimensional trap with an arbitrary
power-law single-particle energy spectrum \cite{ww,gh97b,Groot}
\begin{equation}
\varepsilon_{\mathbf{k}} = \hbar \sum^d_{j=1} \omega_j k^{\sigma}_{j} ,
\qquad \mathbf{k} = \{ k_j ; \quad j=1,2, \dots , d \} ,  \label{II84}
\end{equation}
where $k_j \geq 0$ is a non-negative integer and $\sigma >0$ is an index of
the energy spectrum. We assume $0< \omega_1 \leq \omega_2 \leq \dots \leq
\omega_d$, so that the energy gap between the ground state and the first
excited state in the trap is $\varepsilon_1 = \hbar \omega_1$. We then have
\begin{equation}
\mathcal{H} = \sum_{\mathbf{k} >0} \frac{1}{e^{\varepsilon_{\mathbf{k}}/T}-1}
, \qquad \eta \mathcal{H} = \sum_{\mathbf{k} >0} \frac{1}{(e^{\varepsilon_{%
\mathbf{k}}/T}-1)^2} .  \label{II85}
\end{equation}

In the case $\varepsilon_1 \ll T$, the sum can be replaced by the integral
only for the parameter $\mathcal{H}$ (Eq. (\ref{II85})) if $d > \sigma$,
\begin{equation}
\mathcal{H} = A \zeta \bigl( \frac{d}{\sigma} \bigr) T^{d/\sigma} = \Bigl(
\frac{T}{T_c} \Bigr)^{d/\sigma} N , \qquad d > \sigma ,  \label{II87}
\end{equation}
where the critical temperature is
\begin{equation}
T_c = \Bigl[ \frac{N}{A\zeta(d/\sigma)} \Bigr]^{\sigma /d} , \qquad A =
\frac{[\Gamma (\frac{1}{\sigma}+1)]^d}{\bigl( \Pi^d_{j=1} \hbar \omega_j %
\bigr)^{1/\sigma}} .  \label{II88}
\end{equation}
The second parameter can be calculated by means of this continuum
approximation only if $0 < \sigma < d/2$,
\begin{equation}
\eta \mathcal{H} = AT^{d/\sigma} \Bigl( \zeta (\frac{d}{\sigma} -1) - \zeta (%
\frac{d}{\sigma}) \Bigr) = \Bigl( \frac{T}{T_c} \Bigr)^{d/\sigma} N \frac{%
\zeta (\frac{d}{\sigma} -1) - \zeta (\frac{d}{\sigma})}{ \zeta (\frac{d}{%
\sigma})} , \qquad 0< \sigma < d/2.  \label{II89}
\end{equation}
If $\sigma > d/2$, it has a formal infrared divergence and should be
calculated via a discrete sum,
\begin{equation}
\eta \mathcal{H} = \Bigl( \frac{T}{T_c} \Bigr)^2 N^{2 \sigma /d} \frac{%
a_{\sigma ,d}}{\bigl[ \Gamma (\frac{1}{\sigma}+1) \bigr]^{2\sigma} \bigl[ %
\zeta (\frac{d}{\sigma}) \bigr]^{2 \sigma /d}} , \qquad \sigma > d/2 ,
\label{II90}
\end{equation}
where
\[
a_{\sigma ,d} = \sum_{\mathbf{k} >0} \frac{\bigl( \Pi^d_{j=1} \hbar \omega_j %
\bigr)^{2/d}}{\varepsilon^2_{\mathbf{k}}} .
\]
The traps with the dimension lower than the critical value, $d \leq \sigma$,
can be analyzed on the basis of Eqs.~(\ref{II85}) as well. We omit this
analysis here since there is no phase transition in this case.

The cross-excitation parameter $\eta$ has different dependence on the number
of atoms for high, $d >2 \sigma$, or low, $d <2 \sigma$, dimensions,
\begin{equation}
\eta = \frac{\zeta (\frac{d}{\sigma} -1) -\zeta (\frac{d}{\sigma})}{ \zeta (%
\frac{d}{\sigma})} \quad , \qquad d>2\sigma>0 ,  \label{II91}
\end{equation}
\begin{equation}
\eta = \Bigl( \frac{T}{T_c} \Bigr)^{2-d/\sigma} N^{2\sigma /d -1} \frac{%
a_{\sigma ,d}}{\bigl[ \Gamma (\frac{1}{\sigma}+1) \bigr]^{2\sigma} \bigl[ %
\zeta (\frac{d}{\sigma}) \bigr]^{2 \sigma /d}} \quad, \qquad d<2\sigma .
\label{II92}
\end{equation}
Therefore, the traps with small index of the energy spectrum, $0<\sigma <d/2$,
are similar to the harmonic trap. The traps with larger index of the
energy spectrum, $\sigma >d/2$, are similar to the box with ``homogeneous"
Bose gas. For the latter traps, the cross-excitation parameter $\eta$ goes
to infinity in the thermodynamic limit, proportionally to $N^{2\sigma /d -1}$.
The ratio $\mathcal{H}/\eta$ goes to infinity in the thermodynamic limit
only for $0<\sigma <d$. In the opposite case, $\sigma >d$, it goes to zero.
We obtain
\begin{equation}
\frac{\mathcal{H}}{\eta} = \Bigl( \frac{T}{T_c} \Bigr)^{d/\sigma} N \frac{%
\zeta (\frac{d}{\sigma})}{\zeta (\frac{d}{\sigma}-1) - \zeta (\frac{d}{\sigma%
})} , \qquad d>2\sigma >0 ,  \label{II93}
\end{equation}
\begin{equation}
\frac{\mathcal{H}}{\eta} = \Bigl( \frac{T}{T_c} \Bigr)^{2(d/\sigma -1)}
N^{2(1- \sigma /d)} \Bigl[ \Gamma \bigl( \frac{1}{\sigma}+1 \bigr) \Bigr%
]^{2\sigma} \Bigl[ \zeta \bigl( \frac{d}{\sigma} \bigr) \Bigr]^{2\sigma /d}
a^{-1}_{\sigma ,d} \quad , \qquad d<2\sigma .  \label{II94}
\end{equation}
It is remarkable that BEC occurs only for those spatial dimensions, $%
d>\sigma $, for which $\mathcal{H}/\eta \to \infty$ at $N \to \infty$. (We
do not consider here the case of the critical dimension $d=\sigma$, e.g.,
one-dimensional harmonic trap, where a quasi-condensation occurs at a
temperature $T_c \sim \hbar \omega_1 N / \log N$.) For spatial dimensions
lower than the critical value, $d<\sigma$, BEC does not occur (see, e.g.,
\cite{ww} ). Interestingly, even for the latter case there still exists a
well-defined single peak in the probability distribution $p_{ n_0 }$ at low
enough temperatures. With the help of the explicit formulas in Section III
we can describe this effect as well.

In the opposite case of very low temperatures, $T \ll \varepsilon_1$, the
parameters
\begin{equation}
\mathcal{H} \approx \sum^d_{j=1} e^{- \frac{\hbar \omega_j}{T}} , \quad \eta
\mathcal{H} \approx \sum^d_{j=1} e^{- \frac{2\hbar \omega_j}{T}} , \quad
\eta \sim e^{- \frac{\varepsilon_1}{T}}  \label{II95}
\end{equation}
are exponentially small. The ratio $\frac{\mathcal{H}}{\eta} \sim
\sum^d_{j=1} e^{-(\hbar \omega_j - \varepsilon_1)/T} \sim d$ becomes a
constant.

Formulas (\ref{II84})-(\ref{II95}) for the arbitrary power-law trap contain
all particular formulas for the three-dimensional harmonic trap ($d=3, \quad
\sigma =1$) and the box, i.e., the  ``homogeneous gas'' with $d=3$ and
$\quad \sigma =2$, as the particular cases.

In Fig. \ref{VVKIIm1234}, numerical comparison of the results obtained from
the exact recursion relations in Eqs. (\ref{III76})-(\ref{III77}) and from
our approximate explicit formulas from Section IV in the particular case of
the ideal Bose gas in the three-dimensional isotropic harmonic trap for
various temperatures is demonstrated. The results indicate an excellent
agreement between the exact results and the results based on quasithermal
approximation, including the mean value $\langle n_{0}\rangle $, the squared
variance $\Delta n_{0}^{2}$ as well as the third and fourth central moments.
The low temperature approximation, Eq. (\ref{II30}), is good only at low
temperatures. That is expected since it neglects by the cross-excitation
parameter $\eta $.

\subsection{Condensate statistics in the thermodynamic limit}

The thermodynamic, or bulk \cite{Ziff} limit implies an infinitely large
number of atoms, $N\rightarrow \infty $, in an infinitely large trap under
the condition of a fixed critical temperature, i.e., $N\omega _{x}\omega
_{y}\omega _{z}=const$ in the harmonic trap, $L^{3}N=const$ in the box, and $%
N^{\sigma }\Pi _{j=1}^{d}\omega _{j}=const$ in an arbitrary $d$-dimensional
power-law trap with an energy spectrum index $\sigma $. Then, BEC takes
place at the critical temperature $T_{c}$ (for $d>\sigma $) as a phase
transition, and for some lower temperatures the factor $p_{0}$ is
negligible. As a result, we have the following mean value and the variance
for the number of condensed atoms
\begin{equation}
\langle n_{0}\rangle =N-\mathcal{H}\equiv N-\sum_{k>0}\frac{1}{%
e^{\varepsilon _{k}/T}-1},  \label{II99}
\end{equation}
\begin{equation}
\Delta n_{0}^{2}=(1+\eta )\mathcal{H}\equiv \sum_{k>0}\frac{1}{%
e^{\varepsilon _{k}/T}-1}+\sum_{k>0}\frac{1}{(e^{\varepsilon _{k}/T}-1)^{2}},
\label{II100}
\end{equation}
which agree with the results obtained for the ideal Bose gas for different
traps in the canonical ensemble by other authors \cite
{Ziff,HKK,ww,Groot,Hauge,Dingle,Fraser,Reif}. In particular, we find the
following scaling of the fluctuations of the number of condensed atoms:
\begin{equation}
\Delta n_{0}^{2}\sim C\times {\binom{\Bigl(\frac{T}{T_{c}}\Bigr)^{d/\sigma
}N\quad ,\quad d>2\sigma >0}{\Bigl(\frac{T}{T_{c}}\Bigr)^{2}N^{2\sigma
/d}\quad ,\quad d<2\sigma }},\quad \varepsilon _{1}\ll T<T_{c},  \label{II101}
\end{equation}
\begin{equation}
\Delta n_{0}^{2}\approx \langle n\rangle \approx \sum_{i=1}^{d}\exp \{-\frac{%
\omega _{i}}{\bigl[\Pi _{j=1}^{d}\omega _{j}\bigr]^{1/d}}\Bigl[\zeta \bigl(%
\frac{d}{\sigma }\bigr)\Bigr]^{\sigma /d}\Bigl[\Gamma \bigl(\frac{1}{\sigma }%
+1\bigr)\Bigr]^{\sigma }\frac{T_{c}}{TN^{\sigma /d}}\},\quad T\ll\varepsilon
_{1},  \label{II102}
\end{equation}
where $C$ is a constant. From Eq. (\ref{II101}), we see that in the high
dimensional traps, $d>2\sigma $, e.g., in the three-dimensional harmonic
trap, fluctuations display the proper thermodynamic behavior, $\Delta
n_{0}^{2}\propto N$. However, fluctuations become anomalously large \cite
{ww,gh97b,Hauge,Pit98}, $\Delta n_{0}^{2}\propto N^{2\sigma /d} \gg N$, in the
low dimensional traps, $\sigma <d<2\sigma $. In the quantum regime, when the
temperature is less than the energy gap between the ground and the first
excited level in the trap, it follows from Eq. (\ref{II102}) that condensate
fluctuations become exponentially small. For all temperatures, when BEC
exists ($d>\sigma $), the root-mean-square fluctuations normalized to the
mean number of condensed atoms vanishes in the thermodynamic limit: $\sqrt{%
\Delta n_{0}^{2}}/\langle n_{0}\rangle \rightarrow 0$ as $N\rightarrow
\infty $.

Another remarkable property of the distribution function obtained in Section
IV is that it yields the proper mean value and variance of the number of
atoms in the ground level of the trap even for temperatures higher than the
critical temperature. In particular, it can be shown that its asymptotic for
high temperatures, $T\gg T_{c}$, yields the standard thermodynamic relation
$\Delta n_{0}\approx \langle n_{0}\rangle$ known from the analysis of the
grand canonical ensemble~\cite{Ziff}. This nice fact indicates that the
present master equation approach to the statistics of the cool Bose gas is
valuable in the study of mesoscopic effects as well, both at $T<T_{c}$ and
$T>T_{c}$. Note that, in contrast, the validity of the Maxwell's demon
ensemble approach~\cite{Navez} to the statistics of BEC remains restricted
to temperatures well below the onset of BEC, $T<T_{c}$.

\subsection{Mesoscopic and dynamical effects in BEC}

In recent experiments on BEC in ultracold gases~\cite{bec,morebec,miesner,
kleppner,Sant01,Robe01}, the number of condensed atoms in the trap is finite,
i.e., mesoscopic rather than macroscopic, $N \sim 10^3 - 10^6$. Therefore,
it is interesting to analyze mesoscopic effects associated with the BEC
statistics.

The mean number of atoms in the ground state of the trap with a finite
number of atoms is always finite, even at high temperatures. However, it
becomes macroscopically large only at temperatures lower than some critical
temperature, $T_{c}$, that can be defined via the standard relation
\begin{equation}
\sum \eta _{k}(T_{c})\equiv \mathcal{H}(T_{c})=N.  \label{II105}
\end{equation}
This equation has an elementary physical meaning, namely it determines the
temperature at which the total average number of thermal excitations at all
energy levels of the trap becomes equal to the total number of atoms in the
trap. The results (\ref{II40}), (\ref{II51}), (\ref{II54}) shown in Fig. \ref
{VVKIIm1234} explicitly describe a smooth transition from a mesoscopic
regime (finite number of atoms in the trap, $N<\infty $) to the
thermodynamic limit ($N=\infty $) when the threshold of the BEC becomes very
sharp so that we have a phase transition to the Bose-Einstein condensed
state at the critical temperature given by Eq. (\ref{II105}). This can be
viewed as a specific demonstration of the commonly accepted resolution to
the Uhlenbeck dilemma in his famous criticism of Einstein's pioneering
papers on BEC \cite{Einstein25,Uhlenbeck,Kahn,London}.

Although for systems containing a finite number of atoms there is no
sharp critical point, as is obvious from Figs.~\ref{n0}, \ref{FigII6}, and
\ref{FigIII2}, it is useful to define a critical characteristic value of a
temperature in such a case as well. It should coincide with the standard
definition (\ref{II105}) in the thermodynamic limit. Different definitions
for $T_{c}$ were proposed and discussed in \cite
{kd,gh95,gps96,kt,hhr,b,hsc,gcl,Pathria}. We follow a hint from laser
physics. There we know that fluctuations dominate near threshold. However,
we define a threshold inversion as that for which gain (in photon number for
the lasing mode) equals loss. Let us use a similar dynamical approach for
BEC on the basis of the master equation, see also~\cite{scully40}.

We note that, for a laser operating near the threshold where $B/A\ll 1$, the
equation (\ref{master eqt nn laser}) of motion for the probability $p_{n}$
of having $n$ photons in the cavity implies the following rate of the change
for the average photon number:
\begin{equation}
\frac{d}{dt}\langle n\rangle =(A-C)\langle n\rangle -B\langle
(n+1)^{2}\rangle +A.  \label{II107}
\end{equation}
Here $A,B$, and $C$ are the linear gain, nonlinear saturation, and linear
loss coefficients, respectively. On neglecting the spontaneous emission term
$A$ and noting that the saturation term $B\langle (n+1)^{2}\rangle $ is
small compared to $(A-C)\langle n\rangle $ near threshold, we define the
threshold (critical) inversion to occur when the linear gain rate equals the
linear loss rate, i.e., $A=C$.

Similar to laser physics, the condensate master equation (\ref{II20})
implies a coupled hierarchy of moment equations which are useful in the
analysis of time evolution. In the quasithermal approximation (\ref{II39}),
we find
\[
\frac{d\langle n_{0}^{l}\rangle }{dt}=\kappa \sum_{i=0}^{l-1}{\binom{l}{i}}%
\{(1+\eta )[N(\langle n_{0}^{i}\rangle +\langle n_{0}^{i+1}\rangle )-\langle
n_{0}^{i+1}\rangle -\langle n_{0}^{i+2}\rangle ]
\]
\begin{equation}
+(-1)^{l-i}(\mathcal{H}+\eta N)\langle n_{0}^{i+1}\rangle -(-1)^{l-i}\eta
\langle n_{0}^{i+2}\rangle \}.  \label{II108}
\end{equation}

Similar moment equations in the low-temperature approximation (\ref{II28})
follow from Eq. (\ref{II108}) with $\eta =0$,
\begin{equation}
\frac{d\langle n_{0}^{l}\rangle }{dt}=\kappa \sum_{i=0}^{l-1}{\binom{l}{i}}%
\{N\langle n_{0}^{i}\rangle +(N-1)\langle n_{0}^{i+1}\rangle -\langle
n_{0}^{i+2}\rangle +(-1)^{l-i}\mathcal{H}\langle n_{0}^{i+1}\rangle \}.
\label{II109}
\end{equation}
The dynamical equation for the first moment, as follows from Eq. (\ref{II108}%
), has the following form
\begin{equation}
\frac{d\langle n_{0}\rangle }{dt}=\kappa \{(1+\eta )N+(N-1-\eta -\mathcal{H}%
)\langle n_{0}\rangle -\langle n_{0}^{2}\rangle \}.  \label{II110}
\end{equation}
Near the critical temperature, $T\approx T_{c}$, the mean number of the
condensed atoms is small, $\langle n_{0}\rangle \ll N$, and it is reasonable
to neglect the second moment $\langle n_{0}^{2}\rangle $ compared to $%
N\langle n_{0}\rangle $ and the spontaneous cooling (spontaneous emission in
lasers) term $\kappa (1+\eta )N$ compared to $\kappa N\langle n_{0}\rangle $%
. In this way, neglecting fluctuations, we arrive at a simple equation for
the competition between cooling and heating processes,
\begin{equation}
\frac{d\langle n_{0}\rangle }{dt}\approx \kappa (N-\mathcal{H}-\eta )\langle
n_{0}\rangle .  \label{II111}
\end{equation}
In analogy with the laser threshold we can define the critical temperature, $%
T=T_{c}$, as a point where cooling equals heating, i.e., $d\langle
n_{0}\rangle /dt=0$. This definition of the critical temperature
\begin{equation}
\mathcal{H}(T_{c})+\eta (T_{c})=N,  \label{II112}
\end{equation}
is valid even for mesoscopic systems and states that at $T=T_{c}$ the rate
of the removal of atoms from the ground state equals to the rate of the
addition, in the approximation neglecting fluctuations. In the thermodynamic
limit it corresponds to the standard definition, Eqs. (\ref{II64}) and (\ref
{II88}). For a mesoscopic system, e.g., of $N=10^{3}$ atoms in a trap, the
critical temperature as given by Eq. (\ref{II112}) is about few per cent
shifted from the thermodynamic-limit value given by Eqs. (\ref{II64}) and (%
\ref{II88}). Other definitions for $T_{c}$ also describe the effect of an
effective-$T_{c}$ shift \cite{kd,gh95,gps96,kt,hhr,b,hsc,gcl,Pathria}, which
is clearly seen in Fig.~\ref{FigIII2}, and agree qualitatively with our
definition.

Note that precisely the same definition of the critical temperature follows
from a statistical mechanics point of view, which in some sense is
alternative to the dynamical one. We may define the critical temperature as
the temperature at which the mean number of condensed atoms in the steady-
state solution to the master equation vanishes when neglecting fluctuations
and spontaneous cooling. We make the replacement $\langle n_0^2 \rangle
\approx \langle n_0 \rangle^2$ in Eq. (\ref{II110}) and obtain the
steady-state solution to this nonlinear equation, $\langle n_0 \rangle = N -
\mathcal{H} - \eta$. Now we see that $\langle n_0 \rangle$ vanishes at the
same critical temperature (\ref{II112}). Finally, we remind again that a
precise definition of the critical temperature is not so important and
meaningful for the mesoscopic systems as it is for the macroscopic systems
in the thermodynamic limit since for the mesoscopic systems, of course,
there is not any sharp phase transition and an onset of BEC is dispersed
over a whole finite range of temperatures around whatever $T_c$, as
is clearly seen in Figs.~\ref{n0}, \ref{FigII6}, and \ref{FigIII2}.

\section{Quasiparticle approach and Maxwell's demon ensemble}

In order to understand relations between various approximate schemes, we
formulate a systematic analysis of the equilibrium canonical-ensemble
fluctuations of the Bose-Einstein condensate based on the particle number
conserving operator formalism of Girardeau and Arnowitt \cite{ga}, and the
concept of the canonical-ensemble quasiparticles \cite{KKS-PRL,KKS-PRA}. The Girardeau-Arnowitt operators can be interpreted as the creation and annihilation operators of the canonical-ensemble quasiparticles which are essentially different from the standard quasiparticles in the grand canonical ensemble. This is so because these operators create and annihilate particles in the properly reduced many-body Fock subspace. In this way, we satisfy the $N$-particle constraint of the canonical-ensemble problem in Eq. (\ref{CEconstraint}) from the very beginning. Furthermore, we do this while taking into account all possible correlations in the $N$-boson system in addition to what one has in the grand canonical ensemble. These canonical-ensemble quasiparticles fluctuate independently in the ideal Bose gas and form dressed canonical-ensemble quasiparticles in the dilute weakly interacting Bose gas due to Bogoliubov coupling (see Section VI below).

Such an analysis was elaborated in \cite{KKS-PRL,KKS-PRA} and resulted in
the explicit expressions for the characteristic function and all cumulants
of the ground-state occupation statistics both for the dilute weakly
interacting and ideal Bose gases. We present it here, including the
analytical formulas for the moments of the ground-state occupation
fluctuations in the ideal Bose gases in an arbitrary power-law trap, and,
in particular, in a box (``homogeneous gas'') and in an arbitrary harmonic
trap. In Section VI we extend this analysis to the interacting Bose gas. In
particular, we calculate the effect of Bogoliubov coupling between
quasiparticles on suppression of the ground-state occupation fluctuations at
moderate temperatures and their enhancement at very low temperatures and
clarify a crossover between ideal-gas and weakly-interacting-gas statistics
which is governed by a pair-correlation, squeezing mechanism. The important
conclusion is that in most cases the ground-state occupation fluctuations
are anomalously large and are not Gaussian even in the thermodynamic limit.

Previous studies focused mainly on the mean value, $\bar{n}_0$, and squared
variance, $\langle (n_0 -\bar{n}_0)^2 \rangle$, of the number of condensed
atoms~\cite{exceptions}. Higher statistical moments are more difficult to
calculate, and it was often assumed that the condensate fluctuations have
vanishing higher cumulants (semi-invariants). That is, it was assumed that
the condensate fluctuations are essentially Gaussian with all central
moments determined by the mean value and the variance. We here show that
this is not true even in the thermodynamic limit. In, particular, we prove
that in the general case the third and higher cumulants normalized by the
corresponding power of the variance do not vanish even in the thermodynamic
limit.

The results of the canonical-ensemble quasiparticle approach are valid for
temperatures a little lower than a critical temperature, namely, when the
probability of having zero atoms in the ground state of the trap is
negligibly small and the higher order effects of the interaction between
quasiparticles are not important. We outline also the Maxwell's demon
ensemble approximation introduced and studied for the ideal Bose gas
in \cite{Fierz,pol,Navez,ww,gh97b} and show that it can be justified on the
basis of the method of the canonical-ensemble quasiparticles, and for the
case of the ideal Bose gas gives the same results.

This Section is organized as follows: We start with the reduction of the
Hilbert space and the introduction of the canonical-ensemble quasiparticles
appropriate to the canonical-ensemble problem in subsection~A. Then, in
subsection B, we analytically calculate the characteristic function and all
cumulants of the ground-state occupation distribution for the ideal Bose gas
in a trap with an arbitrary single-particle energy spectrum. We also discuss
the Maxwell's demon ensemble approach and compare it with the canonical
ensemble quasiparticle approach. In subsection~C we apply these results to
the case of an arbitrary $d$-dimensional power-law trap which includes a
three-dimensional box with periodic boundary conditions (``homogeneous gas'')
and a three-dimensional asymmetric harmonic trap as the particular cases.

\subsection{Canonical-ensemble quasiparticles in the reduced Hilbert space}

In principle, to study the condensate fluctuations, we have to fix only the
external macroscopical and global, topological parameters of the system,
like the number of particles, temperature, superfluid flow pattern
(``domain" or vortex structure), boundary conditions, etc. We then proceed
to find the condensate density matrix via a solution of the von Neumann
equation with general initial conditions admitting all possible quantum
states of the condensate. In particular, this is a natural way to approach
the linewidth problem for the atom laser~\cite{at-la,pw}. Obviously, this is
a complicated problem, especially for the interacting finite-temperature
Bose gas; because of the need for an efficient technique to account for the
additional correlations introduced by the constraint in such realistic
ensembles. The latter is the origin of the difficulties in the theory of the
canonical or microcanonical ensembles (see discussion in the Section III.A).
According to~\cite{hm}, a calculation of equilibrium statistical properties
using the grand canonical ensemble and a perturbation series will be
impossible since the series will have zero radius of convergence.

One way out of this problem is to develop a technique which would allow us
to make calculations in the constrained many-body Hilbert space, e.g., on
the basis of the master equation approach as discussed in Section IV.
Another possibility is to solve for the constraint from the very beginning
by a proper reduction of the many-body Hilbert space so that we can work
with the new, already unconstrained quasiparticles. This approach is
demonstrated in the present Section. Working in the canonical ensemble, we
solve for the fluctuations of the number of atoms in the ground state in the
ideal Bose gas in a trap (and similarly in the weakly interacting Bose gas
with the Bogoliubov coupling between excited atoms, see Section VI). More
difficult problems involving phase fluctuations of the condensate with an
accurate account of the quasiparticle renormalization due to interaction at
finite temperatures and the dynamics of BEC will be discussed elsewhere.

We begin by defining an occupation number operator in the many-body Fock
space as usual,
\begin{equation}
\hat{n}_{\mathbf{k}}=\hat{a}_{\mathbf{k}}^{+}\hat{a}_{\mathbf{k}},\quad \hat{%
n}_{\mathbf{k}}|\psi _{\mathbf{k}}^{(n)}\rangle =n|\psi _{\mathbf{k}%
}^{(n)}\rangle ,\quad \hat{a}_{\mathbf{k}}^{+}|\psi _{\mathbf{k}%
}^{(n)}\rangle =\sqrt{n+1}|\psi _{\mathbf{k}}^{(n+1)}\rangle .  \label{III3}
\end{equation}
The particle number constraint (\ref{CEconstraint}) determines a
canonical-ensemble (CE) subspace of the Fock space. Again we would like to
work with the particle-number conserving creation and annihilation
operators. The latter are given in the Girardeau and Arnowitt paper
\cite{ga},
\begin{equation}
\hat{\beta}_{\mathbf{k}}^{+}=\hat{a}_{\mathbf{k}}^{+}\hat{\beta}_{0},\qquad
\hat{\beta}_{\mathbf{k}}=\hat{\beta}_{0}^{+}\hat{a}_{\mathbf{k}},\qquad \hat{%
\beta}_{0}=(1+\hat{n}_{0})^{-1/2}\hat{a}_{0}.  \label{III4}
\end{equation}
These operators for $\mathbf{k}\neq 0$ can be interpreted as describing new
canonical-ensemble quasiparticles which obey the Bose canonical commutation
relations on the subspace $n_{0}\neq 0$,
\begin{equation}
\lbrack \hat{\beta}_{\mathbf{k}},\hat{\beta}_{\mathbf{k^{\prime }}%
}^{+}]=\delta _{\mathbf{k},\mathbf{k^{\prime }}}.  \label{III5}
\end{equation}
We are interested in the properties of the fraction of atoms condensed in
the ground level of the trap, $\mathbf{k}=0$. We focus on the important
situation when the ground-state occupation distribution is relatively well
peaked, i.e., its variance is much less than the mean occupation of the
ground level of a trap,
\begin{equation}
\langle (n_{0}-\bar{n}_{0})^{2}\rangle ^{1/2} \ll \bar{n}_{0}.  \label{III6}
\end{equation}
In such a case, the relative role of the states with zero ground-state
occupation, $n_{0}=0$, is insignificant, so that we can approximate the
canonical-ensemble subspace $\mathcal{H}^{CE}$ by the subspace $\mathcal{H}%
_{n_{0}\neq 0}^{CE}$. Obviously, this approximation is valid only for
temperatures $T<T_{c}$.

The physical meaning of the canonical-ensemble quasiparticles, $\hat{\beta}_{%
\mathbf{k}} = \hat{\beta}_0^+ \hat{a}_{\mathbf{k}}$, is that they describe
transitions between ground $( \mathbf{k} = 0 )$ and excited $( \mathbf{k}
\neq 0 )$ states. All quantum properties of the condensed atoms have to be
expressed via the canonical-ensemble quasiparticle operators in Eq. (\ref
{III4}). In particular, we have the identity
\begin{equation}
\hat{n}_0 = N - \sum_{\mathbf{k} \neq 0} \hat{n}_{\mathbf{k}} ,  \label{III7}
\end{equation}
where the occupation operators of the excited states are
\begin{equation}
\hat{n}_{\mathbf{k}} = \hat{a}^+_{\mathbf{k}} \hat{a}_{\mathbf{k}} = \hat{%
\beta}^+_{\mathbf{k}} \hat{\beta}_{\mathbf{k}} .  \label{III8}
\end{equation}

Note that in Refs.~\cite{Gardiner,gz} quasiparticle operators similar in
spirit to those of Ref.~\cite{ga} were introduced which, unlike $\hat{\beta}%
_{\mathbf{k}}$, did not obey the Bose commutation relations (\ref{III5}) exactly, if non-commutation of the ground-state occupation operators $\hat{a}_0$ and $\hat{a}_0^+$ is important. As was shown by Girardeau~\cite{g}, this is important because the commutation corrections can accumulate in a perturbation series for quantities like an $S$-matrix. Warning concerning a similar subtlety was stressed some time ago~\cite{agd}.

We are interested in fluctuations in the number of atoms condensed in the
ground state of a trap, $n_0$. This is equal to the difference between the
total number of atoms in a trap and the number of excited atoms, $n_0 = N -
n $. In principle, a condensed state can be defined via the bare trap states
as their many-body mixture fixed by the interaction and external conditions.
Hence, occupation statistics of the ground as well as excited states of a
trap is a very informative feature of the BEC fluctuations. Of course, there
are other quantities that characterize BEC fluctuations, e.g., occupations
of collective, dressed or coherent excitations and different phases.

\subsection{Cumulants of BEC fluctuations in an ideal Bose gas}

Now we can use the reduced Hilbert space and the equilibrium canonical-ensemble density matrix $\hat{\rho}$ to conclude that the occupation numbers of the canonical-ensemble quasiparticles, $n_{\mathbf{k}} , \quad \mathbf{k} \neq 0$, are independent stochastic variables with the equilibrium distribution
\begin{equation}
\rho_{\mathbf{k}} ( n_{\mathbf{k}} ) = \exp (- n_{\mathbf{k}} \varepsilon_{%
\mathbf{k}} / T) ( 1 - \exp ( - \varepsilon_{\mathbf{k}} / T) ) .
\label{III9}
\end{equation}
The statistical distribution of the number of excited atoms, $n=\sum_{%
\mathbf{k} \neq 0} n_{\mathbf{k}}$, which is equal, according to Eq. (\ref
{III7}), to the number of non-condensed atoms, is a simple ``mirror" image
of the distribution of the number of condensed atoms,
\begin{equation}
\rho ( n ) \quad = \quad \rho_0 ( n_0 = N - n ) .  \label{III10}
\end{equation}

A useful way to find and to describe it is via the characteristic function
\begin{equation}
\Theta_n ( u ) = Tr \{ e^{i u \hat{n}} \hat{\rho} \} .  \label{III11}
\end{equation}
Thus upon taking the Fourier transform of $\Theta_n (u)$ we obtain the
probability distribution
\begin{equation}
\rho ( n ) = \frac{1}{2 \pi} \int^{\pi}_{- \pi} e^{- i u n} \Theta_n ( u ) d
u .  \label{III12}
\end{equation}

Taylor expansions of $\Theta _{n}(u)$ and $\log \Theta _{n}(u)$ give
explicitly initial (non-central) moments and cumulants, or semi-invariants
\cite{a,cumulants}:
\begin{equation}
\Theta _{n}(u)=\sum_{m=0}^{\infty }\alpha _{m}\frac{u^{m}}{m!},\qquad \alpha
_{m}\equiv \langle n^{m}\rangle =\frac{d^{m}}{du^{m}}\Theta _{n}(u)|_{u=0},
\label{III13}
\end{equation}
\begin{equation}
\log \Theta _{n}(u)=\sum_{m=1}^{\infty }\kappa _{m}\frac{(iu)^{m}}{m!}%
,\qquad \kappa _{m}=\frac{d^{m}}{d(iu)^{m}}\log \Theta _{n}(u)|_{u=0},\qquad
\Theta _{n}(u=0)=1.  \label{III14}
\end{equation}
The cumulants $\kappa _{r}$, initial moments $\alpha _{m}$, and central
moments $\mu _{m}\equiv \langle (n-\bar{n})^{m}\rangle $ are related to each
other by the simple binomial formulas \cite{a,cumulants} via the mean number
of the non-condensed atoms ${\bar{n}}=N-{\bar{n}}_{0}$,
\[
\mu _{r}=\sum_{k=0}^{r}(-1)^{k}{\binom{r}{k}}\alpha _{r-k}\bar{n}^{k},\qquad
\alpha _{r}=\sum_{k=0}^{r}{\binom{r}{k}}\mu _{r-k}\bar{n}^{k},
\]
\[
\bar{n}=\kappa _{1},\quad \langle (n-\bar{n})^{2}\rangle \equiv \mu
_{2}=\kappa _{2},\quad \langle (n-\bar{n})^{3}\rangle \equiv \mu _{3}=\kappa
_{3},\langle (n-\bar{n})^{4}\rangle \equiv \mu _{4}=\kappa _{4}+3\kappa
_{2}^{2},
\]
\begin{equation}
\langle (n-\bar{n})^{5}\rangle \equiv \mu _{5}=\kappa _{5}+10\kappa
_{2}\kappa _{3},\qquad \langle (n-\bar{n})^{6}\rangle \equiv \mu _{6}=\kappa
_{6}+15\kappa _{2}(\kappa _{4}+\kappa _{2}^{2})+10\kappa _{3}^{2},\dots
\label{III16}
\end{equation}

Instead of calculation of the central moments,
$\mu_{m}=\langle (n -\bar{n} )^{m}\rangle$, it is more convenient,
in particular so for the analysis of the non-Gaussian properties, to solve
for the cumulants $\kappa_{m}$, which are related to the moments by simple
binomial expressions. The first six are
\[
\kappa _{1}=\bar{n},\qquad \kappa _{2}=\mu _{2},\qquad \kappa _{3}=\mu
_{3},\qquad \kappa _{4}=\mu _{4}-3\mu _{2}^{2},
\]
\begin{equation}
\kappa _{5}=\mu _{5}-10\mu _{2}\mu _{3},\qquad \kappa _{6}=\mu _{6}-15\mu
_{2}(\mu _{4}-2\mu _{2}^{2}).  \label{III1}
\end{equation}
As discussed in detail below, the essence of the BEC fluctuations and the
most simple formulas are given in terms of the ``generating cumulants'' $%
\tilde{\kappa}_{m}$ which are related to the cumulants $\kappa _{m}$ by the
combinatorial formulas in Eq. (\ref{III19}),
\begin{equation}
\kappa _{1}=\tilde{\kappa}_{1},\quad \kappa _{2}=\tilde{\kappa}_{2}+\tilde{%
\kappa}_{1},\quad \kappa _{3}=\tilde{\kappa}_{3}+3\tilde{\kappa}_{2}+\tilde{%
\kappa}_{1},\quad \kappa _{4}=\tilde{\kappa}_{4}+6\tilde{\kappa}_{3}+7\tilde{%
\kappa}_{2}+\tilde{\kappa}_{1},\dots  \label{III2}
\end{equation}

The main advantage of the cumulant analysis of the probability distribution $%
\rho (n)$ is the simple fact that the cumulant of a sum of independent
stochastic variables is equal to a sum of the partial cumulants, $\kappa_r =
\sum_{\mathbf{k} \neq 0} \kappa^{(\mathbf{k})}_r$. This is a consequence of
the equalities $\log \Theta_n (u) = \log \Pi_{\mathbf{k} \neq 0} \Theta_{n_{%
\mathbf{k}}} (u) = \sum_{\mathbf{k} \neq 0} \log \Theta_{n_{\mathbf{k}}} (u)$%
. For each canonical-ensemble quasiparticle, the characteristic function can
be easily calculated from the equilibrium density matrix as follows
\begin{equation}
\Theta_{n_{\mathbf{k}}} (u) = Tr \{e^{iu\hat{n}_{\mathbf{k}}} \hat{\rho}_{%
\mathbf{k}} \} = Tr \{e^{iu\hat{n}_{\mathbf{k}}} e^{-\varepsilon_{\mathbf{k}%
} \hat{n}_{\mathbf{k}} /T} \} \Bigl( 1 - e^{-\varepsilon_{\mathbf{k}} /T}
\Bigr) = \frac{z_{\mathbf{k}} -1}{z_{\mathbf{k}} - z} .  \label{III18}
\end{equation}
Here we introduced the exponential function of the single-particle energy
spectrum $\varepsilon_{\mathbf{k}}$, namely $z_{\mathbf{k}}=\exp(\varepsilon_{%
\mathbf{k}} /T)$, and a variable $z=\exp(iu)$ which has the character of a
``fugacity''. As a result, we obtain an explicit formula for the
characteristic function and all cumulants of the number of excited (and,
according to the equation $n_0 = N-n$, condensed) atoms in the ideal Bose
gas in an arbitrary trap as follows:
\[
\log \Theta_n (u) = \sum_{\mathbf{k} \neq 0} \log \Bigl( \frac{z_{\mathbf{k}%
} - 1}{z_{\mathbf{k}} -z} \Bigr) = \sum_{m=1}^{\infty} \tilde{\kappa}_m
\frac{(e^{iu} -1)^m}{m!} = \sum_{r=1}^{\infty} \kappa_r \frac{(iu)^r}{r!} ,
\]
\begin{equation}
\tilde{\kappa}_m = (m-1)! \sum_{\mathbf{k} \neq 0} \bigl(e^{\varepsilon_{%
\mathbf{k}} /T} -1 \bigr)^{-m} ; \qquad \kappa_r = \sum_{m=1}^r
\sigma_r^{(m)} \tilde{\kappa}_m .  \label{III19}
\end{equation}
Here we use the Stirling numbers of the 2nd kind \cite{a},
\begin{equation}
\sigma_r^{(m)} = \frac{1}{m!} \sum_{k=0}^m (-1)^{m-k} {\binom{ m }{k }} k^r
, \quad \bigl( e^x -1 \bigr)^k = k! \sum_{n=k}^{\infty} \sigma^{(k)}_n \frac{%
x^n}{n!} ,  \label{III20}
\end{equation}
that yield a simple expression for the cumulants $\kappa_r$ via the
generating cumulants $\tilde{\kappa}_m$. In particular, the first four
cumulants are given in Eq. (\ref{III2}).

Thus, due to the standard relations (\ref{III16}), the result (\ref{III19})
yields all moments of the condensate fluctuations. Except for the average
value, all cumulants are independent of the total number of atoms in the
trap; they depend only on the temperature and energy spectrum of the trap.
This universal temperature dependence of the condensate fluctuations was
observed and used in \cite{Fierz,pol,Navez,ww,gh} to study the condensate
fluctuations in the ideal Bose gas on the basis of the so-called Maxwell's
demon ensemble approximation. The method of the canonical-ensemble
quasiparticles also agrees with and provides further justification for the
``demon'' approximation. The main point is that the statistics is determined
by numbers and fluctuations of the excited, non-condensed atoms which behave
independently of the total atom number $N$ for temperatures well below the
critical temperature since all ``excess'' atoms stay in the ground state of
the trap. Therefore, one can calculate statistics in a formal limit as if we
have an infinite number of atoms in the condensate. That is we can say that
the condensate plays the part of an infinite reservoir for the excited atoms,
in agreement with previous works~\cite{HKK,Fierz,pol,Navez,ww,gh}.

Obviously, our approximation in Eq. (\ref{III6}) as well as the Maxwell's
demon ensemble approximation does not describe all mesoscopic effects that
can be important very close to the critical temperature or for a very small
number of atoms in the trap. However, it takes into account the effect of a
finite size of a trap via the discreteness of the energy levels $%
\varepsilon_{\mathbf{k}}$, i.e., in this sense the approximation (\ref{III6}%
) describes not only the thermodynamic limit but also systems with a finite
number of atoms $N$. In addition, the mesoscopic effects can be partially
taken into account by a ``grand'' canonical approximation for the probability
distribution of the canonical-ensemble quasiparticle occupation numbers
\begin{equation}
{\tilde \rho}_{\mathbf{k}} ( n_{\mathbf{k}} ) = \exp (- n_{\mathbf{k}} {%
\tilde \varepsilon}_{\mathbf{k}}/ T) ( 1 - \exp ( - {\tilde \varepsilon}_{%
\mathbf{k}} / T)), \qquad {\tilde \varepsilon}_{\mathbf{k}}=\varepsilon_{%
\mathbf{k}} - \mu ,  \label{quasiGC}
\end{equation}
where the chemical potential is related to the mean number of the condensed
atoms ${\bar n}_0 = 1/(1-\exp(-\beta \mu ))$ and should be found
self-consistently from the grand-canonical equation
\begin{equation}
N-{\bar n}_0 = \sum_{\mathbf{k} \neq 0} (e^{{\tilde \varepsilon}_{\mathbf{k}%
}/ T} -1)^{-1}.  \label{quasiGCn0}
\end{equation}
The canonical-ensemble quasiparticle result for all cumulants remains the
same as is given by Eq. (\ref{III19}) above, with the only difference that
now all quasiparticle energies are shifted by a negative chemical potential $%
({\tilde \varepsilon}_{\mathbf{k}}=\varepsilon_{\mathbf{k}} - \mu)$,
\begin{equation}
\tilde{\kappa}_m = (m-1)! \sum_{\mathbf{k} \neq 0} \bigl(e^{\tilde{%
\varepsilon}_{\mathbf{k}} /T} -1 \bigr)^{-m} , \quad m=1, 2, ...; \qquad
\kappa_r = \sum_{m=1}^r \sigma_r^{(m)} \tilde{\kappa}_m .
\label{III19quasiGC}
\end{equation}
The first, $m=1$, equation in Eq.~(\ref{III19quasiGC}) is a self-consistency
equation (\ref{quasiGCn0}). The way it takes into account the mesoscopic
effects (within this mean-number ``grand'' canonical approximation) is
similar to the way in which the self-consistency equation (\ref{III72}) of
the mean-field Popov approximation takes into account the effects of weak
atomic interaction. The results of this canonical-ensemble quasiparticle
approach within the ``grand'' canonical approximation for the quasiparticle
occupations (\ref{quasiGC}) were discussed in Section III for the case of
the isotropic harmonic trap. Basically, the ``grand'' canonical approximation
improves only the result for the mean number of condensed atoms
${\bar n}_0 (T)$, but not for the fluctuations.

\subsection{Ideal gas BEC statistics in arbitrary power-law traps}

The explicit formulas for the cumulants demonstrate that the BEC fluctuations
depend universally and only on the single-particle energy spectrum of the
trap, $\varepsilon_{\mathbf{k}}$. There are three main parameters that enter
this dependence, namely,
\begin{description}
\item[(a)] the ratio of the energy gap between the ground level and the first
excited level in the trap to the temperature, $\varepsilon_1 /T$,
\item[(b)] the exponent of the energy spectrum in the infrared limit,
$\varepsilon_{\mathbf{k}} \propto k^{\sigma} \quad \mbox{at} \quad k
\rightarrow 0$, and
\item[(c)] the dimension of the trap, $d$.
\end{description}
The result~(\ref{III19}) allows to easily analyze the condensate fluctuations
in a general case of a trap with an arbitrary dimension $d \geq1$ of the space
and with an arbitrary power-law single-particle energy spectrum
\cite{Groot,ww,gh97b}
\begin{equation}
\varepsilon_{\mathbf{l}} = \hbar \sum^d_{j=1} \omega_j l^{\sigma}_{j} ,
\qquad \mathbf{l} = \{ l_j ; \quad j=1,2, \dots , d \} ,  \label{III44}
\end{equation}
where $l_j \geq 0$ is a non-negative integer and $\sigma >0$ is an index of
the energy spectrum. The results for the particular traps with a trapping
potential in the form of a box or harmonic potential well can be immediately
deduced from the general case by setting the energy spectrum exponent to be
equal to $\sigma =2$ for a box and $\sigma =1$ for a harmonic trap. We assume
that the eigenfrequencies of the trap are ordered,
$0< \omega_1 \leq \omega_2 \leq \dots \leq \omega_d$, so that the energy gap
between the ground state and the first excited state in the trap is
$\varepsilon_1 = \hbar \omega_1$. All cumulants (\ref{III19}) of the
condensate occupation fluctuations are given by the following formula:
\begin{equation}
\tilde{\kappa}_m = ( m - 1 ) ! \sum_{\mathbf{l} \equiv (l_1,\dots ,l_d ) >
0} [ \exp \Bigl( \frac{\hbar}{T} \sum_{j=1}^d \omega_j l_j^{\sigma} \Bigr) -
1 ]^{-m} , \qquad \kappa_r = \sum_{m=1}^r \sigma_r^{(m)} \tilde{\kappa}_m .
\label{III45}
\end{equation}

Let us consider first again the case of moderate temperatures larger than
the energy gap, $\varepsilon _{1} \ll T<T_{c}$. The first cumulant, i.e.,
the mean number of non-condensed atoms, can be calculated by means of a
continuum approximation of the discrete sum by an integral if the space
dimension of a trap is higher than a critical value, $d>\sigma $. Namely,
one has an usual BEC phase transition with the mean value
\begin{equation}
\kappa _{1}\equiv \bar{n}\equiv N-\bar{n}_{0}=A\zeta \Bigl(\frac{d}{\sigma }%
\Bigr)T^{d/\sigma }=\Bigl(\frac{T}{T_{c}}\Bigr)^{d/\sigma }N,\qquad d>\sigma
,\quad \varepsilon _{1} \ll T<T_{c},  \label{III46}
\end{equation}
where the standard critical temperature is
\begin{equation}
T_{c}=\Bigl[\frac{N}{A\zeta (d/\sigma )}\Bigr]^{\sigma /d},\qquad A=\frac{%
[\Gamma (\frac{1}{\sigma }+1)]^{d}}{\bigl(\Pi _{j=1}^{d}\hbar \omega _{j}%
\bigr)^{1/\sigma }}.  \label{III47}
\end{equation}
The second-order generating cumulant can be calculated by means of this
continuum approximation only if $d>2\sigma $,
\begin{equation}
\tilde{\kappa}_{2}=AT^{d/\sigma }\Bigl(\zeta (\frac{d}{\sigma }-1)-\zeta (%
\frac{d}{\sigma })\Bigr)=\Bigl(\frac{T}{T_{c}}\Bigr)^{d/\sigma }N\frac{\zeta
(\frac{d}{\sigma }-1)-\zeta (\frac{d}{\sigma })}{\zeta (\frac{d}{\sigma })}%
,\qquad d>2\sigma .  \label{III48}
\end{equation}
In the opposite case it has to be calculated via a discrete sum because of a
formal infrared divergence of the integral. Keeping only the main term in
the expansion of the exponent in Eq. (\ref{III45}), $\exp (\frac{\hbar }{T}%
\sum_{j=1}^{d}\omega _{j}l_{j}^{\sigma })-1\approx \frac{\hbar }{T}%
\sum_{j=1}^{d}\omega _{j}l_{j}^{\sigma }$, we find
\begin{equation}
\tilde{\kappa}_{2}=\{\frac{T}{T_{c}}N^{\sigma /d}/\bigl[\Gamma (\frac{1}{%
\sigma }+1)\bigr]^{\sigma }\bigl[\zeta (\frac{d}{\sigma })\bigr]^{\sigma
/d}\}^{2}a_{\sigma ,d}^{(2)},\qquad \sigma <d<2\sigma ,  \label{III49}
\end{equation}
where $a_{\sigma ,d}^{(2)}=\sum_{\mathbf{l}>0}(\Pi _{j=1}^{d}\hbar \omega
_{j})^{2/d}/\varepsilon _{\mathbf{l}}^{2}.$ The ratio of the variance to the
mean number of non-condensed atoms is equal to $\sqrt{\kappa _{2}}/\kappa
_{1}=\sqrt{\tilde{\kappa}_{1}^{-1}+\tilde{\kappa}_{2}/\tilde{\kappa}_{1}^{2}}
$, i.e.,
\begin{equation}
\frac{\sqrt{\langle (n-\bar{n})^{2}\rangle }}{\bar{n}}=N^{-1/2}\Bigl(\frac{%
T_{c}}{T}\Bigr)^{d/(2\sigma )}\sqrt{\zeta (\frac{d}{\sigma }-1)/\zeta (\frac{%
d}{\sigma })},\qquad d>2\sigma ,  \label{III50}
\end{equation}
\begin{equation}
\frac{\sqrt{\langle (n-\bar{n})^{2}\rangle }}{\bar{n}}=\sqrt{\frac{1}{N}\Bigl%
(\frac{T_{c}}{T}\Bigr)^{d/\sigma }+N^{2(\frac{\sigma }{d}-1)}\Bigl(\frac{%
T_{c}}{T}\Bigr)^{2(\frac{d}{\sigma }-1)}a_{\sigma ,d}^{(2)}\bigl[\Gamma (%
\frac{1}{\sigma }+1)\bigr]^{-2\sigma }\bigl[\zeta (\frac{d}{\sigma })\bigr]%
^{-2\sigma /d}},\qquad \sigma <d<2\sigma .  \label{III51}
\end{equation}

We see that the traps with a relatively high dimension of the space, $%
d>2\sigma$, produce normal thermodynamic fluctuations (\ref{III50}) $\propto
N^{-1/2}$ and behave similar to the harmonic trap. However, the traps with a
relatively low dimension of the space, $\sigma <d<2\sigma$, produce
anomalously large fluctuations (\ref{III51}) in the thermodynamic limit,
$\propto N^{\sigma /d -1} \gg N^{-1/2}$ and behave similar to the box with
a homogeneous Bose gas, where there is a formal infrared divergence in the
continuum-approximation integral for the variance.

The third and higher-order central moments $\langle (n-\bar{n})^{m}\rangle $,
or the third and higher-order cumulants $\kappa _{m}$, provide further
parameters to distinguish different traps with respect to their fluctuation
behavior. The $m$-th order generating cumulant can be calculated by means of
the continuous approximation only if $d>m\sigma $,
\begin{equation}
\tilde{\kappa}_{m}=\frac{AT^{d/\sigma }}{\Gamma (\frac{d}{\sigma })}%
\int_{0}^{\infty }\frac{t^{\frac{d}{\sigma }-1}}{(e^{t}-1)^{m}}dt=\Bigl(%
\frac{T}{T_{c}}\Bigr)^{d/\sigma }\frac{N}{\Gamma (\frac{d}{\sigma })\zeta (%
\frac{d}{\sigma })}\int_{0}^{\infty }\frac{t^{\frac{d}{\sigma }-1}}{%
(e^{t}-1)^{m}}dt,\qquad d>m\sigma .  \label{III52}
\end{equation}
In the opposite case we have to use a discrete sum because of a formal
infrared divergence of the integral. Again, keeping only the main term in
the expansion of the exponent in Eq. (\ref{III45}), $\exp (\frac{\hbar }{T}%
\sum_{j=1}^{d}\omega _{j}l_{j}^{\sigma })-1\approx \frac{\hbar }{T}%
\sum_{j=1}^{d}\omega _{j}l_{j}^{\sigma }$, we find
\begin{equation}
\tilde{\kappa}_{m}=\{\frac{T}{T_{c}}N^{\sigma /d}/\bigl[\Gamma (\frac{1}{%
\sigma }+1)\bigr]^{\sigma }\bigl[\zeta (\frac{d}{\sigma })\bigr]^{\sigma
/d}\}^{m}a_{\sigma ,d}^{(m)},\qquad \sigma <d<m\sigma ,  \label{III53}
\end{equation}
where $a_{\sigma ,d}^{(m)}=\sum_{\mathbf{l}>0}(\Pi _{j=1}^{d}\hbar \omega
_{j})^{m/d}/\varepsilon _{\mathbf{l}}^{m}.$ (For the sake of simplicity, as
in Eq. (\ref{III36}), we again omit here a discussion of an obvious
logarithmic factor that suppresses the ultraviolet divergence in the latter
sum $\sum_{\mathbf{l}>0}$ for the marginal case $d=m\sigma $; see, e.g., Eq.
(\ref{III40}).)

We conclude that all cumulants up to the order $m<d/\sigma $ have normal
behavior, $\kappa _{m}\propto N$, but for the higher orders, $m>d/\sigma $,
they acquire an anomalous growth in the thermodynamic limit, $\kappa
_{m}\simeq \tilde{\kappa}_{m}\propto N^{m\sigma /d}$. This result provides a
simple classification of the relative strengths of the higher order
fluctuation properties of the condensate in different traps. In particular,
it makes it obvious that for all power-law traps with $1<d/\sigma <2$ the
condensate fluctuations are not Gaussian, since
\begin{equation}
\frac{\kappa _{m}}{\kappa _{2}^{m/2}}\propto N^{0}\rightarrow const\neq
0\qquad \mbox{at}\quad N\rightarrow \infty ,\qquad m\geq 3,  \label{III54}
\end{equation}
so that the asymmetry coefficient, $\gamma _{1}\equiv \langle (n_{0}-\bar{n}%
_{0})^{3}\rangle /\langle (n_{0}-\bar{n}_{0})^{2}\rangle ^{3/2}\neq 0$, and
the excess coefficient, $\gamma _{2}\equiv \langle (n_{0}-\bar{n}%
_{0})^{4}\rangle /\langle (n_{0}-\bar{n}_{0})^{2}\rangle ^{2}-3\neq 0$, are
not zero. Traps with $d/\sigma >2$ show Gaussian condensate fluctuations,
since all higher-order cumulant coefficients $\kappa _{m}/\kappa _{2}^{m/2}$
vanish, namely,
\begin{equation}
\frac{\kappa _{m}}{\kappa _{2}^{m/2}}\propto N^{1-m/2}\rightarrow 0\qquad %
\mbox{at}\quad N\rightarrow \infty \quad \mbox{if}\quad 3\leq m<\frac{d}{%
\sigma },  \label{III55}
\end{equation}
\begin{equation}
\frac{\kappa _{m}}{\kappa _{2}^{m/2}}\propto N^{m(\frac{\sigma }{d}-\frac{1}{%
2})}\rightarrow 0\qquad \mbox{at}\quad N\rightarrow \infty \quad \mbox{if}%
\quad m>\frac{d}{\sigma }.  \label{III56}
\end{equation}
(For the sake of simplicity, we omit here an analysis of the special cases
when $d/\sigma $ is an integer. It also can be done straightforwardly on the
basis of the result (\ref{III45}).) Very likely, a weak interaction also
violates this non-robust property and makes properties of the condensate
fluctuations in the traps with a relatively high dimension of the space, $%
d/\sigma >2$, similar to that of the box with the homogeneous Bose gas (see
Section VI below), as it is stated below for the particular case of the
harmonic traps.

For traps with a space dimension lower than the critical value, $d<\sigma$,
it is known that a BEC phase transition does not exist (see, e.g., \cite{ww}).
Nevertheless, even in this case there still exists a well-peaked
probability distribution $\rho_0 ( n_0 )$ at low enough temperatures, so
that the condition (\ref{III6}) is satisfied and our general result (\ref
{III19}) describes this effect as well. In this case there is a formal
infrared divergence in the corresponding integrals for all cumulants (\ref
{III19}), starting with the mean value. Hence, all of them should be
calculated as discrete sums. For moderate temperatures we find approximately
\begin{equation}
\tilde{\kappa}_m \simeq (m-1)! \Bigl( \frac{T}{\hbar} \Bigr)^m \sum_{\mathbf{%
l}>0} \Bigl( \sum_{j=1}^d \omega_j l_j^{\sigma} \Bigr)^{-m} \sim (m-1)!
\sum_{j=1}^d \Bigl( \frac{T}{\hbar \omega_j} \Bigr)^m ,  \label{III57}
\end{equation}
so that higher cumulants have larger values. In particular, the mean number
of non-condensed atoms is of the same order as the variance,
\[
\kappa_1 \equiv \bar{n} \equiv N-\bar{n}_0 \simeq \frac{T}{\hbar} \sum_{%
\mathbf{l}>0} \Bigl( \sum_{j=1}^d \omega_j l_j^{\sigma} \Bigr)^{-1} \sim
\sum_{j=1}^d \frac{T}{\hbar \omega_j}
\]
\begin{equation}
\sim \sqrt{\Delta n_0^2} \sim \sqrt{\sum_{j=1}^d \Bigl( \frac{T}{\hbar
\omega_j} \Bigr)^2} , \qquad d < \sigma .  \label{III58}
\end{equation}
Therefore, until
\begin{equation}
\bar{n} \sim \sqrt{\Delta n_0^2} \ll N , \quad \mbox{i.e.}, \quad T \ll T_c =
N / \sum_{\mathbf{l}>0} \Bigl( \sum_{j=1}^d \hbar \omega_j l_j^{\sigma}
\Bigr)^{-1} \sim N / \sum_{j=1}^d \frac{1}{\hbar \omega_j} ,  \label{III59}
\end{equation}
there is a well-peaked condensate distribution, $\sqrt{\Delta n_0^2} \ll \bar{%
n}_0$.

The marginal case of a trap with the critical space dimension, $d=\sigma$,
is also described by our result (\ref{III45}), but we omit its discussion in
the present paper. We mention only that there is also a formal infrared
divergence, in this case a logarithmic divergence, and, at the same time, it
is necessary to keep the whole exponent in Eq. (\ref{III45}); because
otherwise in an approximation like (\ref{III57}) there appears an
ultraviolet divergence. The physical result is that in such traps, e.g.,
in a one-dimensional harmonic trap, a quasi-condensation of the ideal Bose
gas takes place at the critical temperature \cite{gh96,ww} $T_c \sim \hbar
\omega_1 N / \ln N$.

For very low temperatures, $T\ll \varepsilon_1$, the second and higher energy
levels in the trap are not thermally excited and atoms in the ideal Bose gas
in any trap go to the ground level with an exponential accuracy. This
situation is also described by Eq. (\ref{III45}) that yields $\bar{n}_0
\simeq N$ and proves that all cumulants of the number-of-non-condensed-atom
distribution become the same, since all higher-order generating cumulants
exponentially vanish faster than $\kappa_1$,
\begin{equation}
\kappa_m \simeq \kappa_1 \equiv \tilde{\kappa}_1 , \qquad \tilde{\kappa}_m
\simeq (m-1)! \sum_{j=1}^d e^{-m\hbar \omega_j /T} , \qquad T \ll
\varepsilon_1.  \label{III60}
\end{equation}
The conclusion is that for the ideal Bose gas in any trap the distribution
of the number of non-condensed atoms becomes Poissonian at very low
temperatures. It means that the distribution of the number of condensed
atoms is not Poissonian, but a ``mirror" image of the Poisson's
distribution. We see, again, that the complementary number of non-condensed
atoms, $n=N-n_0$, is more convenient for the characterization of the
condensate statistics. A physical reason for this is that the non-condensed
atoms in different excited levels fluctuate independently.

The formulas (\ref{III45})--(\ref{III60}) for the power-law traps contain all
corresponding formulas for a box ($d=3, \quad \sigma =2$) and for a harmonic
trap ($d=3, \quad \sigma =1$) in a 3-dimensional space as particular
cases. It is worth to stress that BEC fluctuations in the ideal gas for the latter two cases are very different. In the box, if the temperature is larger than the trap energy gap, $\varepsilon_1 \ll T < T_c$, all cumulants, starting with the variance, $m \geq 2$, are anomalously large and dominated by the lowest energy modes, i.e., formally infrared divergent,
\begin{equation}
\kappa_m \approx \tilde{\kappa}_m \propto (T/T_c )^m N^{2 m / 3} ,
\qquad m \geq 2 .
\label{III24}
\end{equation}
Only the mean number of condensed atoms,
\begin{equation}
\bar{n}_0 = N - \bar{n} = N - \kappa_1 = N \Bigl( 1 - \bigl( T /
T_c )^{3/2} \Bigr) , \qquad T_c = \frac{2 \pi \hbar^2}{m} \Bigl( \frac{N}{L^3 \zeta(3/2)} \Bigr)^{2/3} , \quad
\label{III28}
\end{equation}
can be calculated correctly via replacement of the discrete sum in Eq. (\ref{III19}) by an integral $\int_0^{\infty} ...d^3 k$. The correct value of the squared variance,
\begin{equation}
\kappa_2 \equiv <(n_0 - \bar{n}_0 )^2 > = N^{4/3} \left (\frac{T}{T_c}
\right )^2 \frac{s_4}{\pi^2 (\zeta(3/2))^{4/3}} , \qquad
s_4 = \sum_{{\bf l} \neq 0} \frac{1}{{\bf l}^4} = 16.53,
\label{III25}
\end{equation}
can be calculated from Eq.~(\ref{III19}) only as a discrete sum.
Thus, for the box the condensate fluctuations are anomalous and non-Gaussian
even in the thermodynamic limit. To the contrary, for the harmonic trap with
temperature much larger than the energy gap, $\varepsilon_1 \ll T < T_c$,
the condensate fluctuations are Gaussian in the thermodynamic limit. This is
because, contrary to the case of the homogeneous gas, in the harmonic
trap only the third and higher order cumulants, $m \geq 3$, are
lowest-energy-mode dominated, i.e., formally infrared divergent,
$$\kappa_3 \approx \tilde{\kappa}_3 = 2 \sum_{{\bf l} > 0}
\Bigl( e^{\hbar {\bf \omega} {\bf l} / T} - 1 \Bigr)^{-3} ,$$
\begin{equation}
\kappa_m \approx \tilde{\kappa}_m \approx ( m - 1 ) ! \Bigl(
\frac{T}{\hbar} \Bigr)^m \sum_{{\bf l} > 0} \bigl( {\bf \omega}
{\bf l} \bigr)^{-m} \propto N^{m / 3} , \qquad m > 3 ,
\label{III36}
\end{equation}
and they are small compared with an appropriate power of the variance squared
\begin{equation}
\kappa_2 \equiv < ( n_0 -\bar{n}_0 )^2 > =
\frac{\zeta ( 2 )}{\zeta ( 3 )} \Bigl( \frac{T}{T_c} \Bigr)^3 N .
\label{III37}
\end{equation}
The asymmetry coefficient behaves as
\begin{equation}
\gamma_1
\equiv \frac{\kappa_3}{\kappa_2^{3/2}} \equiv \frac{< ( n_0 - \bar{n}_0 )^3
>}{< ( n_0 - \bar{n}_0 )^2 >^{3/2}} \propto \frac{\log N}{N^{1/2}}
\rightarrow 0 ,
\label{III40}
\end{equation}
and all higher normalized cumulants ($m \geq 3$)
vanish in the thermodynamic limit, $N \to \infty$, as follows:
\begin{equation}
\frac{\kappa_m}{\kappa_2^{m/2}} \propto \frac{1}{N^{m/6}} \rightarrow 0 .
\label{III41}
\end{equation}

It is important to realize that a weak interaction violates this non-robust
property and makes properties of the condensate fluctuations in a harmonic
trap similar to that of the homogeneous gas in the box (see Section VI below).
For the variance, the last fact was first pointed out in \cite{Pit98}.

\subsection{Equivalent formulation in terms of the poles of the generalized
Zeta function}

Cumulants of the BEC fluctuations in the ideal Bose gas, Eq. (\ref{III19}),
can be written in an equivalent form which is quite interesting
mathematically (see~\cite{PhysicaA} and references therein). Namely,
starting with the cumulant generating function $\ln \Xi _{ex}(\beta ,z)$,
where $\beta =1/k_{B}T$ and $z=e^{\beta \mu }$,
\begin{equation}
\ln \Xi _{ex}(\beta ,z) =-\sum_{\nu =1}^{\infty }\ln (1-z\exp [-\beta
(\varepsilon _{\nu }-\varepsilon _{0})]) = \sum_{\nu =1}^{\infty
}\sum_{n=1}^{\infty }\frac{z^{n}\exp [-n\beta (\varepsilon _{\nu
}-\varepsilon _{0})]}{n} \; ,
\end{equation}
we use the Mellin-Barnes transform
\[
e^{-a}=\frac{1}{2\pi i}\int_{\tau -i\infty }^{\tau +i\infty
}dt \, a^{-t}\Gamma(t)
\]
to write
\[
\ln \Xi _{ex}(\beta ,z) =\sum_{\nu =1}^{\infty }\sum_{n=1}^{\infty }\frac{%
z^{n}}{n}\frac{1}{2\pi i}\int_{\tau -i\infty }^{\tau +i\infty }dt \, \Gamma (t)%
\frac{1}{[n\beta (\varepsilon _{\nu }-\varepsilon _{0})]^{t}} =
\]
\begin{equation}
\frac{1}{2\pi i}\int_{\tau -i\infty }^{\tau +i\infty }dt \,  \Gamma (t)\sum_{\nu
=1}^{\infty }\frac{1}{[\beta (\varepsilon _{\nu }-\varepsilon _{0})]^{t}}
\sum_{n=1}^{\infty }\frac{z^{n}}{n^{t+1}} .
\end{equation}
Recalling the series representation of the Bose functions $g_{\alpha
}(z)=\sum_{n=1}^{\infty } z^{n}/n^{\alpha }$ and introducing the
generalized, ``spectral'' Zeta function $Z(\beta ,t)=[\beta (\varepsilon
_{\nu }-\varepsilon _{0})]^{-t}$, we arrive at the convenient (and exact)
integral representation
\begin{equation}
\ln \Xi _{ex}(\beta ,z)=\frac{1}{2\pi i}\int_{\tau -i\infty }^{\tau +i\infty
}dt\,\Gamma (t)Z(\beta ,t)g_{t+1}(z) .  \label{p1}
\end{equation}

Utilizing the well-known relations $z\frac{d}{dz}g_{\alpha }(z)=g_{\alpha
-1}(z)$ and $g_{\alpha }(1)=\zeta (\alpha )$, where $\zeta (\alpha )$
denotes the original Riemann Zeta function, Eq. (\ref{p1}) may be written in
the appealing compact formula
\begin{equation}
\kappa _{k}(\beta ) =\left( z\frac{\partial }{\partial z}\right) ^{k}\ln \Xi
_{ex}(\beta ,z)|_{z=1} =\frac{1}{2\pi i}\int_{\tau -i\infty }^{\tau +i\infty
}dt\,\Gamma (t)Z(\beta ,t)\zeta (t+1-k) .  \label{p2}
\end{equation}

Thus, by means of the residue theorem, Eq. (\ref{p2}) links all cumulants of
the canonical distribution in the condensate regime to the poles of the
generalized Zeta function $Z(\beta ,t)$, which embodies all the system's
properties, and to the pole of a system-independent Riemann Zeta function,
the location of which depends on the order $k$ of the respective cumulant.
The formula (\ref{p2}) provides a systematic asymptotic expansion of the
cumulants $\kappa _{k}(\beta )$ through the residues of the analytically
continued integrands, taken from right to left. The large-system behavior is
extracted from the leading pole, finite-size corrections are encoded in the
next-to-leading poles, and the non-Gaussian nature of the condensate
fluctuations is definitely seen. The details and examples of such analysis
can be found in \cite{PhysicaA}.

Concluding the Section V, it is worthwhile to mention that previously only
first two moments, $\kappa_1$ and $\kappa_2$, were analyzed for the ideal
gas \cite{Dingle,Groot,Fraser,Reif,Hauge,Ziff,ww,gh,KSZZ}, and the known
results coincide with ours. Our explicit formulas provide a complete answer
to the problem of all higher moments of the condensate fluctuations in the
ideal gas. The canonical-ensemble quasiparticle approach, taken together
with the master equation approach gives a fairly complete picture of the
central moments.

For the interacting Bose gas this problem becomes much more involved. We
address it in the next, last part of this review within a simple
approximation that takes into account one of the main effects of the
interaction, namely, the Bogoliubov coupling.

\section{Why condensate fluctuations in the interacting Bose gas are
anomalously large, non-Gaussian, and governed by universal infrared
singularities ?}

In this Section, following the Refs. \cite{KKS-PRL,KKS-PRA}, the analytical
formulas for the statistics, in particular, for the characteristic function
and all cumulants, of the Bose-Einstein condensate in the dilute, weakly
interacting gases in the canonical ensemble are derived using the canonical
ensemble quasiparticle method. We prove that the ground-state occupation
statistics is not Gaussian even in the thermodynamic limit. We calculate the
effect of Bogoliubov coupling on suppression of ground-state occupation
fluctuations and show that they are governed by a pair-correlation,
squeezing mechanism.

It is shown that the result of Giorgini, Pitaevskii, and Stringari (GPS)
\cite{Pit98} for the variance of condensate fluctuations is correct, and the
criticism of Idziaszek and others \cite{Navez99,Yukalov} is incorrect.
A crossover between the interacting and ideal Bose gases is described. In
particular, it is demonstrated that the squared variance of the condensate
fluctuations for the interacting Bose gas, Eq. (\ref{III72a}), tends to a
half of that for the ideal Bose gas, Eq. (\ref{III25}), because the atoms
are coupled in strongly correlated pairs such that the number of independent
degrees of freedom contributing to the fluctuations of the total number of
excited atoms is only 1/2 the atom number $N$. This pair correlation
mechanism is a consequence of two-mode squeezing due to Bogoliubov coupling
between $\mathbf{k} $ and $-\mathbf{k}$ modes. Hence, the fact that the
fluctuation in the interacting Bose gas is 1/2 of that in the ideal Bose
gas is not an accident, contrary to the conclusion of GPS. Thus, there is a
deep (not accidental) parallel between the fluctuations of ideal and
interacting bosons.

Finally, physics and universality of the anomalies and infrared
singularities of the order parameter fluctuations for different systems with
a long range order below a critical temperature of a second order phase
transition, including strongly interacting superfluids and ferromagnets, is
discussed. In particular, an effective nonlinear $\sigma$ model for the
systems with a broken continuous symmetry is outlined and the crucial role
of the Goldstone modes fluctuations combined with an inevitable geometrical
coupling between longitudinal and transverse order parameter fluctuations
and susceptibilities in the constrained systems is demonstrated.

\subsection{Canonical-ensemble quasiparticles in the atom-number-conserving
Bogoliubov approximation}

We consider a dilute homogeneous Bose gas with a weak interatomic scattering described by the well-known Hamiltonian \cite{LL}
\begin{equation}
H=\sum_{\mathbf{k}}\frac{\hbar ^{2}\mathbf{k}^{2}}{2M}\hat{a}_{\mathbf{k}%
}^{+}\hat{a}_{\mathbf{k}}+\frac{1}{2V}\sum_{\{\mathbf{k}_{i}\}}\langle
\mathbf{k}_{3}\mathbf{k}_{4}|U|\mathbf{k}_{1}\mathbf{k}_{2}\rangle \hat{a}_{%
\mathbf{k}_{4}}^{+}\hat{a}_{\mathbf{k}_{3}}^{+}\hat{a}_{\mathbf{k}_{2}}%
\hat{a}_{\mathbf{k}_{1}},  \label{III61}
\end{equation}
where $V=L^{3}$ is a volume of a box confining the gas with periodic
boundary conditions.
The main effect of the weak interaction is the Bogoliubov coupling between
bare canonical-ensemble quasiparticles, $\hat{\beta}_{\mathbf{k}} = \hat{%
\beta}_0^+ \hat{a}_{\mathbf{k}}$, via the condensate. It may be described,
to a first approximation, by a quadratic part of the Hamiltonian (\ref{III61}%
), i.e., by the atom-number-conserving Bogoliubov Hamiltonian \cite{g}
\begin{equation}
H_B = \frac{N ( N - 1 ) U_0}{2 V} + \sum_{\mathbf{k} \neq 0} \Bigl( \frac{%
\hbar^2 \mathbf{k}^2}{2 M} + \frac{(\hat{n}_0 +1/2) U_{\mathbf{k}}}{V}
\Bigr) \hat{\beta}_{\mathbf{k}}^+ \hat{\beta}_{\mathbf{k}} + \frac{1}{2 V}
\sum_{\mathbf{k} \neq 0} \Bigl( U_{\mathbf{k}} \sqrt{( 1 + \hat{n}_0 ) ( 2 +
\hat{n}_0 )} \hat{\beta}_{\mathbf{k}}^+ \hat{\beta}_{\mathbf{-k}}^+ + h.c.
\Bigr) ,  \label{III62}
\end{equation}
where we will make an approximation $\hat{n}_0 \simeq \bar{n}_0 \gg1$, which
is consistant with our main assumption (\ref{III6}) of the existence of a
well peaked condensate distribution function. Then, the Bogoliubov canonical
transformation,
\[
\hat{\beta}_{\mathbf{k}} = u_{\mathbf{k}} \hat{b}_{\mathbf{k}} + v_{\mathbf{k%
}} \hat{b}_{-\mathbf{k}}^+ ; \qquad \qquad u_{\mathbf{k}} = \frac{1}{\sqrt{1
- A_{\mathbf{k}}^2}} , \quad v_{\mathbf{k}} = \frac{A_{\mathbf{k}}}{\sqrt{1
- A_{\mathbf{k}}^2}} ,
\]
\begin{equation}
\hat{\beta}_{\mathbf{k}}^+ = u_{\mathbf{k}} \hat{b}_{\mathbf{k}}^+ + v_{%
\mathbf{k}} \hat{b}_{-\mathbf{k}} ; \qquad \qquad A_{\mathbf{k}} = \frac{V}{%
\bar{n}_0 U_{\mathbf{k}}} \Bigl( \varepsilon_{\mathbf{k}} - \frac{\hbar^2
\mathbf{k}^2}{2 M} - \frac{\bar{n}_0 U_{\mathbf{k}}}{V} \Bigr) ,
\label{III63}
\end{equation}
describes the condensate canonical-ensemble quasiparticles which have a
``gapless" Bogoliubov energy spectrum and fluctuate independently in the
approximation (\ref{III62}), since
\begin{equation}
H_B = E_0 + \sum_{\mathbf{k} \neq 0} \varepsilon_{\mathbf{k}} \hat{b}^+_{%
\mathbf{k}} \hat{b}_{\mathbf{k}} , \qquad \varepsilon_{\mathbf{k}} = \sqrt{%
\Bigl( \frac{\hbar^2 \mathbf{k}^2}{2 M} + \frac{\bar{n}_0 U_{\mathbf{k}}}{V}
\Bigr)^2 - \Bigl( \frac{\bar{n}_0 U_{\mathbf{k}}}{V} \Bigr)^2} .
\label{III65}
\end{equation}
In other words, we again have an ideal Bose gas although now it
consists of the dressed quasiparticles which are different both
from the atoms and bare (canonical-ensemble) quasiparticles
introduced in Section V. Hence, the analysis of fluctuations can
be carried out in a similar fashion to the case of the
non-interacting, ideal Bose gas of atoms. This results in a
physically transparent and analytical theory of BEC fluctuations
that was suggested and developed in \cite{KKS-PRL,KKS-PRA}.

The only difference with the ideal gas is that now the equilibrium density
matrix,
\begin{equation}
\hat{\rho} = \Pi_{\mathbf{k} \neq 0} \hat{\rho}_{\mathbf{k}} , \qquad \qquad
\hat{\rho}_{\mathbf{k}} = e^{- \varepsilon_{\mathbf{k}} \hat{b}_{\mathbf{k}%
}^+ \hat{b}_{\mathbf{k}} / T} \Bigl( 1 - e^{- \varepsilon_{\mathbf{k}} / T}
\Bigr) ,  \label{III66}
\end{equation}
is not diagonal in the bare atomic occupation numbers, the statistics of
which we are going to calculate. This feature results in the well-known
quantum optics effect of squeezing of the fluctuations. The number of atoms
with coupled momenta $\mathbf{k}$ and $-\mathbf{k}$ is determined by the
Bogoliubov coupling coefficients according to the following equation:
\begin{equation}
\hat{a}_{\mathbf{k}}^+ \hat{a}_{\mathbf{k}} + \hat{a}_{-\mathbf{k}}^+ \hat{a}%
_{-\mathbf{k}} = \hat{\beta}_{\mathbf{k}}^+ \hat{\beta}_{\mathbf{k}} + \hat{%
\beta}_{-\mathbf{k}}^+ \hat{\beta}_{-\mathbf{k}} = ( u_{\mathbf{k}}^2 + v_{%
\mathbf{k}}^2 ) ( \hat{b}_{\mathbf{k}}^+ \hat{b}_{\mathbf{k}} + \hat{b}_{-%
\mathbf{k}}^+ \hat{b}_{-\mathbf{k}} ) + 2 u_{\mathbf{k}} v_{\mathbf{k}} (
\hat{b}_{\mathbf{k}}^+ \hat{b}_{-\mathbf{k}}^+ + \hat{b}_{\mathbf{k}} \hat{b}%
_{-\mathbf{k}} ) + 2 v_{\mathbf{k}}^2 .  \label{III67}
\end{equation}

\subsection{Characteristic function and all cumulants of BEC fluctuations}

The characteristic function for the total number of atoms in the two, $%
\mathbf{k}$ and $-\mathbf{k}$, modes squeezed by Bogoliubov coupling is
calculated in \cite{KKS-PRL,KKS-PRA} as
\[
\Theta_{\pm \mathbf{k}} ( u ) \equiv Tr \Bigl( e^{i u ( \hat{\beta}_{\mathbf{%
k}}^+ \hat{\beta}_{\mathbf{k}} + \hat{\beta}_{-\mathbf{k}}^+ \hat{\beta}_{-%
\mathbf{k}} )} e^{- \varepsilon_{\mathbf{k}} ( \hat{b}_{\mathbf{k}}^+ \hat{b}%
_{\mathbf{k}} + \hat{b}_{-\mathbf{k}}^+ \hat{b}_{-\mathbf{k}} ) / T} \Bigl(
1 - e^{- \varepsilon_{\mathbf{k}} / T} \Bigr)^2 \Bigr)
\]
\begin{equation}
= \frac{( z ( A_{\mathbf{k}} ) - 1 ) ( z ( - A_{\mathbf{k}} ) - 1 )}{( z (
A_{\mathbf{k}} ) - e^{i u} ) ( z ( - A_{\mathbf{k}} ) - e^{i u} )} , \qquad
z ( A_{\mathbf{k}} ) = \frac{A_{\mathbf{k}} - e^{\varepsilon_{\mathbf{k}} /
T}}{A_{\mathbf{k}} e^{\varepsilon_{\mathbf{k}} / T} - 1} .  \label{III68}
\end{equation}
The term ``squeezing" originates from the studies of a noise reduction in
quantum optics (see the discussion after Eq.~(\ref{III74})).

The characteristic function for the distribution of the total number of the
excited atoms is equal to the product of the coupled-mode characteristic
functions, $\Theta_n ( u ) \quad = \quad \Pi_{\mathbf{k} \neq 0 , mod\{ \pm
\mathbf{k} \}} \Theta_{\pm \mathbf{k}} ( u )$, since different pairs of $(%
\mathbf{k},-\mathbf{k})$-modes are independent to the first approximation (%
\ref{III62}). The product $\Pi$ runs over all different pairs of $(\mathbf{k}%
,-\mathbf{k})$-modes.

It is worth noting that by doing all calculations via the canonical-ensemble
quasiparticles (Section V) we automatically take into account all
correlations introduced by the canonical-ensemble constraint. As a result,
similar to the ideal gas (Eq. (\ref{III19})), we obtain the explicit formula
for all cumulants in the dilute weakly interacting Bose gas,
\begin{equation}
\tilde{\kappa}_m = \frac 12 ( m - 1 ) ! \sum_{\mathbf{k} \neq 0} \Bigl[
\frac{1}{\bigl( z ( A_{\mathbf{k}} ) - 1 \bigr)^m} + \frac{1}{\bigl( z ( -
A_{\mathbf{k}} ) - 1 \bigr)^m} \Bigr] , \qquad \kappa_r = \sum_{m=1}^r
\sigma_r^{(m)} \tilde{\kappa}_m .  \label{III71}
\end{equation}
In comparison with the ideal Bose gas, Eq. (\ref{III19}), we have
effectively a mixture of two species of atom pairs with $z ( \pm A_{\mathbf{k%
}} )$ instead of $\exp ( \varepsilon_{\mathbf{k}} / T )$.

\begin{figure}[tbp]
\center\epsfxsize=5.5cm\epsffile{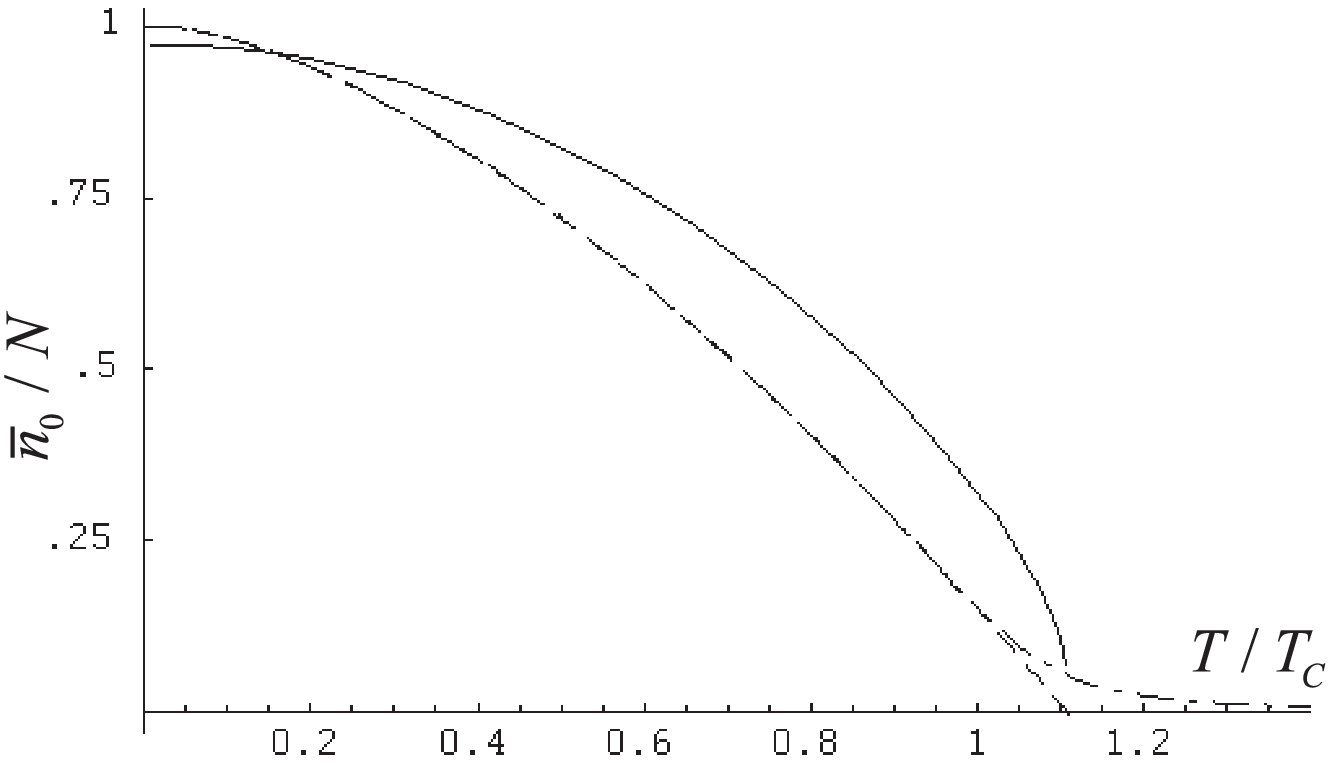} \center\epsfxsize=5.5cm%
\epsffile{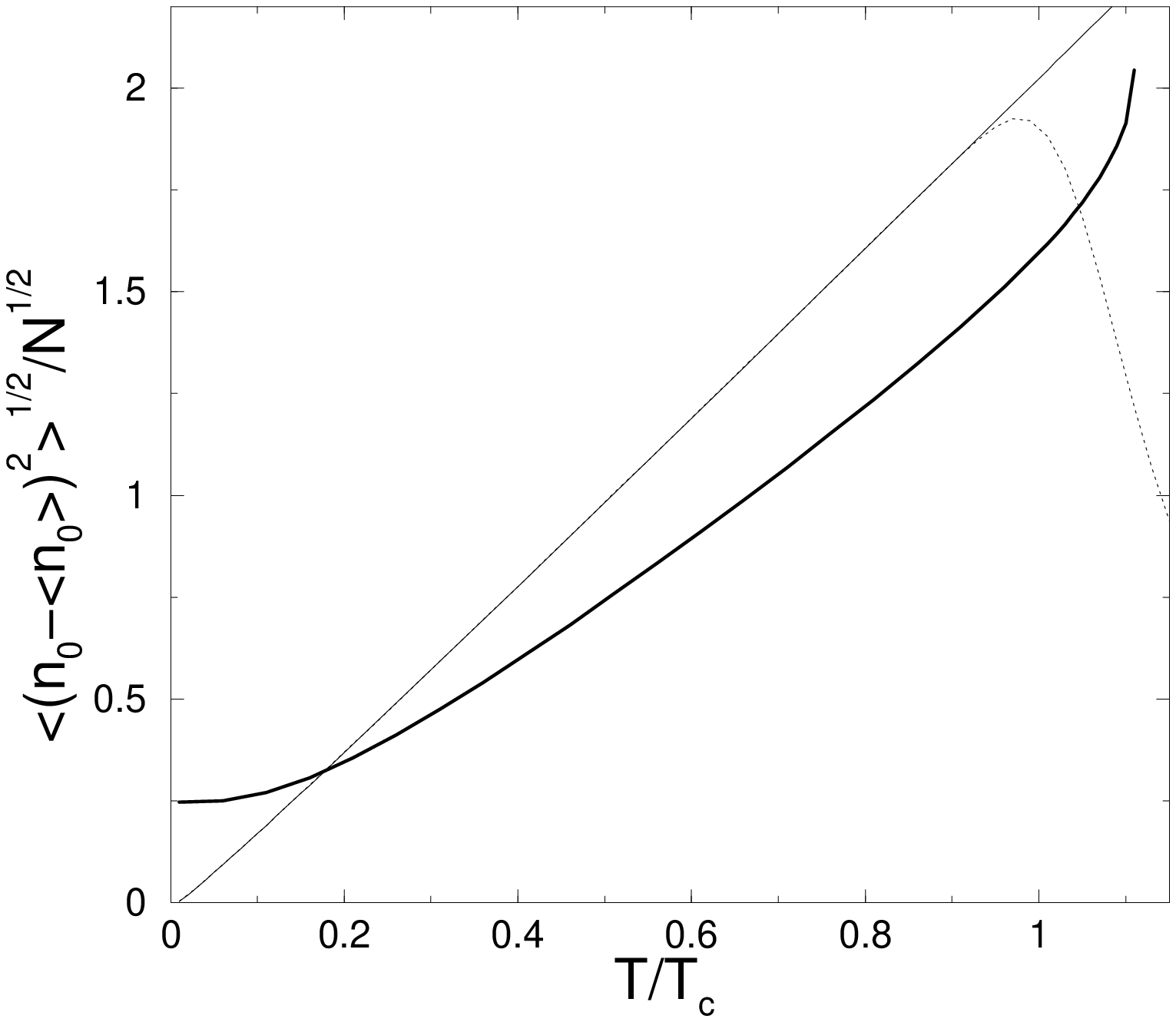} \center\epsfxsize=5.5cm\epsffile{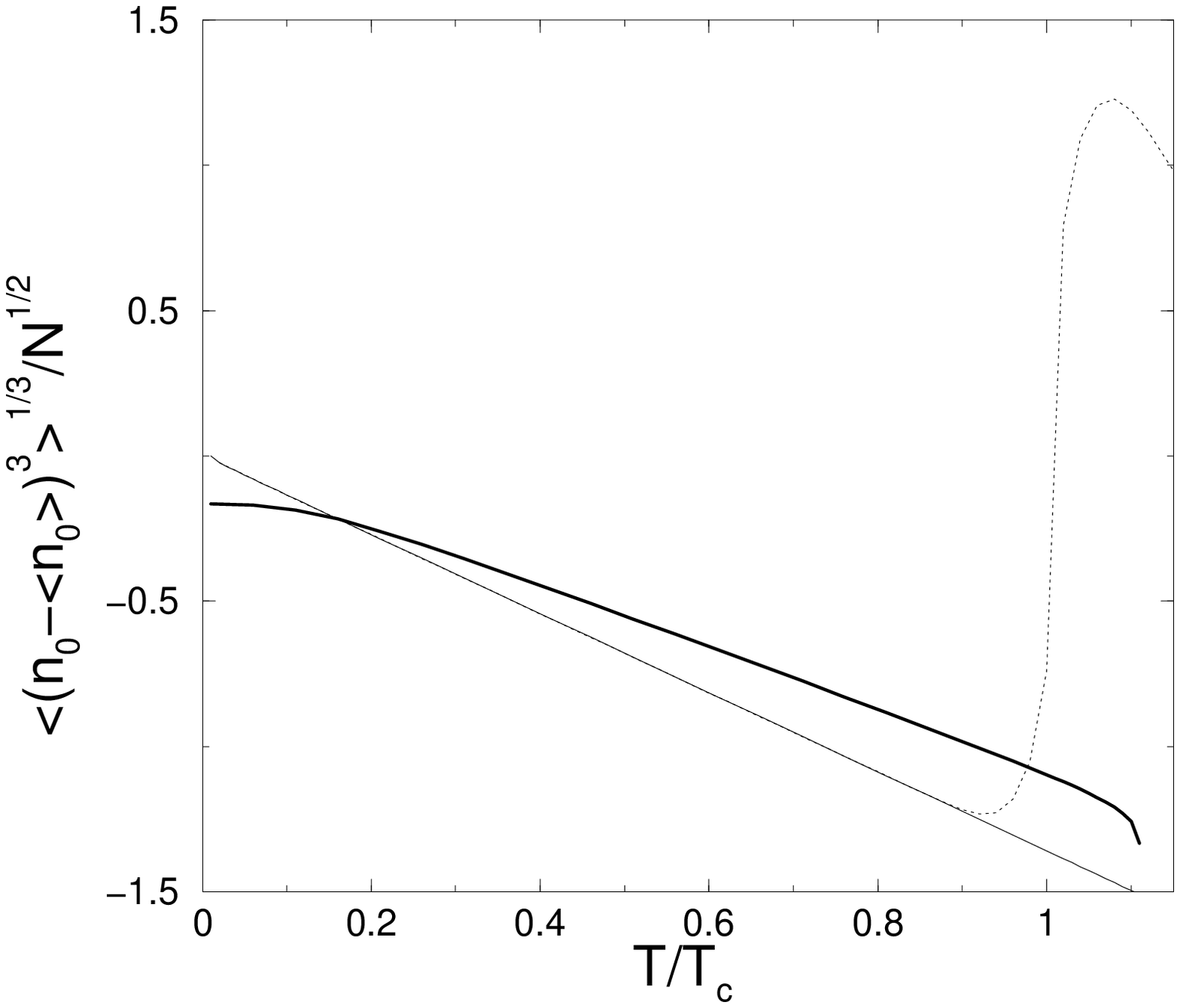} \center%
\epsfxsize=5.5cm\epsffile{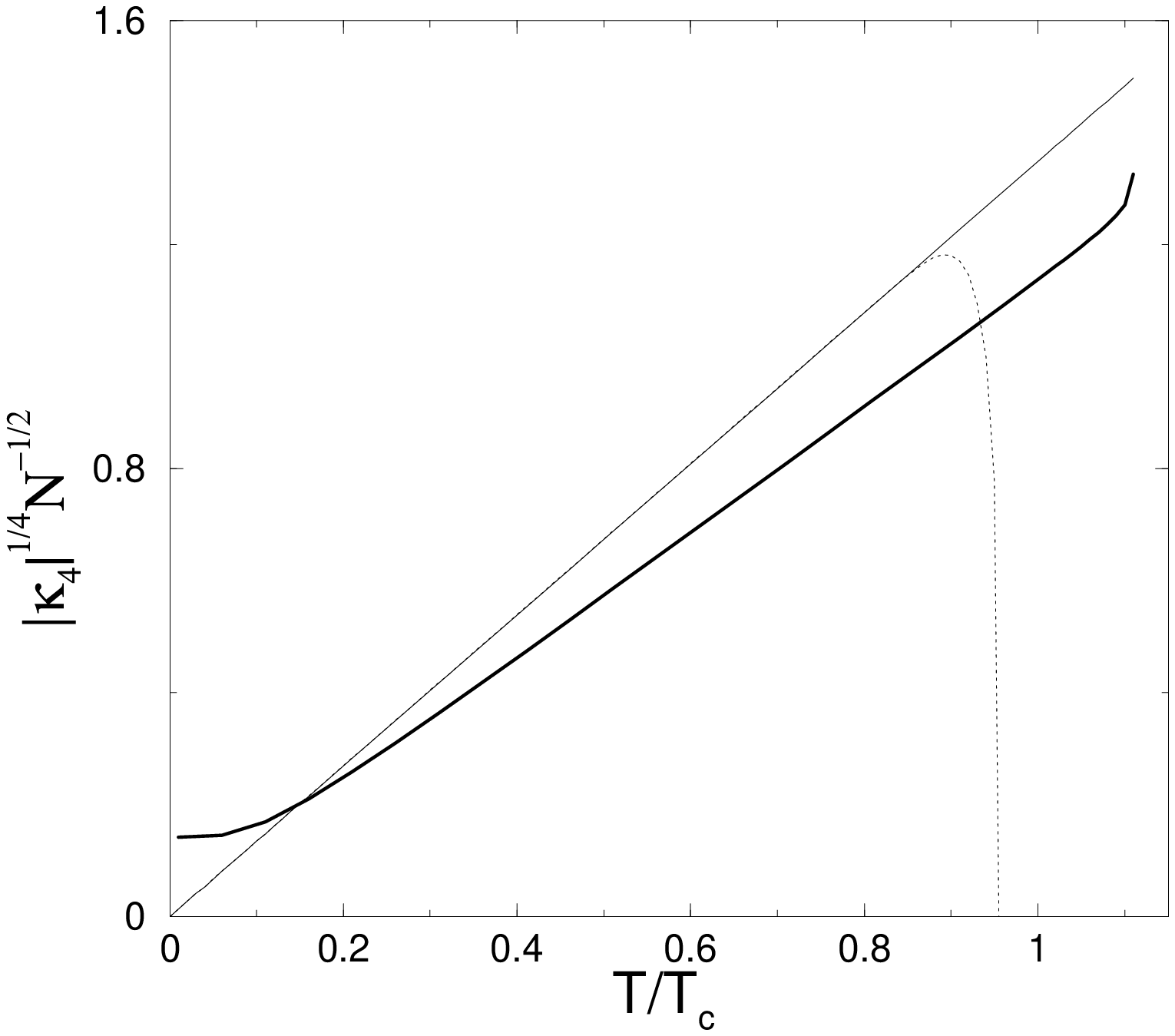}
\caption{Temperature scaling of the first four cumulants, the mean value
$\bar{n}_0/N = N-\protect\kappa_1/N$, the variance
$\protect\sqrt{\protect\kappa_2/N} =
\langle(n_0-\bar{n}_0)^2\rangle^{1/2}/N^{1/2}$, the third central moment
$-\protect\kappa_3^{1/3}/N^{1/2} =
\langle(n_0-\bar{n}_0)^3\rangle^{1/3}/N^{1/2}$, the fourth cumulant
$|\protect\kappa_4|^{1/4}/N^{1/2} =
|\langle(n_0-\bar{n}_0)^4\rangle-3\protect\kappa_2^2|/N^2$,
of the ground-state occupation fluctuations for the dilute weakly interacting
Bose gas (Eq. (\ref{III71})), with
$U_0 N^{1/3} /\protect\varepsilon_1 V = 0.05$ (thick solid lines), as compared
with Eq. (\ref{III19}) (thin solid lines) and with the exact recursion relation (\ref{III77})
(dot-dashed lines) for the ideal gas in the box; $N=1000$. For the ideal gas
our results (thin solid lines) are almost indistinguishable from the exact
recursion calculations (dot-dashed lines) in the condensed region,
$T<T_c(N)$. Temperature is normalized by the standard thermodynamic-limit
critical value $T_c ( N=\infty )$ that differs from the finite-size value
$T_c (N)$, as is clearly seen in graphs.}
\label{FigIII2}
\end{figure}

It is important to emphasize that the first equation in (\ref{III71}), $m=1$,
is a non-linear self-consistency equation,
\begin{equation}
N - \bar{n}_0 = \kappa_1 (\bar{n}_0) \equiv \sum_{\mathbf{k} \neq 0} \frac{1
+ A_{\mathbf{k}}^2 e^{\varepsilon_{\mathbf{k}} / T}}{(1-A_{\mathbf{k}}^2)
(e^{\varepsilon_{\mathbf{k}} / T} - 1)} ,  \label{III72}
\end{equation}
to be solved for the mean number of ground-state atoms $\bar{n}_0(T)$, since
the Bogoliubov coupling coefficient (\ref{III63}), $A_{\mathbf{k}}$, and the energy spectrum (\ref{III65}), $\varepsilon_{\mathbf{k}}$, are themselves functions of the mean value $\bar{n}_0$. Then, all the other equations in (\ref{III71}), $m \geq 2$, are nothing else but explicit expressions for all cumulants, $m \geq 2$, if one substitutes the solution of the self-consistency equation (\ref{III72}) for the mean value $\bar{n}_0$. The Eq. (\ref{III72}), obtained here for the interacting Bose gas (\ref{III62}) in the canonical-ensemble quasiparticle approach, coincides precisely with the self-consistency
equation for the grand-canonical dilute gas in the so-called first-order
Popov approximation (see a review in \cite{Griffin}). The latter is well
established as a reasonable first approximation for the analysis of the
finite-temperature properties of the dilute Bose gas and is not valid only
in a very small interval near $T_c$, given by
$T_c - T < a(N/V)^{1/3}T_c \ll T_c$,
where $a=MU_0/4\pi \hbar^2$ is a usual $s$-wave scattering length. The
analysis of the Eq.~(\ref{III72}) shows that in the dilute gas the
self-consistent value $\bar{n}_0 (T)$ is close to that given by the ideal
gas model, Eq. (\ref{III28}), and for very low temperatures goes smoothly to
the value given by the standard Bogoliubov theory \cite{LL,agd,fw} for a
small condensate depletion, $N-\bar{n}_0 \ll N$. This is illustrated by Fig.
\ref{FigIII2}a. (Of course, near the critical temperature $T_c$ the number
of excited quasiparticles is relatively large, so that along with the
Bogoliubov coupling (\ref{III62}) other, higher order effects of interaction
should be taken into account to get a complete theory.) Note that the effect
of a weak interaction on the condensate fluctuations is very significant
(see Fig. \ref{FigIII2}b-d) even if the mean number of condensed atoms
changes by relatively small amount.

\subsection{Surprises: BEC fluctuations are anomalously large and
non-Gaussian even in the thermodynamic limit}

According to the standard textbooks on statistical physics, e.g.,
\cite{LL,la,Reif,kittel}, any extensive variable $\hat{C}$ of a thermodynamic
system has vanishing relative root-mean-square fluctuations. Namely, in the
thermodynamic limit, a relative squared variance
$\langle (\hat{C}-\bar{C})^{2}\rangle /{\bar{C}}^{2}=
\langle \hat{C}^{2}\rangle /\langle \hat{C}\rangle ^{2}-1\propto V^{-1}$
goes inversely proportional to the system volume $V$, or total number of
particles~$N$. This fundamental property originates from the presence of a
finite correlation length $\xi$ that allows us to partition a large system
into an extensive number $V/\xi ^{3}$ of statistically independent subvolumes,
with a finite variance in each subvolume. As a result, the central limit
theorem of probability theory yields a Gaussian distribution for the variable
$\hat{C}$ with a standard scaling for variance,
$\sqrt{\langle (\hat{C}-\bar{C})^{2}\rangle}\sim V^{1/2}$. Possible deviations
from this general rule are especially interesting. It turns out that BEC in a
trap is one of the examples of such peculiar systems. A physical reason for
the anomalously large and non-Gaussian BEC fluctuations is the existence of
the long range order below the critical temperature of the second order phase
transition.

Let us show in detail that the result (\ref{III71}) implies, similar to the
case of the ideal homogeneous gas (Section V), that the ground-state
occupation fluctuations in the weakly interacting gas are not Gaussian in
the thermodynamic limit, and anomalously large. The main fact is that the
anomalous contribution to the BEC fluctuation cumulants comes from the modes
which have the most negative Bogoliubov coupling coefficient $A_{\mathbf{k}%
}\approx -1$ since in this case one has $z(A_{\mathbf{k}})\rightarrow 1$,
so that this function produces a singularity in the first term in
Eq.~(\ref{III71}). (The second term in Eq.~(\ref{III71}) never makes a
singular contribution since always $z(-A_{\mathbf{k}})<-1$ or $z(-A_{\mathbf{k}})>\exp(\varepsilon_{\mathbf{k}}/T)$.) These modes,
with $A_{\mathbf{k}}\approx -1$, exist only if the interaction energy
$g=\bar{n}_{0}U_{\mathbf{k}}/V$ in the Bogoliubov Hamiltonian (\ref{III62})
is larger than the energy gap between the ground state and the first excited
states in the trap,
$\varepsilon _{1}=\Bigl(\frac{2\pi \hbar }{L}\Bigr)^{2}\frac{1}{2M}$, i.e.
\begin{equation}
\frac{g}{\varepsilon_1} \equiv \frac{2a{\bar n}_0}{\pi L} \gg 1 \quad ,
\label{g}
\end{equation}
and are the infrared modes with the longest wavelength of the trap energy
spectrum. In terms of the scattering length for atom-atom collisions,
$a=MU_{\mathbf{k}}/4\pi \hbar ^{2}$, the latter condition (\ref{g})
coincides with a familiar condition for the Thomas-Fermi regime, in which
the interaction energy is much larger then the atom's kinetic energy.

Let us use a representation which is obvious from Eqs.~(\ref{III63}) and
(\ref{III65}),
\begin{equation}
z(A_{\mathbf{k}})-1\equiv \frac{2\varepsilon _{\mathbf{k}}}{g(1-A_{\mathbf{k}%
})}\Bigl[1+\frac{A_{\mathbf{k}}(1-A_{\mathbf{k}})}{1+e^{\varepsilon _{%
\mathbf{k}}/T}}\Bigr]\approx \frac{\varepsilon _{\mathbf{k}}}{g}\tanh \Bigl(%
\frac{\varepsilon _{\mathbf{k}}}{2T}\Bigr)\quad ,  \label{z}
\end{equation}%
where in the last, approximate equality we set $A_{\mathbf{k}}\approx -1$,
and neglect the contribution from the second term in Eq. (\ref{III71}),
assuming that the singular contribution from the modes with $A_{\mathbf{k}%
}\approx -1$ via the first term in Eq. (\ref{III71}) is dominant. Then, for
all infrared-dominated cumulants of higher orders $m\geq 2$, the
result (\ref{III71}) reduces to a very transparent form
\begin{equation}
\tilde{\kappa}_{m}\approx \frac{1}{2}(m-1)!\sum_{\mathbf{k}\neq 0}\frac{1}{%
\Bigl[\frac{\varepsilon _{\mathbf{k}}}{g}\tanh \Bigl(\frac{\varepsilon _{%
\mathbf{k}}}{2T}\Bigr)\Bigr]^{m}}\quad .  \label{cc}
\end{equation}

Finally, using the Bogoliubov energy spectrum in Eq. (\ref{III65}), $%
\varepsilon _{\mathbf{k}}=\varepsilon _{1}\sqrt{\mathbf{l}%
^{4}+(2g/\varepsilon_{1})\mathbf{l}^{2}}$ with a set of integers $\mathbf{l}
=(l_{x},l_{y},l_{z})$, we arrive to the following simple formulas for the
higher order generating cumulants in the Thomas-Fermi regime~(\ref{g}):
\begin{equation}
\tilde{\kappa}_{m}\approx \frac{1}{2}(m-1)!\Bigl(\frac{T}{\varepsilon_{1}}
\Bigr)^{m}\sum_{\mathbf{l}\neq 0}\frac{1}{\mathbf{l}^{2m}},\qquad
g \ll 2T^{2}/\varepsilon_{1};\qquad \varepsilon_{1} \ll g \ll T,
\label{cc1}
\end{equation}
\begin{equation}
\tilde{\kappa}_{m}\approx \frac{1}{2}(m-1)!\Bigl(\frac{g}{2\varepsilon_{1}}%
\Bigr)^{m/2}\sum_{\mathbf{l}\neq 0}\frac{1}{\Bigl[|\mathbf{l}|\tanh (|%
\mathbf{l}|\sqrt{2g\varepsilon_{1}}/T)\Bigr]^{m}},
\qquad 2T^{2}/\varepsilon_{1}\geq g\geq T;\quad g\gg\varepsilon_{1},
\label{cc2}
\end{equation}
\begin{equation}
\tilde{\kappa}_{m}\approx \frac{1}{2}(m-1)!\Bigl(\frac{g}{2\varepsilon_{1}}
\Bigr)^{m/2}\sum_{\mathbf{l}\neq 0}\frac{1}{|\mathbf{l}|^{m}},\qquad
g \gg 2T^{2}/\varepsilon_{1}, \qquad g \gg \varepsilon_{1}, \qquad  \label{cc3}
\end{equation}
for high, moderate, and very low temperatures $T$ respectively, as compared
to the geometrical mean of the interaction and gap energies in the trap,
$\sqrt{g\varepsilon_{1}/2}$.

In particular, for the Thomas-Fermi regime (\ref{g}) and relatively high
temperatures, the squared variance, as given by Eq. (\ref{cc1}),
\begin{equation}
\langle (n_{0}-\bar{n}_{0})^{2}\rangle =\frac{1}{2}\sum_{\mathbf{k}\neq 0}%
\Bigl[\frac{1}{(z(A_{\mathbf{k}})-1)^{2}}+\frac{1}{(z(-A_{\mathbf{k}})-1)^{2}%
}+\frac{1}{z(A_{\mathbf{k}})-1}+\frac{1}{z(-A_{\mathbf{k}})-1}\Bigr]%
\rightarrow \frac{N^{4/3}(T/T_{c})^{2}s_{4}}{2\pi ^{2}(\zeta (3/2))^{4/3}}
\label{III72a}
\end{equation}
scales as $\langle (n_{0}-\bar{n}_{0})^{2}\rangle \propto N^{4/3}$. Here the arrow indicates the limit of sufficiently strong interaction, $g \gg \varepsilon_1$ and $T \gg g \gg \varepsilon_1$.

The behavior (\ref{III72a}), (\ref{III24}), and (\ref{idgas}) is essentially
different from that of the normal fluctuations of most extensive physical
observables, which are Gaussian with the squared variance proportional to~$N$.
The only exception for the Eqs. (\ref{cc})--(\ref{cc3}) is the low temperature
limit of the variance that is not infrared-dominated and should be calculated
not via the Eq.~(\ref{cc3}), but directly from Eq. (\ref{III71}) using the
fact that all modes are very poorly occupied at low temperatures, i.e.,
$\exp(-\varepsilon _{\mathbf{k}}/T) \ll 1$, if $g \gg 2T^{2}/\varepsilon_{1},
\varepsilon _{1}$. Thus, we immediately find
\begin{equation}
\kappa _{2}=\tilde\kappa_2 + \tilde\kappa_1 \approx
\sum_{\mathbf{k}\neq 0}\frac{2A_{\mathbf{k}}^{2}}{(1-A_{\mathbf{k}}^{2})^{2}}
=\frac{1}{2}\sum_{\mathbf{k}\neq 0}\frac{g^{2}}{\varepsilon_{\mathbf{k}}^{2}}
\approx 2\pi \frac{g^{2}}{\varepsilon_{1}^{2}}
\int_{0}^{\infty }\frac{dr}{r^{2}+2g/\varepsilon_{1}}
=\frac{\pi ^{2}}{\sqrt{2}}\Bigl(\frac{g}{\varepsilon _{1}}\Bigr)^{3/2}
=2\sqrt{\pi }\Bigl(\frac{a{\bar{n}}_{0}}{L}\Bigr)^{3/2}.  \label{kappa2}
\end{equation}

The above results extend and confirm the result of the pioneering
paper~\cite{Pit98} where only the second moment,
$\langle (n_{0}-\bar{n}_{0})^{2}\rangle $, was calculated.
(The result of \cite{Pit98} was rederived by a different way in \cite{Meier},
and generalized in \cite{Zwerger}.) The higher-order cumulants $\kappa _{m}$,
$m>2$, given by Eqs. (\ref{III71}) and (\ref{cc})--(\ref{cc3}), are not zero,
do not go to zero in the thermodynamic limit and, moreover, are relatively
large compared with the corresponding exponent of the variance
$(\kappa _{m})^{m/2}$ that proves and measures the non-Gaussian character of
the condensate fluctuations. For the Thomas-Fermi regime (\ref{g}) and
relatively high temperatures, the relative values of the higher order
cumulants are given by Eq. (\ref{cc1}) as
\begin{equation}
\frac{\kappa_m}{(\kappa_2 )^{m/2}} \approx
2^{m/2 -1} (m-1)! \frac{s_{2m}}{(s_4)^{m/2}} , \qquad
s_{2m} = \sum_{{\bf l} \neq 0} \frac{1}{{\bf l}^{2m}} , \qquad
\label{cc1m}
\end{equation}
where ${\bf l}=(l_x , l_y , l_z )$ are integers, $s_4 \approx 16.53$, $s_6 \approx 8.40$, and $s_{2m} \approx 6$ for $m \gg 1$. In particular, the asymmetry coefficient of the ground-state occupation probability distribution
\begin{equation}
\gamma_1 = \langle (n_0 - {\bar n}_0 )^3 \rangle / \langle (n_0 - {\bar n}_0 )^2 \rangle^{3/2} = -\kappa_3 /\kappa_2^{3/2} \approx 2\sqrt{2} s_6 /(s_4 )^{3/2} \approx 0.35 \qquad
\label{gamma1}
\end{equation}
is not very small at all.

This non-Gaussian statistics stems from an infrared singularity that exists
for the fluctuation cumulants $\kappa _{m},$ $m\geq 2$, for the weakly
interacting gas in a box despite of the acoustic (i.e., linear, like for
the ideal gas in the harmonic trap) behavior of the Bogoliubov-Popov
energy spectrum (\ref{III65}) in the infrared limit. The reason is that the
excited mode squeezing (i.e., linear mixing of atomic modes) via Bogoliubov
coupling affects the BEC statistics (\ref{III71}) for the interacting gas
also directly (not only via a modification of the quasiparticle energy
spectrum), namely, via the function (\ref{III68}), $z(A_{\mathbf{k}})$,
which is different from a simple exponent of the bare energy,
$\exp (\varepsilon_{\mathbf{k}}/T)$, that enters into the corresponding
formula for the non-interacting gas (\ref{III19}).

\subsection{Crossover between ideal and interaction-dominated BEC:
quasiparticles squeezing and pair correlation}

Now we can use the analytical formula (\ref{III71}) to describe explicitly
a crossover between the ideal-gas and interaction-dominated regimes of the
BEC fluctuations. Obviously, if the interaction energy
$g=\bar{n}_{0} U_{\mathbf{k}}/V$ is less than the energy gap between the
ground state and the first excited state in the trap, $g < \varepsilon_1$,
the Bogoliubov coupling (\ref{III63}) becomes small for all modes,
$|A_{\bf k}| \ll 1$, so that both terms in Eq. (\ref{III71}) give similar
contributions and all fluctuation cumulants $\tilde\kappa_m$ tend to their
ideal gas values in the limit of vanishing interaction, $g\ll \varepsilon_1$.
Namely, in the near-ideal gas regime $\bar{n}_{0}a/L\ll 1$ the squared
variance linearly decreases from its ideal-gas value with an increase of the
weak interaction as follows:
\begin{equation}
\langle (n_{0}-\bar{n}_{0})^{2}\rangle \approx \frac{N^{4/3}(T/T_{c})^{2}s_{4}}{\pi ^{2}(\zeta (3/2))^{4/3}}\left[ 1-\pi^2 \frac{s_{6}}{s_{4}}\frac{g}{\varepsilon_1}\right] =\frac{N^{4/3}(T/T_{c})^{2}s_{4}}{\pi ^{2}(\zeta (3/2))^{4/3}}\left[ 1-3.19\frac{\bar{n}_{0}a}{L}\right] .
\label{idgas}
\end{equation}

With a further increase of the interaction energy over the energy gap in
the trap, $g > \varepsilon_1$, the essential differences between the
weakly interacting and ideal gases appear. First, the energy gap is increased by the interaction, that is
\begin{equation}
\tilde{\varepsilon}_{1}=\sqrt{\varepsilon _{1}^{2}+2\varepsilon _{1}\bar{n}%
_{0}(T)U_{0}/V}>\varepsilon _{1}=\Bigl(\frac{2\pi \hbar }{L}\Bigr)^{2}\frac{1%
}{2M}\quad ,  \label{e1}
\end{equation}
so that the border $T\sim \tilde{\varepsilon}_{1}$ between the moderate
temperature and very low temperature regimes is shifted to a higher
temperature, $T \sim \sqrt{g\varepsilon_1 /2}$. Thus, the interaction strength
$g$ determines also the temperature $T \sim \sqrt{g\varepsilon_1 /2}$ above
which another important effect of the weak interaction comes into play.
Namely, according to Eqs.~(\ref{cc})--(\ref{cc3}), the suppression of all
condensate-fluctuation cumulants by a factor of $1/2$, compared with the
ideal gas values (see Eq. (\ref{III25})), takes place for moderate
temperatures, $\tilde{\varepsilon}_{1} \ll T<T_{c}$, when a strong coupling
$(A_{\mathbf{k}}\approx -1)$ contribution dominates in Eq. (\ref{III71}).
The factor $1/2$ comes from the fact that, according to Eq. (\ref{III68}),
\[
z(A_{\mathbf{k}}=-1)=1,\qquad z(A_{\mathbf{k}}=1)=-1,
\]
so that the first term in Eq. (\ref{III71}) is resonantly large but the
second term is relatively small. In this case, the effective energy
spectrum, which can be introduced for the purpose of comparison with the
ideal gas formula (\ref{III19}), is
\begin{equation}
\varepsilon _{\mathbf{k}}^{eff}=T\ln (z(A_{\mathbf{k}}))\simeq \frac{1}{2}%
(1+A_{\mathbf{k}})\varepsilon _{\mathbf{k}}\simeq \frac{1}{2}\varepsilon _{%
\mathbf{k}}^{2}V/U_{0}\bar{n}_{0}\simeq \frac{\mathbf{k}^{2}}{2M}\;.
\label{III73}
\end{equation}
That is, the occupation of a pair of strongly coupled modes in the weakly
interacting gas can be characterized by the same effective energy spectrum
as that of a free atom. It is necessary to emphasize that the effective
energy $\varepsilon _{\mathbf{k}}^{eff}=T\ln (z(A_{\mathbf{k}}))$,
introduced in Eq. (\ref{III73}), describes only the occupation of a pair of
\textit{bare} atom excitations $\mathbf{k}$ and $\mathbf{-k}$ (see Eqs.
(\ref{III66})--(\ref{III68})) and, according to Eq. (\ref{III71}), the
ground-state occupation. That is, it would be wrong to reduce the analysis
of the thermodynamics and, in particular, the entropy of the interacting gas
to this effective energy. Thermodynamics is determined by the original
energy spectrum of the dressed canonical-ensemble quasiparticles,
Eq. (\ref{III65}).

\begin{figure}[tbp]
\center\epsfxsize=7cm\epsffile{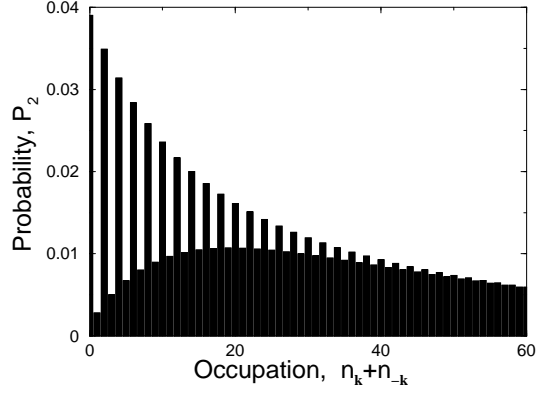}
\caption{The probability distribution (\ref{III74}) $P_2$ as a function of
the number of atoms in $\mathbf{k}$ and $\mathbf{-k}$ modes, $n_{\mathbf{k}}
+ n_{\mathbf{-k}}$, for the interaction energy $U_0 \bar{n}_0 /V = 10^3
\protect\varepsilon_{\mathbf{k}}$ and temperature $T = \protect\varepsilon_{%
\mathbf{k}}$. The pair correlation effect due to Bogoliubov coupling in the
weakly interacting Bose gas is clearly seen for low occupation numbers,
i.e., even occupation numbers $n_{\mathbf{k}}+n_{\mathbf{-k}}$ are more
probable than odd ones.}
\label{FigIII1}
\end{figure}

This remarkable property explains why the ground-state occupation
fluctuations in the interacting gas in this case are anomalously large to
the same extent as in the non-interacting gas except factor of $1/2$
suppression in the cumulants of all orders. These facts were considered in
\cite{Pit98} to be an accidental coincidence. We see now that, roughly
speaking, this is so because the atoms are coupled in strongly correlated
pairs such that the number of independent stochastic occupation variables
(``degrees of freedom") contributing to the fluctuations of the total number
of excited atoms is only $1/2$ the atom number $N$. This strong pair
correlation effect is clearly seen in the probability distribution of the
total number of atoms in the two coupled $\mathbf{k}$ and $\mathbf{-k}$
modes,
\begin{equation}
P_2 (n_{\mathbf{k}} + n_{\mathbf{-k}}) = \frac{[z(A_{\mathbf{k}})-1][z(-A_{%
\mathbf{k}})-1]}{z(-A_{\mathbf{k}}) - z(A_{\mathbf{k}})} \Bigl[ \Bigl( \frac{%
1}{z(A_{\mathbf{k}})} \Bigr)^{n_{\mathbf{k}}+n_{\mathbf{-k}}+1} - \Bigl(
\frac{1}{z(-A_{\mathbf{k}})} \Bigr)^{n_{\mathbf{k}}+n_{\mathbf{-k}}+1} \Bigr]
\label{III74}
\end{equation}
(see Fig.~\ref{FigIII1}). The latter formula follows from Eq.~(\ref{III68}).
Obviously, a higher probability for even occupation numbers $n_{\mathbf{k}}
+ n_{\mathbf{-k}}$ as compared to odd numbers at low occupations means that
the atoms in the $\mathbf{k}$ and $-\mathbf{k}$ modes have a tendency to
appear or disappear simultaneously, i.e., in pairs. This is a particular
feature of the well-studied in quantum optics phenomenon of two-mode
squeezing (see, e.g., \cite{squeezing,Walls}). This squeezing means a
reduction in the fluctuations of the population difference $n_{\mathbf{k}} -
n_{\mathbf{-k}}$, and of the relative phases or so-called quadrature-phase
amplitudes of an interacting state of two bare modes $\hat{a}_{\mathbf{k}}$
and $\hat{a}_{\mathbf{-k}}$ compared with their appropriate uncoupled state,
e.g., coherent or vacuum state. The squeezing is due to the quantum
correlations which build up in the bare excited modes via Bogoliubov
coupling (\ref{III63}) and is very similar to the noise squeezing in a
non-degenerate parametric amplifier studied in great details by many authors
in quantum optics in 80s \cite{squeezing,Walls}. We note that fluctuations
of individual bare excited modes are not squeezed, but there is a high
degree of correlation between occupation numbers in each mode.

It is very likely that in the general case of an arbitrary power-law trap the
interaction also results in anomalously large fluctuations of the number of
ground-state atoms, and a formal infrared divergence due to excited mode
squeezing via Bogoliubov coupling and renormalization of the energy
spectrum. In the particular case of the isotropic harmonic trap this was
demonstrated in \cite{Pit98} for the variance of the condensate
fluctuations. Therefore, the ideal gas model for traps with a low
spectral index $\sigma < d/2$ (such as a three-dimensional harmonic trap
where $\sigma = 1 < d/2 = 3/2$), showing Gaussian, normal thermodynamic
condensate fluctuations with the squared variance proportional to~$N$
instead of anomalously large fluctuations (see Eqs.~(\ref{III48}),
(\ref{III50}) and (\ref{III55}), (\ref{III56})), is not robust with respect
to the introduction of a weak interatomic interaction.

At the same time, the ideal gas model for traps with a high spectral index
$\sigma > d/2$ (e.g., for a three-dimensional box with $\sigma =2 > d/2=3/2$)
exhibits non-Gaussian, anomalously large ground-state occupation fluctuations
with a squared variance proportional to $N^{2\sigma /d} \gg N$
(see Eqs.~(\ref{III49}), (\ref{III50})) similar to those found for the
interacting gas. Fluctuations in the ideal Bose gas and in the Bogoliubov
Bose gas differ by a factor of the order of 1, which, of course, depends
on the trap potential and is equal to $1/2$ in the particular case of the
box, where $\Delta n^2_0 \propto N^{4/3}$. We conclude that, contrary to the
interpretation formulated in \cite{Pit98}, similar behavior of the
condensate fluctuations in the ideal and interacting Bose gases in a
box is \textit{not} accidental, but is a general rule for all traps with a
high spectral index $\sigma >d/2$, or a relatively low dimension of space,
$d < 2\sigma$.

As follows from Eq. (\ref{III71}), the interaction essentially modifies
the condensate fluctuations also at very low temperatures,
$T \ll \tilde{\varepsilon}_1$ (see Figs.~\ref{FigIII2}).
Namely, in the interacting Bose gas a temperature-independent quantum noise,
\begin{equation}
\tilde{\kappa}_m \rightarrow \tilde{\kappa}_m ( T = 0 ) \neq 0 , \qquad m
\geq 2 ,  \label{III75}
\end{equation}
additional to vanishing (at $T \to 0$) in the ideal Bose gas noise, appears due to quantum fluctuations
of the excited atoms, which are forced by the interaction to occupy the
excited levels even at $T = 0$, so that $\bar{n}_{\mathbf{k}}(T = 0) \neq 0$.
Thus, in the limit of very low temperatures the results of the ideal gas
model (Section V) are essentially modified by weak interaction and do not describe condensate statistics in the realistic weakly interacting Bose gases.

The temperature scaling of the condensate fluctuations described above
is depicted in Fig.~\ref{FigIII2} both for the weakly interacting and
ideal gases. A comparison with the corresponding quantities calculated
numerically from the exact recursion relation in Eqs.~(\ref{III76}) and
(\ref{III77}) for the ideal gas in a box is also indicated.
It is in good agreement with our approximate analytical formula (\ref{III71})
for all temperatures in the condensed phase, $T<T_{c}$, except of a region
near to the critical temperature, $T\approx T_{c}$. It is worth stressing
that the large deviations of the asymmetry coefficient,
$\gamma _{1}=\langle (n_{0}-\bar{n}_{0})^{3}\rangle/
\langle (n_{0}-\bar{n}_{0})^{2}\rangle ^{3/2}$,
and of the excess coefficient,
$\gamma_{2}=\langle(n_{0}-\bar{n}_{0})^{4}\rangle/
\langle (n_{0}-\bar{n}_{0})^{2}\rangle ^{2}-3$ from zero, which are
of the order of 1 at $T\sim T_{c}/2$ or even more at $T\approx 0$ and
$T\approx T_{c}$, indicate how far the ground-state occupation fluctuations
are from being Gaussian. (In the theory of turbulence, the coefficients
$\gamma_{1}$ and $\gamma _{2}$ are named as skewness and flatness,
respectively.) This essentially non-Gaussian behavior of the ground-state
occupation fluctuations remains even in the thermodynamic limit.

Mesoscopic effects near the critical temperature are also clearly seen in
Fig.~\ref{FigIII2}, and for the ideal gas are taken into account exactly
by the recursion relations (\ref{III76}), (\ref{III77}). The analytical
formulas (\ref{III71}) take them into account only via a finite-size effect,
$L \propto N^{-1/3}$, of the discreteness of the single-particle spectrum
$\varepsilon_{\mathbf{k}}$. This finite-size (discreteness) effect produces,
in particular, some shift of the characteristic BEC critical temperature
compared to its thermodynamic-limit value, $T_c$. For the case shown in
Fig.~\ref{FigIII2}, it is increased by a few per cent. Similarly to the
mean-number ``grand'' canonical approximation described at the end of
Section~III.B for the case of the ideal gas, the canonical-ensemble
quasiparticle approach can partially accommodates for the mesoscopic
effects by means of the grand-canonical shift of all quasiparticle energies
by a chemical potential,
$\tilde{\varepsilon}_{\mathbf{k}}=\varepsilon_{\mathbf{k}} - \mu$.
In this case the self-consistency equation (\ref{III72}) acquires an
additional nonlinear contribution due to the relation
$\exp(-\beta \mu)=1+1/{\bar n}_0$. Obviously, this ``grand'' canonical
approximation works only for $T<T_c$, whereas at $T>T_c$ we can use the
standard grand canonical approach, since the ground-state occupation is not
macroscopic above the critical temperature. However, the ``grand'' canonical
approach takes care only of the mean number of condensed atoms ${\bar n}_0$,
and does not improve the results of the canonical-ensemble quasiparticle
approach for BEC fluctuations.

\subsection{Universal anomalies and infrared singularities of the order
parameter fluctuations in the systems with a broken continuous symmetry}

The result that there are singularities in the central moments of the
condensate fluctuations, emphasized in~\cite{KKS-PRL,KKS-PRA} and discussed
above in detail for the BEC in a trap, can be generalized for other
long-range ordered systems below the critical temperature of a second-order
phase transition, including strongly interacting systems. That universality
of the infrared singularities was discussed in~\cite{Meier,Zwerger} and can
be traced back to a well known property of an infrared singularity in a
longitudinal susceptibility $\chi_{||}(k)$ of such systems \cite
{LL,gps96,Pokrovski}. The physics of these phenomena is essentially determined
by long wavelength phase fluctuations, which describe the non-condensate
statistics, and is intimately related to the fluctuation-dissipation
theorem, Bogoliubov's $1/k^2$ theorem for the static susceptibility
$\chi_{\varphi \varphi}(k)$ of superfluids, and the presence of Goldstone
modes, as will be detailed in the following.

\subsubsection{Long wavelength phase fluctuations, fluctuation-dissipation
theorem, and Bogoliubov's $1/k^2$ theorem}

First, let us refer to a microscopic derivation given in~\cite{Popov} that
demonstrates the fact that the phase fluctuations alone dominate the
low-energy physics. Also, let us assume that, as was shown by Feynman, the
spectrum of excitations in the infrared limit $k \to 0$ is exhausted by
phonon-like modes with a linear dispersion $\omega_k = ck$, where $c$ is an
actual velocity of sound. Then, following a textbook~\cite{LL}, we can
approximate an atomic field operator via a phase fluctuation operator as
follows:
\begin{equation}
{\hat \Psi} (\mathbf{x}) = \sqrt{{\tilde n}_0} e^{i{\hat \varphi} (\mathbf{x}%
)} , \quad {\hat \varphi}(\mathbf{x}) = (mc/2Vn_{tot})^{1/2} \sum_{\mathbf{k}
\neq 0} (\hbar k)^{-1/2} ({\hat c}_{\mathbf{k}} e^{i\mathbf{kx}} + {\hat c}_{%
\mathbf{k}}^{+} e^{-i\mathbf{kx}}) ,  \label{phase}
\end{equation}
where ${\tilde n}_0$ is the bare condensate density, $n_{tot}$ the mean
particle density, $m$ the particle mass, ${\hat c}_{\mathbf{k}}$ and
${\hat c}_{\mathbf{k}}^{+}$ are phonon annihilation and creation operators,
respectively. An omission of the $k=0$ term in the sum in Eq. (\ref{phase})
can be rigorously justified on the basis of the canonical-ensemble
quasiparticle approach (see Section V) and is related to the fact that a
global phase factor is irrelevant to any observable gauge-invariant
quantity. In a homogeneous superfluid, the renormalized condensate density
is determined by the long range order parameter,
\begin{equation}
\bar n_0 = \lim_{\Delta x \to \infty} \langle {\hat \Psi}^{+} (\Delta x) {%
\hat \Psi} (0) \rangle \approx {\tilde n}_0 \exp (-\langle {\hat \varphi}^2
(0) \rangle ) ,  \label{ODLRO}
\end{equation}
if we use the approximation (\ref{phase}) and neglect phonon interaction.
Thus, the mean number of condensate particles is depleted with an increase
of temperature in accordance with an increase of the variance of the phase
fluctuations, ${\bar n}_0 (T) \propto \exp (-\langle {\hat \varphi}^2 (0)
\rangle_T )$, which yields a standard formula \cite{LL} for the thermal
depletion, ${\bar n}_0 (T)-{\bar n}_0 (T=0) = - {\bar n}_0 (T=0) m(k_B T)^2
/12n_{tot}c\hbar^3$.

Similar to the mean value, higher moments of the condensate fluctuations of
the homogeneous superfluid in the canonical ensemble can be represented in
terms of the phase fluctuations of the non-condensate via the operator
\begin{equation}
{\hat n}= N/V - {\hat n}_0 = \int \Delta {\hat \Psi}^{+} (x) \Delta {\hat
\Psi} (x) d^3 x ,  \label{fluct-delta}
\end{equation}
where
\begin{equation}
\Delta {\hat \Psi} (x) = {\hat \Psi} (x) - \sqrt{{\bar n}_0 (T)} = \sqrt{{%
\tilde n}_0} (e^{i{\hat \varphi}} (x) - e^{-\langle {\hat \varphi}^2 (0)
\rangle /2} ).  \label{deltaPsi}
\end{equation}
In particular, the variance of the condensate occupation is determined by
the correlation function of the phase fluctuations as follows~\cite{Meier}:
\begin{equation}
\langle ({\hat n}_0 - {\bar n}_0 )^2 \rangle \approx 2 {\tilde n}_0^2 e^{-2
\langle {\hat \varphi}^2 (0) \rangle} \int \langle {\hat \varphi} (x) {\hat
\varphi} (y) \rangle^2 d^3 x d^3 y.  \label{var-phase}
\end{equation}
The last formula can be evaluated with the help of a classical form of the
fluctuation-dissipation theorem,
\begin{equation}
\int \langle {\hat \varphi} (x) {\hat \varphi} (y) \rangle^2 d^3 x d^3 y =
\sum_{\mathbf{k} \neq 0} (k_B T \chi_{\varphi \varphi}(\mathbf{k}))^2 .
\label{fdt}
\end{equation}
For a homogeneous superfluid, the static susceptibility in the infrared
limit $k \to 0$ does not depend on the interaction strength and, in
accordance with Bogoliubov's $1/k^2$-theorem for the static susceptibility
of superfluids, is equal to~\cite{Meier}
\begin{equation}
\chi_{\varphi \varphi}(\mathbf{k}) =m/n_{tot} \hbar^2 k^2 , \qquad k \to 0 .
\label{fdt1}
\end{equation}
Then Eq.~(\ref{var-phase}) immediately yields the squared variance of the
condensate fluctuations in the superfluid with arbitrary strong interaction
in exactly the same form as is indicated by the arrow in Eq.~(\ref{III72a}),
only with the additional factor $({\bar n}_0 (0)/n_{tot})^2$. The latter
factor is almost 1 for the BEC in dilute Bose gases but can be much less
in liquids, for example in $^4 He$ superfluid it is about 0.1. Thus, indeed,
the condensate fluctuations at low temperatures can be calculated via the
long wavelength phase fluctuations of the non-condensate.

\subsubsection{Effective nonlinear $\protect\sigma$ model, Goldstone modes,
and universality of the infrared anomalies}

Following~\cite{Zwerger}, let us use an effective nonlinear $\sigma$ model
\cite{Parisi} to demonstrate that the infrared singularities and anomalies
of the order parameter fluctuations exist in all systems with a broken
continuous symmetry, independently on the interaction strength, and are
similar to that of the BEC fluctuations in the Bose gas. That model
describes directional fluctuations of an order parameter $\vec \Psi (x) =
m_s^{(0)} \vec \Omega (x)$ with a fixed magnitude $m_s^{(0)}$ in terms of an
$N_\Omega$-component unit vector $\vec \Omega (x)$. It is inspired by the
classical theory of spontaneous magnetization in ferromagnets. The
constraint $|\vec \Omega (x)|=1$ suggests a standard decomposition of the
order parameter
\begin{equation}
\vec \Omega (x) = \{ \Omega_0 (x), \Omega_i (x) ; i=1,..., N_\Omega -1 \} ,
\quad \Omega_0 (x) = \sqrt{1- \sum_{i=1}^{N_\Omega -1} \Omega_i^2 (x)},
\label{Goldstone}
\end{equation}
into a longitudinal component $\Omega_0 (x)$ and $N_\Omega -1$ transverse
Goldstone fields $\Omega_i (x)$. The above $\sigma$ model constraint, $%
\Omega_0^2 (x) = 1 - \sum_{i} \Omega_i^2 (x)$, resembles the particle-number
constraint $\hat{n}_0 = N - \sum_{\mathbf{k} \neq 0} \hat{n}_{\mathbf{k}}$
in Eq. (\ref{III7}) for the many-body atomic Bose gas in a trap. Although
the former is a local, more stringent, constraint and the latter is only a
global, integral constraint, in both cases it results in the infrared
singularities, anomalies, and non-Gaussian properties of the order parameter
fluctuations. Obviously, in the particular case of a homogeneous system the
difference between the local and integral constraints disappears entirely.
Within the $\sigma$ model, a superfluid can be described as a particular
case of an $N_\Omega =2$ system, with the superfluid (condensate) and normal
(non-condensate) component.

At zero external field, the effective action for the fluctuations of the
order parameter is $S[\Omega] = (\rho_s /2T) \int [\nabla \vec \Omega (x)]^2
d^3 x$, where the spin stiffness $\rho_s$ is the only parameter. Below the
critical temperature $T_c$, a continuous symmetry becomes broken and there
appears an intensive nonzero order parameter with a mean value
\begin{equation}
\int_V \langle \vec \Psi (x) \cdot \vec \Psi (0) \rangle d^3 x = Vm_L^2 \to
Vm_s^2  \; , \label{order}
\end{equation}
which gives the spontaneous magnetization $m_s^2$ of the infinite system in
the thermodynamic limit $V \to \infty$. The leading long distance behavior
of the two-point correlation function $G(x) = \langle \vec \Psi (x) \cdot
\vec \Psi (0) \rangle$ may be obtained from a simple Gaussian spin wave
calculation. Assuming low enough temperatures, we can neglect the spin wave
interactions and consider the transverse Goldstone fields $\Omega_i (\mathbf{%
k})$ as the Gaussian random functions of the momentum $\mathbf{k}$ with the
correlation function $\langle \Omega_i (\mathbf{k}) \Omega_{i^{\prime}} (%
\mathbf{k^{\prime}}) \rangle = \delta_{i,i^{\prime}} \delta_{\mathbf{k},%
\mathbf{-k^{\prime}}} T/\rho_s k^2$. As a result, the zero external field
correlation function below the critical temperature is split into
longitudinal and transverse parts $G(x) = m_s^2 [1+G_{\parallel}(x) +
(N_\Omega -1) G_{\perp}(x)]$, where $m_s^2 = G( \infty )$ now is the
renormalized value of the spontaneous magnetization. To the lowest
nontrivial order in the small fluctuations of the Goldstone fields, the
transverse correlation function decays very slowly with a distance $r$, $%
G_{\perp} \propto T/\rho_s r$, in accordance with Bogoliubov's
$1/k^2$-theorem of the divergence of the transverse susceptibility in the
infrared limit,
\begin{equation}
\chi_{\perp} (k) = m_s^2 G_{\perp} (k)/T =m_s^2 /\rho_s k^2 .
\label{transv-susceptibility}
\end{equation}
The longitudinal correlation function is simply related to the transverse
one~\cite{Parisi},
\begin{equation}
G_{\parallel} (x) \approx \frac{1}{4} \langle \sum \Omega_i^2 (x) \sum
\Omega_{i^{\prime}}^2 (0) \rangle_c = \frac{1}{2} (N_\Omega -1) G_{\perp}^2
(x) ,  \label{longitud-susceptibility}
\end{equation}
and, hence, decays slowly with a $1/r^2$ power law. That means that
contrary to the naive mean field picture, where the longitudinal
susceptibility $\chi_{\parallel} (k \to 0)$ below the critical temperature
is finite, the Eq.~(\ref{longitud-susceptibility}) leads to an infrared
singularity in the longitudinal susceptibility and correlation function
\cite{Pokrovski},
\begin{equation}
\chi_{\parallel}(k \to 0)\sim T/\rho_s^2 k ; \qquad G_{\parallel} \sim 1/r^2
.  \label{IR-singularity}
\end{equation}

Although the Eq.~(\ref{longitud-susceptibility}) is obtained by means of
perturbation theory, the result for the slow power law decay of the
longitudinal correlation function, Eq. (\ref{IR-singularity}), holds for
arbitrary temperatures $T<T_c$~\cite{Parisi}.

Knowing susceptibilities, it is immediately possible to find the variance
of the operator
\begin{equation}
{\hat M}_s = V^{-1}\int_V \int_V \vec \Psi (x) \cdot \vec \Psi (y) d^3 x d^3
y  \label{magnetization}
\end{equation}
that describes the fluctuations of the spontaneous magnetization in a finite
system at zero external field. Its mean value is given by Eq.~(\ref{order})
as $\langle {\hat M}_s \rangle = {\bar M}_s = Vm_L^2$. Its fluctuations are
determined by the connected four-point correlation function $G_4 = \langle
\vec \Psi (x_1) \cdot \vec \Psi (x_2) \vec \Psi (x_3) \cdot \vec \Psi (x_4)
\rangle_c$. In the 4-th order of the perturbation theory for an infinite
system, it may be expressed via the squared transverse susceptibilities as
follows:
\begin{equation}
G_4 = \frac{1}{2} m_s^4 (N_\Omega -1) [G_{\perp} (x_1 - x_3) - G_{\perp}
(x_1 - x_4) - G_{\perp} (x_2 - x_3) + G_{\perp} (x_2 - x_4)]^2 .
\label{4-corr-transverse}
\end{equation}
For a finite system, it is more convenient to do similar calculations in the
momentum representation and replace all integrals by the corresponding
discrete sums, which yields the squared variance
\begin{equation}
\langle ({\hat M}_s - {\bar M}_s )^2 \rangle = 2 m_s^4 (N_\Omega
-1)T^2\rho_s^{-2} \sum_{\mathbf{k} \neq 0} k^{-4} \propto T^2 V^{4/3}
\label{M-variance} \; .
\end{equation}
This expression has the same anomalously large scaling $\propto V^{4/3}$ in
the thermodynamic limit $V \to \infty$ (due to the same infrared singularity)
as the variance of the BEC fluctuations in the ideal or interacting Bose gas
in Eqs.~(\ref{III25}) or (\ref{III72a}), respectively. Again, although the
Eq.~(\ref{M-variance}) is obtained by means of perturbation theory, the
latter scaling is universal below the critical temperature, just like the
$1/k$ infrared singularity of the longitudinal susceptibility. Of course,
very close to $T_c$ there is a crossover to the critical singularities, as
discussed, e.g., in~\cite{Chen}. Thus, the average fluctuation of the order
parameter is still vanishing in the thermodynamic limit $\sqrt{\langle
({\hat M}_s - {\bar M}_s )^2 \rangle}/{\bar M}_s \propto V^{-1/3} \to 0$;
that is, the order parameter (e.g., spontaneous magnetization or macroscopic
wave function) is still a well defined self-averaging quantity. However,
this self-averaging in systems with a broken continuous symmetry is much
weaker than expected naively from the standard Einstein theory of the
Gaussian fluctuations in macroscopic thermodynamics. Note that, in
accordance with the well-known Hohenberg-Mermin-Wagner theorem, in
systems with lower dimensions, $d \leq 2$, the strong fluctuations of the
direction of the magnetization completely destroy the long range order, and
the self-averaging order parameter does not exist anymore. An important
point also is that the scaling result in Eq.~(\ref{M-variance}) holds for
any dynamics and temperature dependence of the average order parameter
${\bar M}_s (T)$ although the function ${\bar M}_s (T)$ is, of course,
different, say, for ferromagnets, anti-ferromagnets, or a BEC in different
traps. It is the constraint, either $|\vec \Omega|=1$ in the $\sigma$ model
or $N = \hat{n}_0 + \sum_{\mathbf{k} \neq 0} \hat{n}_{\mathbf{k}}$ in the
BEC, that predetermines the anomalous scaling in Eq.~(\ref{M-variance}).
The temperature dependence of the variance in Eq.~(\ref{M-variance}) at low
temperatures is also universal, since $\rho_s \to const$ at $T \to 0$. The
reason for this fact is that the dominant finite size dependence is
determined by the leading low energy constant in the effective field theory
for fluctuations of the order parameter, which is precisely $\rho_s$ in the
effective action $S[\Omega]$.

\subsubsection{Universal scaling of condensate fluctuations in superfluids}

In homogeneous superfluids the translational invariance requires the
superfluid density to be equal to the full density $n_{tot}$, so that the
associated stiffness $\rho_s (T \to 0) = \hbar^2 n_{tot}/m$ is independent
of the interaction strength. Thus, Eq.~(\ref{M-variance}) in accord with the
Eq.~(\ref{III72a}) yields the remarkable conclusion that the relative
variance of the ground-state occupation at low temperatures is a universal
function of the density and the thermal wavelength
$\lambda_T = h/\sqrt{2 \pi mT}$, as well as the system size $L=V^{1/3}$ and
the boundary conditions,
\begin{equation}
\sqrt{\langle (n_0 - {\bar n}_0)^2 \rangle}/{\bar n}_0 = B/(n_{tot}
\lambda_T^2 L)^2  \label{universal-variance} \; .
\end{equation}
The boundary conditions determine the low-energy spectrum of quasiparticles
in the trap in the infrared limit and, hence, the numerical prefactor $B$ in
the infrared singularity of the variance, namely, the coefficient $B$ in the
sum $\sum_{\mathbf{k} \neq 0} k^{-4} = BV^{4/3}/8\pi^2$. In particular, in
accord with Eqs. (\ref{III25}) and (\ref{III72a}), one has
$B=0.8375$ for the box with periodic boundary conditions, and
$B=8 E_3(2)/\pi^2 = 0.501$ for the box with Dirichlet boundary conditions.
Here
$E_d (t) = \sum_{n_1 =1,..., n_d =1}^{\infty} (n_1^2 + \ldots + n_d^2 )^{-t}$
is the generalized Epstein zeta function~\cite{PhysicaA}, convergent for
$d<2t$. Of course, in a finite trap at temperatures of the order of or less
than the energy of the first excited quasiparticle, $T< \varepsilon_1$, the
scaling law (\ref{M-variance}) is no longer valid and the condensate
fluctuations acquire a different temperature scaling, due to the
temperature-independent quantum noise produced by the excited atoms, which
are forced by the interaction to occupy the excited energy levels even at
zero temperature, as it was discussed for the Eqs. (\ref{III75}) and
(\ref{III60}).

\subsubsection{Constraint mechanism of anomalous order-parameter
fluctuations and susceptibilities \\
versus instability in the systems with a broken continuous symmetry}

It is important to realize that the anomalously large order-parameter
fluctuations and susceptibilities have a simple geometrical nature, related
to the fact that the direction of the order parameter is only in a neutral, rather than in a stable equilibrium, and does not violate an overall stability of the system with a broken continuous symmetry at any given temperature below
phase transition, $T<T_c$. On one hand, on the basis of a well-known
relation between the longitudinal susceptibility and the variance $\Delta_M$
of the order parameter fluctuations,
\begin{equation}
\chi_{\alpha \alpha} \equiv \frac{1}{N} \frac{\partial M_{\alpha}}{\partial
B_{\alpha}} = \frac{\Delta_M^2}{Nk_B T} , \qquad \Delta_M^2 = \langle ({\hat
M}_{\alpha} - {\bar M}_{\alpha} )^2 \rangle ,  \label{s-v}
\end{equation}
it is obvious that if the fluctuations are anomalous, i.e., go as
$\Delta_M^2 \sim N^{\gamma}$ with $\gamma > 1$ instead of the standard
macroscopic thermodynamic scaling $\Delta_M^2 \sim N$, then the longitudinal
susceptibility diverges in the thermodynamic limit $N \to \infty$ as $%
\Delta_M^2 /N \sim N^{\gamma-1} \to \infty$. On the other hand, the
susceptibilities in stable systems should be finite, since otherwise any
spontaneous perturbation will result in an infinite response and a transition
to another phase. One could argue (as it was done in a recent series of
papers~\cite{Yukalov}) that the anomalous fluctuations cannot exist since
they break the stability condition and make the system unstable. However,
such an argument is not correct.

First of all, the anomalously large transverse susceptibility in the
infrared limit $\chi_{\perp} \sim k^{-2} \sim L^2$ in Bogoliubov's
$1/k^2$-theorem (\ref{transv-susceptibility}) originates from the obvious
property of a system with a broken continuous symmetry that it is
infinitesimally easy to change the direction of the order parameter and,
of course, does not violate stability of the system. However, it implies
anomalously large fluctuations in the direction of the order parameter.
Therefore, an anomalously large variance of the fluctuations comes from the
fluctuations $({\hat {\mathbf{M}}} - {\bar {\mathbf{M}}}) \perp {\bar {%
\mathbf{M}}}$, which are perpendicular to the mean value of the order
parameter. This is the key issue. It means that a longitudinal external
field $\mathbf{B} \parallel \mathbf{\bar M}$, that easily rotates these
transverse fluctuations towards $\mathbf{\bar M}$ when $\chi_{\perp}B \sim
\Delta_M$, will change the longitudinal order parameter by a large increment,
$\chi_{\parallel} B \sim |{\bar {\mathbf{M}}}|-\sqrt{{\bar {\mathbf{M}}}^2 -
\Delta_M^2} \approx \Delta_M^2 /2{\bar M}$. It is obvious that this pure
geometrical rotation of the order parameter has nothing to do with an
instability of the system, but immediately reveals the anomalously large
longitudinal susceptibility and relates its value to the anomalous
transverse susceptibility and variance of the fluctuations,
\begin{equation}
\chi_{\parallel} \sim \Delta_M^2 /2B{\bar M} \sim \Delta_{M} \chi_{\perp} /2{%
\bar M} .  \label{longit-chi}
\end{equation}
This basically geometrical mechanism of an anomalous behavior of constrained
systems constitutes an essence of \textit{the constraint mechanism of the
infrared anomalies} in fluctuations and susceptibilities of the order
parameter for all systems with a broken continuous symmetry. The latter
qualitative estimate yields the anomalous scaling of the longitudinal
susceptibility discussed above, $\chi_{\parallel} \sim L$ with increase of
the system size $L$, Eq. (\ref{IR-singularity}), since $\chi_{\perp} \sim
L^2, \quad {\bar M} \sim L^3$, and $\Delta_M \sim L^2$.

A different question is whether the Bose gas in a trap is unstable in the
grand canonical ensemble when an actual exchange of atoms with a reservoir
is allowed, and only the mean number of atoms in the trap is fixed,
${\bar N} = const$. In this case, for example, the isothermal compressibility
is determined by the variance of the number-of-atoms fluctuations $\kappa_T
\equiv -V^{-1} (\partial V/\partial P)_T = V\langle ({\hat N} - {\bar N})^2
\rangle /{\bar N}^2 k_B T$, and diverges in the thermodynamic limit if
fluctuations are anomalous. In particular, the ideal Bose gas in the grand
canonical ensemble does not have a well-defined condensate order parameter,
since the variance is of the order of the mean value,
$\sqrt{\langle ({\hat N} - {\bar N})^2 \rangle} \sim {\bar N}$, and it is
unstable against a collapse \cite{Ziff,DterHaar,Yukalov}.

In summary, one could naively expect that the order parameter fluctuations
below $T_c$ are just like that of a standard thermodynamic variable, because
there is a finite restoring force for deviations from the equilibrium value.
However, in all systems with a broken continuous symmetry the universal
existence of infrared singularities in the variance and higher moments
ensures anomalously large and non-Gaussian fluctuations of the order
parameter. This effect is related to the long-wavelength phase fluctuations
and the infrared singularity of the longitudinal susceptibility originating
from the inevitable geometrical coupling between longitudinal and transverse
order parameter fluctuations in constrained systems.

\section{Conclusions}

It is interesting to note that the first results for the average and
variance of occupation numbers in the ideal Bose gas in the canonical
ensemble were obtained about fifty years ago by standard statistical
methods~\cite{Fierz, Dingle,Fraser} (see also \cite{Hauge,Reif} and
the review \cite{Ziff}). Only later, in the 60s, laser physics and its
byproduct, the master equation approach, was developed (see, e.g.,
\cite{sl,ssl}). In this paper we have shown that the latter approach provides
very simple and effective tools to calculate statistical properties of an
ideal Bose gas in contact with a thermal reservoir. In particular, the
results (\ref{II51}) and (\ref{II54}) reduce to the mentioned old results
in the ``condensed region'' in the thermodynamic limit.

However, the master equation approach gives even more. It yields simple
analytical expressions for the distribution function of the number of
condensed atoms (\ref{II40}) and for the canonical partition function
(\ref{II41}). In terms of cumulants, or semi-invariants \cite{a,cumulants},
for the stochastic variables $n_0$ or $n=N-n_0$, it was shown \cite{KKS-PRA}
that the quasithermal approximation (\ref{II35}), with the results (\ref{II51})
and (\ref{II54}), gives correctly both the first and the second cumulants.
The analysis of the higher-order cumulants is more complicated and includes,
in principle, a comparison with more accurate calculations of the conditioned
average number of non-condensed atoms (\ref{II19}) as well as higher-order
corrections to the second order master equation (\ref{II12}). It is clear
that the master equation approach is capable of giving the correct answer
for higher-order cumulants and, therefore, moments of the condensate
fluctuations. Even without these complications, the approximate result (\ref
{II40}) reproduces the higher moments, calculated numerically via the exact
recursion relation (\ref{III77}), remarkably well for all temperatures
$T<T_c $ and $T \sim T_c$ (see Fig. \ref{FigII6}).

As we demonstrated in Section IV, the simple formulas yielded by the master
equation approach allow us to study mesoscopic effects in BECs for a
relatively small number of atoms that is typical for recent experiments
\cite{bec,morebec,miesner,kleppner,Sant01,Robe01}. Moreover, it is interesting
in the study of the dynamics of BEC. This technique for studying statistics
and dynamics of BEC shows surprisingly good results even within the simplest
approximations. Thus, the analogy with phase transitions and quantum
fluctuations in lasers (see, e.g., \cite{sz,MOS99,ds,ph-tra}) clarifies some
problems in BEC. The equilibrium properties of the number-of-condensed-atom
statistics in the ideal Bose gas are relatively insensitive to the details
of the model. The origin of dynamical and coherent properties of the
evaporatively cooling gas with an interatomic interaction is conceptually
different from that in the present ``ideal gas + thermal reservoir'' model.
The present model is rather close to the dilute $He^4$ gas in porous gel
experiments~\cite{rep} in which phonons in the gel play the role of the
external thermal reservoir. Nevertheless, the non-condensed atoms always
play a part of some internal reservoir, and the condensate master equation
probably contains terms similar to those in Eq. (\ref{II20}) for any cooling
mechanism.

For the ideal Bose gas in the canonical ensemble the statistics of the
condensate fluctuations below $T_c$ in the thermodynamic limit is
essentially the statistics of the sum of the non-condensed modes of a trap,
$\hat{n}_{\mathbf{k}}$, that fluctuate independently, $\hat{n}_0=N-\sum_{%
\mathbf{k} \neq 0} \hat{n}_{\mathbf{k}}$. This is well understood, especially,
due to the Maxwell's demon ensemble approximation elaborated in a series of
papers \cite{HKK,Fierz,pol,Navez,ww,gh97b}, and is completed and justified
to a certain extent in \cite{KKS-PRL,KKS-PRA} by the explicit calculation of the moments (cumulants) of all orders, and by the reformulation of the
canonical-ensemble problem in the properly reduced subspace of the original
many-particle Fock space. The main result (\ref{III71}) of \cite{KKS-PRL,KKS-PRA} explicitly describes the non-Gaussian properties and the crossover between the ideal-gas and interaction-dominated regimes of the BEC fluctuations.

The problem of dynamics and fluctuations of BEC for the interacting gas is
much more involved. The master equation approach provides a powerful tool
for the solution of this problem as well. Of course, to take into account
higher-order effects of interaction between atoms we have to go beyond the
second order master equation, i.e., to iterate Eq. (\ref{II11}) more times
and to proceed with the higher-order master equation similarly to what we
discussed above. It would be interesting to show that the master equation
approach could take into account all higher order effects in a way
generalizing the well-known nonequilibrium Keldysh diagram technique
\cite{Keldysh,LL,Stoof99}. As a result, the second order master equation
analysis presented above can be justified rigorously, and higher-order
effects in condensate fluctuations at equilibrium, as well as nonequilibrium
stages of cooling of both ideal and interacting Bose gases can be calculated.

The canonical-ensemble quasiparticle method, i.e., the reformulation of the
problem in terms of the proper canonical-ensemble quasiparticles, gives even
more. Namely, it opens a way to an effective solution of the canonical
ensemble problems for the statistics and nonequilibrium dynamics of the BEC
in the interacting gas as well. The first step in this direction is done in
Section VI, where the effect of the Bogoliubov coupling between excited
atoms due to a weak interaction on the statistics of the fluctuations of the
number of ground-state atoms in the canonical ensemble was analytically
calculated for the moments (cumulants) of all orders. In this case, the BEC
statistics is essentially the statistics of the sum of the dressed
quasiparticles that fluctuate independently. In particular, a suppression of
the condensate fluctuations at the moderate temperatures and their
enhancement at very low temperatures immediately follow from this picture.

There is also the problem of the BEC statistics in the microcanonical
ensemble, which is closely related to the canonical-ensemble problem.
In particular, the equilibrium microcanonical statistics can be calculated
from the canonical statistics by means of an inversion of a kind of Laplace
transformation from the temperature to the energy as independent variable.
Some results concerning the BEC statistics in the microcanonical ensemble
for the ideal Bose gas were presented
in \cite{HKK,gr,Navez,ww,gh97b,Holt99}. We can calculate all
moments of the microcanonical fluctuations of the condensate from the
canonical moments found in the present paper. Calculation of the
microcanonical statistics starting from the grand canonical ensemble and
applying a saddle-point method twice, first, to obtain the canonical
statistics and, then, to get the microcanonical statistics~\cite{gr}, meets
certain difficulties since the standard saddle-point approximation is not
always good and explicit to restore the canonical statistics from the
grand canonical one with sufficient accuracy~\cite{gh}. The variant of the
saddle-point method discussed in Appendix~\ref{ap:saddle} is not subject
to these restrictions.

Another important problem is the study of mesoscopic effects due to a
relatively small number of trapped atoms ($N \sim 10^3 - 10^6$). The
canonical-ensemble quasiparticle approach under the approximation
(\ref{III6}), i.e., $\mathcal{H}^{CE} \approx \mathcal{H}^{CE}_{n_0 \neq 0}$,
takes into account only a finite-size effect of the discreteness of the
single particle spectrum, but does not include all mesoscopic effects.
Hence, other methods should be used (see, e.g., \cite{HKK,pol,Herzog,ww,
bb,la,bo,br,recursion}). In particular, the master equation approach provides
amazingly good results in the study of mesoscopic effects, as was demonstrated
recently in \cite{MOS99,KSZZ}.

The canonical-ensemble quasiparticle approach also makes it clear how to
extend the Bogoliubov and more advanced diagram methods for the solution of
the canonical-ensemble BEC problems and ensure conservation of the number of
particles. The latter fact cancels the main arguments of Refs. \cite
{Navez99,Yukalov} against the Giorgini-Pitaevskii-Stringari result \cite
{Pit98} and shows that our result (\ref{III71}) and, in particular, the
result of \cite{Pit98} for the variance of the number of ground-state atoms
in the dilute weakly interacting Bose gas, correctly take into account one of
the main effects of the interaction, namely, dressing of the excited atoms
by the macroscopic condensate via the Bogoliubov coupling. If one ignores
this and other correlation effects, as it was done in \cite{Navez99,Yukalov},
the result cannot be correct. This explains a sharp disagreement of the
ground-state occupation variance suggested in \cite{Navez99} with the
predictions of \cite{Pit98} and our results as well. Note also that the
statement from \cite{Navez99} that ``the phonon spectrum plays a crucial
role in the approach of \cite{Pit98}'' should not be taken literally since
the relative weights of bare modes in the eigenmodes (quasiparticles) is, at
least, no less important than eigenenergies themselves. In other words, our
derivation of Eq.~(\ref{III71}) shows that squeezing of the excited states
due to Bogoliubov coupling in the field of the macroscopic condensate is
crucial for the correct calculation of the BEC fluctuations. Besides, the
general conclusion that very long wavelength excitations have an acoustic,
``gapless'' spectrum (in the thermodynamic limit) is a cornerstone fact of
the many-body theory of superfluidity and BEC~\cite{agd}. Contrary to a
pessimistic picture of a mess in the study of the condensate fluctuations in
the interacting gas presented in \cite{Navez99}, we are convinced that the
problem can be clearly formulated and solved by a comparative analysis of
the contributions of the main effects of the interaction in the tradition of
many-body theory. In particular, the result (\ref{III71}) corresponds to a
well established first-order Popov approximation in the diagram technique
for the condensed phase \cite{Griffin}.

We emphasize here an important result of an analytical calculation of all
higher cumulants (moments) \cite{KKS-PRA}. In most cases (except, e.g., for
the ideal gas in the harmonic trap and similar high dimensional traps where
$d>2\sigma $), both for the ideal Bose gas and for the interacting Bose gas,
the third and higher cumulants of the number-of-condensed-atom fluctuations
normalized to the corresponding power of the variance do not tend to the
Gaussian zero value in the thermodynamic limit, e.g., $\langle
(n_{0}-\langle n_{0}\rangle )^{3}\rangle /\langle (n_{0}-\langle
n_{0}\rangle )^{2}\rangle ^{3/2}$ does not vanish in the thermodynamic limit.

Thus, fluctuations in BEC are not Gaussian, contrary to what is usually
assumed following the Einstein theory of fluctuations in the macroscopic
thermodynamics. Moreover, BEC fluctuations are, in fact, anomalously large,
i.e., they are not normal at all. Both these remarkable features originate
from the universal infrared anomalies in the order-parameter fluctuations
and susceptibilities in constrained systems with a broken continuous
symmetry. The infrared anomalies come from a long range order in the phases
below the critical temperature of a second order phase transition and have
a clear geometrical nature, related to the fact that the direction of the
order parameter is only in a neutral, rather than in a stable equilibrium. Hence, the transverse susceptibility and fluctuations are anomalously large and, through an inevitable geometrical coupling between longitudinal and
transverse order parameter fluctuations in constrained systems, produce
the anomalous order parameter fluctuations. In other words, the long
wavelength phase fluctuations of the Goldstone modes, in accordance with
the Bogoliubov $1/k^2$-theorem for the transverse susceptibility, generate
anomalous longitudinal fluctuations in the order parameter of the systems
below the critical temperature of the second order phase transition.
Obviously, this constraint mechanism of the infrared anomalies in
fluctuations and susceptibilities of the order parameter is universal for
all systems with a broken continuous symmetry, including BEC in ideal or
weakly interacting gases as well as superfluids, ferromagnets and other
systems with strong interaction. It would be interesting to extend the
analysis of the order parameter fluctuations presented in this review
from the BEC in gases to other systems.

The next step should be an inclusion of the effects of a finite
renormalization of the energy spectrum as well as the interaction of the
canonical-ensemble quasiparticles at finite temperatures on the statistics
and dynamics of BEC. It can be done on the level of the second-order
Beliaev-Popov approximation, which is considered to be enough for the
detailed account of most many-body effects (for a review, see \cite
{Griffin}). A particularly interesting problem is the analysis of phase
fluctuations of the condensate in the trap, or of the matter beam in the
atom laser~\cite{at-la}, because the interaction is crucial for the
existence of the coherence in the condensate
\cite{LL,hm,fw,Anderson,KKS-PRA,agd,Liu,Giorgini,Graham}.
As far as the equilibrium or quasi-equilibrium properties are concerned, the
problem can be solved effectively by applying either the traditional methods
of statistical physics to the canonical-ensemble quasiparticles, or the
master equation approach, that works surprisingly well even without any
explicit reduction of the many-particle Hilbert space~\cite{MOS99,KSZZ}.
For the dynamical, nonequilibrium properties, the analysis can be based on
an appropriate modification of the well-known nonequilibrium Keldysh diagram
technique \cite{Keldysh,Stoof99,LP,d} which incorporates both the standard
statistical and master equation methods.

Work in the directions mentioned above is in progress and will be presented
elsewhere. Clearly, the condensate and non-condensate fluctuations are
crucially important for the process of the second order phase transition,
and for the overall physics of the Bose-Einstein-condensed interacting gas
as a many-particle system.

\vspace{2mm}

\indent
{\bf Acknowledgments}

We would like to acknowledge the support of the Office of Naval 
Research (Award No. N00014-03-1-0385) and the Robert A. Welch Foundation 
(Grant No. A-1261). One of us (MOS) wishes to thank Micheal Fisher, Joel 
Lebowitz, Elliott Lieb, Robert Seiringer for stimulating discussions and 
Leon Cohen for suggesting this review article.

\appendix

\section{Bose's and Einstein's way of counting microstates}

\label{ap:proof}

When discussing Einstein's 1925 paper~\cite{Einstein25}, we
referred to the identity~(\ref{combinatory}),
\begin{equation}
    \sum_{ \{ p_0, p_1, \ldots, p_N \} }
    \!\!\!\!\!\!\!\!\!\! {'} \quad\;
    \frac{Z!}{p_0! \, \ldots \, p_N!}
    = {N + Z -1 \choose N} \; ,
\label{eq:APci}
\end{equation}
which expresses the number of ways to distribute $N$~Bose
particles over $Z$~quantum cells in two different manners: On the
left-hand side, which corresponds to Bose's way of counting
microstates, numbers~$p_r$ specify how many cells contain $r$
quanta; of course, with only $N$~quanta being available, one has
$p_r = 0$ for $r > N$. As symbolically indicated by the prime on
the summation sign, the sum thus extends only over those sets $\{ p_0,
p_1, \ldots, p_N \}$ with comply with the conditions
\begin{equation}
    \sum_{r=0}^N p_r = Z \; ,
\label{eq:APr1}
\end{equation}
stating that there are $Z$~cells to accommodate the quanta, and
\begin{equation}
    \sum_{r=0}^N r p_r = N \; ,
\label{eq:APr2}
\end{equation}
stating that the number of quanta be~$N$. The right-hand side of
Eq.~(\ref{eq:APci}) gives the total number of microstates, taking into
account all possible sets of occupation numbers in the single expression
already used by Einstein in his final derivation~\cite{Einstein25} of the
Bose--Einstein distribution which can still be found in today's textbooks.

The validity of Eq.~(\ref{eq:APci}) is clear for combinatorial reasons.
Nonetheless, since this identity~(\ref{eq:APci}) constitutes one of the
less known relations in the theory of Bose--Einstein statistics, we give
its explicit proof in this appendix.

As for most mathematical proofs, one needs tools, an idea, and a
conjurer's trick. In the present case, the tools are two
generalizations of the binomial theorem
\begin{equation}
    (a + b)^n = \sum_{k=0}^n {n \choose k} \, a^k \, b^{n-k} \; ,
\label{eq:APbt}
\end{equation}
where
\begin{equation}
    {n \choose k} = \frac{n!}{k! \, (n-k)!}
\end{equation}
denotes the familiar binomial coefficents. The first such
generalization is the {\em multinomial theorem\/}
\begin{equation}
    (a_1 + a_2 + \ldots + a_N)^n
    = \sum_{\sum p_r = n}\frac{n!}{p_1! \, p_2! \ldots p_N!} \,
    a_1^{p_1} a_2^{p_2} \ldots a_N^{p_N} \; ,
\label{eq:APmt}
\end{equation}
which is easily understood: When multiplying out the left-hand
side, every product obtained contains one factor $a_i$ from each
bracket $(a_1 + a_2 + \ldots + a_N)$. Hence, in every product
$a_1^{p_1} a_2^{p_2} \ldots a_N^{p_N}$ the exponents add up to the
number of brackets, which is~$n$. Therefore, for such a product
there are $n!$ permutations of the individual factors $a_i$.
However, if identical factors $a_i$ are permuted among themselves,
for which there are $p_i!$ possibilities, one obtains the same
value. Hence, the coefficient of each product on the right-hand
side in Eq.~(\ref{eq:APmt}) corresponds to the number of possible
arrangements of its factors, divided by the number of equivalent
arrangements. Note that the reasoning here is essentially the same
as for the justification of Bose's expression~(\ref{eq:Bose}),
which is, of course, not accidental.

The second generalization of the binomial theorem~(\ref{eq:APbt})
required for the proof of the identity~(\ref{eq:APci}) emerges
when we replace the exponent~$n$ by a non-natural number~$\gamma$:
One then has
\begin{equation}
    (a + b)^\gamma = \sum_{k=0}^\infty {\gamma \choose k} \,
    a^k b^{\gamma-k} \; ,
\label{eq:APgb}
\end{equation}
with the definition
\begin{equation}
    {\gamma \choose k} =
    \frac{\gamma(\gamma -1)(\gamma-2) \cdot \ldots \cdot (\gamma - k + 1)}
    {k!} \; .
\label{eq:APdf}
\end{equation}
If $\gamma$ is not a natural number, this series~(\ref{eq:APgb})
converges for any complex numbers $a,b$, provided $|a/b| < 1$.
This {\em generalized binomial theorem\/}~(\ref{eq:APgb}), which
is treated in introductory analysis courses, is useful, {\em
e.g.\/}, for writing down the Taylor expansion of $(1+x)^\gamma$.

Given these tools, the idea for proving the
identity~(\ref{eq:APci}) now consists in considering the
expression
\begin{equation}
    P_{NZ}(x) = (1 + x + x^2 + x^3 + \ldots + x^N)^Z \; ,
\end{equation}
where $x$ is some variable which obeys $|x| < 1$, but need not be
specified further. According to the multinomial
theorem~(\ref{eq:APmt}), one has
\begin{equation}
    P_{NZ}(x) = \sum_{\sum_{r=0}^N p_r = Z}
    \frac{Z!}{p_0! \, p_1! \ldots p_N!} \,
    (x^0)^{p_0} (x^1)^{p_1} \ldots (x^N)^{p_N} \; .
\label{eq:APnz}
\end{equation}
Directing the attention then to a systematic ordering of this
series with respect to powers of~$x$, the coefficient of $x^N$
equals the sum of all coefficients encountered here which
accompany terms with $\sum_r r p_r = N$, which is precisely the
left-hand side of the desired equation~(\ref{eq:APci}), keeping in
mind the restrictions~(\ref{eq:APr1}) and~(\ref{eq:APr2}).

Now comes the conjurer's trick: The coefficient of $x^N$ in this
expression~(\ref{eq:APnz}), equaling the left-hand side of
Eq.~(\ref{eq:APci}), also equals the coefficent of $x^N$ in the
expression
\begin{equation}
    P_{\infty}(x) = (1 + x + x^2 + x^3 + \ldots x^N + x^{N+1} + \ldots)^Z
    \; ,
\end{equation}
since this differs from $P_{NZ}(x)$ only by powers of~$x$ higher
than~$N$. But this involves a geometric series, which is
immediately summed:
\begin{eqnarray}
    P_{\infty}(x) & = & \left( \sum_{r=0}^\infty x^r \right)^Z
\nonumber \\
    & = & (1 - x)^{-Z} \; .
\end{eqnarray}
According to the generalized binomial theorem~(\ref{eq:APgb}), one
has
\begin{equation}
    (1 - x)^{-Z} = \sum_{k=0}^\infty {-Z \choose k} \, (-x)^{k} \; ,
\end{equation}
so that, in view of the definition~(\ref{eq:APdf}), the
coefficient of~$x^N$ equals
\begin{eqnarray}
    (-1)^N {-Z \choose N} & = & (-1)^N
    \frac{(-Z)(-Z-1)\cdot \ldots \cdot (-Z-N+1)}{N!}
\nonumber \\
    & = & \frac{(Z+N-1)\cdot \ldots \cdot (Z+1) Z}{N!}
\nonumber \\
    & = & \frac{(Z+N-1)!}{N! \, (Z-1)!}
\nonumber \\
    & = & {Z+N-1 \choose N} \; .
\end{eqnarray}
This is the right-hand side of the identity~(\ref{eq:APci}), which
completes the proof.

\section{Analytical expression for the mean number of condensed atoms}
\label{ap:mean}

We obtain an analytical expression for
$\bar{n}_{0}$ from Eq.~(\ref{N relates n0}). One can reduce the
triple sum into a single sum if we take into account the
degeneracy $g(E)$ of the level with energy $E=\hbar \Omega
(l+m+n)$, which is equal to the number of ways to fill the level.
This number can be calculated from the Einstein's complexion
equation, i.e.
\begin{equation}
g(E)=\frac{\{(l+m+n)+(3-1)\}!}{(l+m+n)!(3-1)!}=\frac{1}{2}\left( \frac{E}{%
\hbar \Omega }+2\right) \left( \frac{E}{\hbar \Omega }+1\right) .
\label{g(E)}
\end{equation}

Now, we have reduced three variables $l,m,n$ to only one. By
letting $E=s\hbar \Omega $ where $s$ is integer, one can write
\begin{equation}
N=\sum\limits_{s=0}^{\infty }\frac{\frac{1}{2}(s+2)(s+1)}{\left( \frac{1}{%
\bar{n}_{0}}+1\right) e^{s\beta \hbar \Omega }-1}\simeq \bar{n}_{0}+\frac{%
\bar{n}_{0}}{(1+\bar{n}_{0})}S,  \label{N
quadratic}
\end{equation}
where for $\bar{n}_{0} \gg 1$
\begin{equation}
S\simeq \sum\limits_{s=1}^{\infty }\frac{\frac{1}{2}(s+2)(s+1)}{e^{s\beta
\hbar \Omega }-1}.  \label{S}
\end{equation}

The root of the quadratic Eq.~(\ref{N quadratic}) yields
\begin{equation}
\bar{n}_{0}=-\frac{1}{2}[(1+S-N)-\sqrt{(1+S-N)^{2}+4N}].  \label{quad1}
\end{equation}

In order to find an analytical expression for $S$, we write
\begin{equation}
S\simeq \frac{1}{2}\sum\limits_{s=0}^{\infty }\frac{s^{2}+3s}{e^{s\beta
\hbar \Omega }-1}+\sum\limits_{s=1}^{\infty }\frac{1}{e^{s\beta \hbar \Omega
}-1}.  \label{S rewrote}
\end{equation}
Converting the summation into integration by replacing $x=s\beta \hbar
\Omega $ yields
\begin{eqnarray}
S &\simeq &\frac{1}{a}\int_{a}^{\infty }\frac{dx}{e^{x}-1}+\frac{1}{2a^{3}}%
\int_{0}^{\infty }\frac{x^{2}dx}{e^{x}-1}+\frac{3}{2a^{2}}\int_{0}^{\infty }%
\frac{xdx}{e^{x}-1}  \nonumber \\
&=&1-\frac{1}{a}\ln \left( e^{a}-1\right) +\frac{\zeta \left( 3\right) }{%
a^{3}}\allowbreak +\left( \frac{\pi }{2a}\right) ^{2},  \label{S
ana}
\end{eqnarray}
where $a=\beta \hbar \Omega =\hbar \Omega /k_{B}T=(\zeta (3)/N)^{1/3}T_{c}/T$%
.

Thus, Eq. (\ref{S}) gives the following analytical expression for $S$:
\[
(1+S-N)
\]
\[
=-N\left\{ 1-\left( \frac{T}{T_{c}}\right) ^{3}\right\} \allowbreak +\frac{%
\pi ^{2}}{4}\left( \frac{N}{\zeta (3)}\right) ^{2/3}\left( \frac{T}{T_{c}}%
\right) ^{2}
\]
\begin{equation}
-\left( \frac{N}{\zeta (3)}\right) ^{1/3}\frac{T}{T_{c}}\ln \left( e^{(\zeta
(3)/N)^{1/3}T_{c}/T}-1\right) +2  \label{Bx}
\end{equation}

\begin{figure}[tbp]
\center\epsfxsize=10cm\epsffile{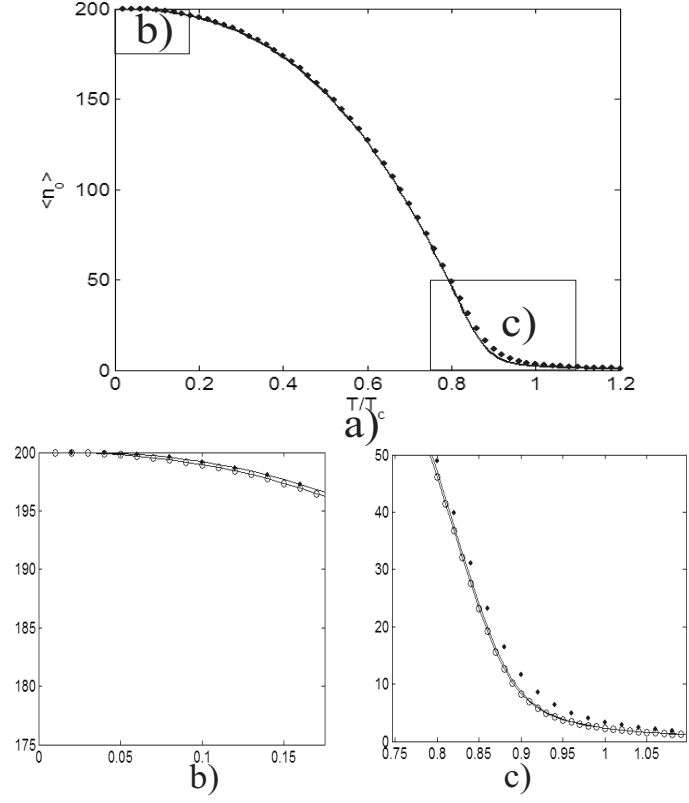}
\caption{Grand canonical result for $\langle n_{0}\rangle $ as a function of
temperature for $N=200$, computed from analytical expressions, Eqs. (\ref
{quad1}) and (\ref{Bx}) (line with circles); semi-analytical expressions
given by Eqs. (\ref{quad1}) and (\ref{S}) (solid line); and exact numerical
solution of Eq. (\ref{N relates n0}) (dots). Expanded views show: b) exact
agreement of the semi-analytical approach at low temperature and c) a small
deviation near $T_{c}$. Also, the analytical and the semi-analytical results
agree quite well.}
\label{n0semanaexa}
\end{figure}

Fig.~\ref{n0semanaexa} compares different approximations for calculating
$\bar{n}_{0}$ within the grand canonical ensemble.

\section{Formulas for the central moments of condensate fluctuations}
\label{ap:fluct}

By using $n_{0}=N-\sum\limits_{k\neq 0}n_{k}$ and
$\langle n_{0}\rangle =N-\sum\limits_{k \neq 0}\langle n_{k}\rangle $,
we have
\begin{equation}
\langle (n_{0}-\langle n_{0}\rangle )^{s}\rangle =(-1)^{s}\langle \lbrack
\sum\limits_{k \neq 0}(n_{k}-\langle n_{k}\rangle )]^{s}\rangle \text{,}
\end{equation}
which shows that the fluctuations of the condensate particles are proportional
to the fluctuations of the non-condensate particles.

As an example we show how to evaluate the fluctuation for the second order
moment, or variance,
\begin{equation}
\langle (n_{0}-\langle n_{0}\rangle )^{2}\rangle
=\sum\limits_{j,k \neq 0}(\langle n_{j}n_{k}\rangle -\langle n_{j}\rangle \langle
n_{k}\rangle ).
\end{equation}
The numbers of particles in different levels are statistically independent,
since $\langle n_{j}n_{k\neq j}\rangle =Tr\{\hat{a}_{j}^{\dagger }\hat{a}_{j}%
\hat{a}_{k}^{\dagger }\hat{a}_{k}\hat{\rho}\}=Tr\{\hat{a}_{j}^{\dagger }\hat{%
a}_{j}\hat{\rho}_{j}\}Tr\{\hat{a}_{k}^{\dagger }\hat{a}_{k}\hat{\rho}%
_{k}\}=\langle n_{j}\rangle \langle n_{k}\rangle$. Thus, we find
\begin{eqnarray}
\langle (n_{0}-\langle n_{0}\rangle )^{2}\rangle
&=&\sum\limits_{k \neq 0}[\langle n_{k}^{2}\rangle -\langle n_{k}\rangle ^{2}]
\label{non-cond mom2} \\
\langle (n_{0}-\langle n_{0}\rangle )^{3}\rangle
&=&\sum\limits_{k \neq 0}[-\langle n_{k}^{3}\rangle +3\langle n_{k}^{2}\rangle
\langle n_{k}\rangle -2\langle n_{k}\rangle ^{3}]  \label{non-cond mom3} \\
\langle (n_{0}-\langle n_{0}\rangle )^{4}\rangle
&=&\sum\limits_{k \neq 0}[\allowbreak \langle n_{k}{}^{4}\rangle -4\langle
n_{k}{}^{3}\rangle \langle n_{k}{}\rangle +6\langle n_{k}{}^{2}\rangle
\langle n_{k}{}\rangle ^{2}-3\langle n_{k}{}\rangle ^{4}] \; .
\label{non-cond mom4}
\end{eqnarray}
In the grand canonical approach, $\langle n_{k}{}^{s}\rangle $ can be evaluated using
\begin{equation}
\langle n_{k}{}^{s}\rangle =\frac{1}{Z_{k}}\sum\limits_{n_{k}}n_{k}^{s}e^{-%
\beta (\epsilon _{k}-\mu )n_{k}}  \label{mean n^s} \; ,
\end{equation}
where $Z_{k}=(1-e^{-\beta (\epsilon _{k}-\mu )})^{-1}$. An alternative way
is to use the formula $\langle n_{k}{}^{s}\rangle =\frac{d^{s}\Theta _{k}}{%
d(iu)^{s}}|_{u=0}$ derived in Section V.B. In particular, one can show that
in this approach
\begin{eqnarray}
\langle n_{k}^{2}\rangle &=&2\langle n_{k}\rangle ^{2}+\langle n_{k}\rangle ,
\label{mean n^2} \\
\langle n_{k}^{3}\rangle &=&6\langle n_{k}\rangle ^{3}+6\langle n_{k}\rangle
^{2}+\langle n_{k}\rangle ,  \label{mean n^3} \\
\langle n_{k}^{4}\rangle &=&24\langle n_{k}{}\rangle ^{4}+36\langle
n_{k}{}\rangle ^{3}+14\langle n_{k}{}\rangle ^{2}+\langle n_{k}{}\rangle .
\label{mean n^4}
\end{eqnarray}
Using Eqs.~(\ref{mean n^2})--(\ref{mean n^4}) we obtain
\begin{eqnarray}
\langle (n_{0}-\langle n_{0}\rangle )^{2}\rangle
&=&\sum\limits_{k \neq 0}[\langle n_{k}\rangle ^{2}+\langle n_{k}\rangle ]
\label{mom2} \\
\langle (n_{0}-\langle n_{0}\rangle )^{3}\rangle
&=&-\sum\limits_{k \neq 0}[2\langle n_{k}\rangle ^{3}+3\langle n_{k}\rangle
^{2}+\langle n_{k}\rangle ]  \nonumber \\
\langle (n_{0}-\langle n_{0}\rangle )^{4}\rangle &=&\sum_{k \neq 0}[9\langle
n_{k}\rangle ^{4}+18\langle n_{k}\rangle ^{3}+10\langle n_{k}\rangle
^{2}+\langle n_{k}\rangle ]  \label{mom4} \; .
\end{eqnarray}

\section{Analytical expression for the variance of condensate fluctuations}
\label{ap:variance}

In a spherically symmetric harmonic trap with trap
frequency $\Omega $ one can convert the triple sums in Eq.~(\ref{a1})
into a single sum by using Eq.~(\ref {g(E)}), and then do
integration upon replacing $\sum\limits_{x=0}^{\infty}...\rightarrow
\frac{1}{\beta \hbar \Omega }\int_{0}^{\infty }dx...$ and $%
\beta E\rightarrow x$, giving
\[
\Delta n_{0}^{2}=\sum\limits_{E\neq 0}^{\infty } \frac{1}{2}\left( \frac{E}{%
\hbar \Omega }+2\right) \left( \frac{E}{\hbar \Omega }+1\right) \left\{
\frac{1}{\left[ \exp (\beta E)\left( 1+\frac{1}{\bar{n}_{0}}\right) -1\right]
^{2}}+\frac{1}{\exp (\beta E)\left( 1+\frac{1}{\bar{n}_{0}}\right) -1}%
\right\}
\]
\[
=\frac{1}{2\beta \hbar \Omega }\int_{a}^{\infty }dx\left[ \left( \frac{x}{%
\beta \hbar \Omega }\right) ^{2}+\frac{3x}{\beta \hbar \Omega }+2\right]
\left\{ \frac{1}{\left[ \exp (x)\left( 1+\frac{1}{\bar{n}_{0}}\right) -1%
\right] ^{2}}+\frac{1}{\exp (x)\left( 1+\frac{1}{\bar{n}_{0}}\right) -1}%
\right\}
\]
\begin{equation}
=\frac{1}{2a}\int_{a}^{\infty }dx\left[ \left( \frac{x}{a}\right) ^{2}+\frac{%
3x}{a}+2\right] \frac{\exp (x)\left( 1+\frac{1}{\bar{n}_{0}}\right) }{\left[
\exp (x)\left( 1+\frac{1}{\bar{n}_{0}}\right) -1\right] ^{2}},  \label{DN023}
\end{equation}
where $\ a=\beta \hbar \Omega =\hbar \Omega /k_{B}T=(T_{c}/T)(\zeta
(3)/N)^{1/3}$ and the density of states
can be written as $\rho (E)=\frac{1}{2\hbar \Omega }\left[ \left( \frac{E}{%
\hbar \Omega }\right) ^{2}+\frac{3E}{\hbar \Omega }+2\right] $. Then we
integrate by parts using the identity
\[
\frac{\exp (x)\left( 1+\frac{1}{\bar{n}_{0}}\right) }{\left[ \exp (x)\left(
1+\frac{1}{\bar{n}_{0}}\right) -1\right] ^{2}}=-\frac{\partial }{\partial x}%
\frac{1}{\left[ \exp (x)\left( 1+\frac{1}{\bar{n}_{0}}\right) -1\right] },
\]
arriving at
\begin{equation}
\Delta n_{0}^{2}=\frac{3}{a\left[ e^{a}\left( 1+\frac{1}{\bar{n}_{0}}\right)
-1\right] }+\frac{1}{2a^{2}}\int_{a}^{\infty }dx\frac{\left( \frac{2x}{a}%
+3\right) }{\left[ \exp (x)\left( 1+\frac{1}{\bar{n}_{0}}\right) -1\right] }.
\label{a4}
\end{equation}
The integral in Eq.~(\ref{a4}) can be calculated analytically, using
\[
\int_{a}^{\infty }\frac{xdx}{\left[ \exp (x)A-1\right] }=\frac{\pi ^{2}}{6}-%
\frac{1}{2}\ln ^{2}A+\ln \left( Ae^{a}-1\right) \ln A+\text{di}\log (Ae^{a})+%
\frac{a^{2}}{2}\text{,}
\]
\[
\int_{a}^{\infty }\frac{dx}{\left[ \exp (x)A-1\right] }=\ln (1+\alpha )-\ln
\left( Ae^{a}-1\right) +a\text{,}
\]
where
\[
\text{di}\log (x)=\int_{1}^{x}\frac{\ln (t)}{1-t}dt.
\]
As a result, we get
\[
\Delta n_{0}^{2}=\frac{1}{a^{3}}\left[ \frac{\pi ^{2}}{6}+\text{di}\log %
\left[ e^{a}\left( 1+\frac{1}{\bar{n}_{0}}\right) \right] -\frac{1}{2}\ln
^{2}\left( 1+\frac{1}{\bar{n}_{0}}\right) +\ln \left[ e^{a}\left( 1+\frac{1}{%
\bar{n}_{0}}\right) -1\right] \ln \left( 1+\frac{1}{\bar{n}_{0}}\right)
\right.
\]
\begin{equation}
\left. + \frac{3}{2}a\ln \left( \frac{\bar{n}_{0}+1}{e^{a}\left( \bar{n}%
_{0}+1\right) -\bar{n}_{0}}\right) \right] +\frac{3}{a\left[ e^{a}\left( 1+%
\frac{1}{\bar{n}_{0}}\right) -1\right] }+\frac{2}{a}.  \label{Dn02 ana}
\end{equation}
Taking into account that $a=(T_{c}/T)(\zeta(3)/N)^{1/3}$, we finally
obtain
\[
\Delta n_{0}^{2}=\left( \frac{T}{T_{c}}\right) ^{3}\frac{N}{\zeta (3)}\left[
\frac{\pi ^{2}}{6}+\text{di}\log \left[ \exp [(T_{c}/T)(\zeta
(3)/N)^{1/3}]\left( 1+\frac{1}{\bar{n}_{0}}\right) \right] -\frac{1}{2}\ln
^{2}\left( 1+\frac{1}{\bar{n}_{0}}\right) \right.
\]
\[
\left. +\ln \left[ \exp [(T_{c}/T)(\zeta (3)/N)^{1/3}]\left( 1+\frac{1}{%
\bar{n}_{0}}\right) -1\right] \ln \left( 1+\frac{1}{\bar{n}_{0}}\right) %
\right]
\]
\[
+\frac{3}{2}\left( \frac{T}{T_{c}}\right) ^{2}\left( \frac{N}{\zeta (3)}%
\right) ^{2/3}\ln \left[ \frac{\bar{n}_{0}+1}{\exp [(T_{c}/T)(\zeta
(3)/N)^{1/3}]\left( \bar{n}_{0}+1\right) -\bar{n}_{0}}\right]
\]
\begin{equation}
+\frac{T}{T_{c}}\left( \frac{N}{\zeta (3)}\right) ^{1/3}\left[ \frac{3}{\exp
[(T_{c}/T)(\zeta (3)/N)^{1/3}]\left( 1+\frac{1}{\bar{n}_{0}}\right) -1}+2%
\right] .  \label{Dn02 analytic}
\end{equation}

\section{Single mode coupled to a reservoir of oscillators}
\label{ap:reservoir}

The derivation of the damping Liouvillean
proceeds from the Liouville-von Neumann equation
\begin{equation}
\frac{\partial }{\partial t}\hat{\rho}(t)=\frac{1}{i\hbar }[\hat{V}_{sr},%
\hat{\rho}(t)]  \; , \label{LvN}
\end{equation}
where $\hat{V}_{sr}$ is the Hamiltonian in the interaction picture for
the system ($s$) coupled to a reservoir ($r$).

We can derive a closed form of the dynamical equation for the
reduced density operator for the system, $\hat{\rho}_{s}(t)=$
Tr$_{r}\{\hat{\rho}(t)\}$ by tracing out the reservoir. This is
accomplished by first integrating Eq.~(\ref{LvN}) for
$\hat{\rho}(t)=\hat{\rho}(0)+\frac{1}{i\hbar }\int_{0}^{t}[%
\hat{V}_{sr}(t^{\prime }),\hat{\rho}(t^{\prime })]dt^{\prime }$
and then substituting it back into the right hand side of Eq.~(\ref{LvN}),
giving
\begin{equation}
\frac{\partial }{\partial t}\hat{\rho}_{s}(t)=\frac{1}{i\hbar }Tr_{r}[\hat{V}%
_{sr},\hat{\rho}(0)]+\frac{1}{(i\hbar )^{2}}Tr_{r}\int_{0}^{t}[\hat{V}%
_{sr}(t),[\hat{V}_{sr}(t^{\prime }),\hat{\rho}(t^{\prime
})]dt^{\prime }] \label{dp/dt} \; .
\end{equation}

We may repeat this indefinitely, but owing to the weaknesses of
the system-reservoir interaction, it is possible to ignore terms
higher than 2nd order in $\hat{V}_{sr}$. Furthermore, we assume
the system and reservoir are approximately uncorrelated in the
past and the reservoir is so large that it
remains practically in thermal equilibrium $\hat{\rho}_{r}^{th}$, so $\hat{%
\rho}(t^{\prime })\simeq $ $\hat{\rho}_{s}(t^{\prime })\otimes $ $\hat{\rho}%
_{r}^{th}$ and \ $\hat{\rho}(0)=\hat{\rho}_{s}(0)\otimes $ $\hat{\rho}%
_{r}^{th}$.

For a single mode field ($f$) coupled to a reservoir of oscillators,
one has
\begin{equation}
\hat{V}_{fr}=\hbar \sum\limits_{\mathbf{k}}g_{\mathbf{k}}(\hat{b}_{\mathbf{k}%
}\hat{a}^{\dagger }e^{i(\nu -\nu _{\mathbf{k}})t}+\hat{b}_{\mathbf{k}%
}^{\dagger }\hat{a}e^{-i(\nu -\nu _{\mathbf{k}})t}) \; . \label{Vsr}
\end{equation}

Since Tr$_{r}\{\hat{a}^{\dagger }\hat{\rho}_{r}^{th}\}=0$ and Tr$_{r}\{\hat{a%
}\hat{\rho}_{r}^{th}\}=0$, the first term vanishes and by using
Eq.~(\ref {Vsr}) we have 16 terms. But secular approximation
reduces the number of
terms by half. We now perform the Markov approximation, $\hat{\rho}%
_{f}(t^{\prime})\simeq \hat{\rho}_{f}(t)$, stating  that the dynamics of
the system is independent of the states in the past.

The thermal average of the radiation operators is
\[
\sum\limits_{\mathbf{k}}g_{\mathbf{k}}^{2}\int_{0}^{t}e^{-i(\nu -\nu _{%
\mathbf{k}})(t-t^{\prime })}Tr_{r}\{\hat{\rho}_{r}^{th}\hat{b}_{\mathbf{k}%
}^{\dagger }\hat{b}_{\mathbf{k}}\}dt^{\prime }\simeq \sum\limits_{\mathbf{k}%
}g_{\mathbf{k}}^{2}\pi \delta (\nu -\nu _{\mathbf{k}})\bar{n}(\nu _{\mathbf{k%
}})=\bar{n}(\nu )\mathcal{G}(\nu )/2 \; ,
\]
where $Tr_{r}\{\hat{\rho}_{r}^{th}\hat{b}_{\mathbf{k}}^{\dagger }\hat{b}_{%
\mathbf{k}}\}=\bar{n}(\nu _{\mathbf{k}})=(e^{\beta \hbar \nu _{\mathbf{k}%
}}-1)^{-1}$. \ Thus, we have
\begin{equation}
\frac{\partial }{\partial t}\hat{\rho}_{f}(t)=-\frac{1}{2}\mathcal{C}[\hat{a}%
^{\dagger }\hat{a}\hat{\rho}_{f}(t)-2\hat{a}\hat{\rho}_{f}(t)\hat{a}%
^{\dagger }+\hat{\rho}_{f}(t)\hat{a}^{\dagger }\hat{a}]-\frac{1}{2}\mathcal{D%
}[\hat{a}\hat{a}^{\dagger }\hat{\rho}_{f}(t)-2\hat{a}^{\dagger }\hat{\rho}%
_{f}(t)\hat{a}+\hat{\rho}_{f}(t)\hat{a}\hat{a}^{\dagger }] \; ,
\label{Louivillean}
\end{equation}
where $\mathcal{D}=\mathcal{G}\bar{n}$ and
$\mathcal{C}=\mathcal{G}(\bar{n}+1)$.

\section{The saddle-point method for condensed Bose gases}
\label{ap:saddle}

The saddle-point method is one of the most
essential tools in statistical physics. Yet, the conventional form
of this approximation fails in the case of condensed ideal Bose
gases \cite{Schu46,Ziff}. The point is that in the condensate
regime the saddle-point of the grand canonical partition function
approaches the ground-state singularity at $z=\exp (\beta \varepsilon _{0})$%
, which is a hallmark of BEC. However, the customary Gaussian approximation
requires that intervals around the saddle-point stay clear of singularities.
Following the original suggestion by Dingle~\cite{Dinglebook}, Holthaus and
Kalinowski~\cite{Holt99} worked out a natural solution to this problem:
One should exempt the ground-state factor of the grand canonical
partition function from the Gaussian expansion and treat that factor
exactly, but proceed as usual otherwise. The success of this refined
saddle-point method hinges on the fact that the emerging integrals with
singular integrands can be done exactly; they lead directly to parabolic
cylinder functions. Here we discuss the refined saddle-point method in
some detail.

We start from the grand canonical partition function
\begin{equation}
\Xi (\beta ,z)=\prod\limits_{\nu =0}^{\infty }\frac{1}{1-z\exp (-\beta
\varepsilon _{\nu })},
\end{equation}
where $\varepsilon _{\nu }$ are single-particle energies, $\beta =1/k_{B}T$
and $z=\exp (\beta \mu )$. The grand canonical partition function $\Xi
(\beta ,z)$ generates the canonical partition functions $Z_{N}(\beta )$\ by
means of the expansion
\begin{equation}
\Xi (\beta ,z)=\sum_{N=0}^{\infty }z^{N}Z_{N}(\beta ).
\end{equation}
Then we treat $z$ as a complex variable and using Cauchy's theorem represent
$Z_{N}(\beta )$ by a contour integral,
\begin{equation}
Z_{N}(\beta )=\frac{1}{2\pi i}\oint dz\frac{\Xi (\beta ,z)}{z^{N+1}}=\frac{1%
}{2\pi i}\oint dz\exp (-F(z)),
\end{equation}
where the path of integration encircles the origin counter-clockwise, and
\begin{equation}
F(z)=(N+1)\ln z-\ln \Xi (\beta ,z)=(N+1)\ln z+\sum_{\nu =0}^{\infty }\ln
[1-z\exp (-\beta \varepsilon _{\nu })].  \label{x4}
\end{equation}
The saddle-point $z_{0}$ is determined by the requirement that $F(z)$
becomes stationary,
\begin{equation}
\left.\frac{\partial F(z)}{\partial z}\right|_{z=z_{0}}=0,
\end{equation}
giving
\begin{equation}
N+1=\sum_{\nu =0}^{\infty }\frac{1}{\exp (\beta \varepsilon _{\nu })/z_{0}-1}%
.
\end{equation}
This is just the grand canonical relation between particle number $N$ and
fugacity $z_{0}$ (apart from the appearance of one extra particle on the
left-hand side). Appendix C discusses an approximate analytic solution of
such equations.

\begin{figure}[tbp]
\center \epsfxsize=12cm\epsffile{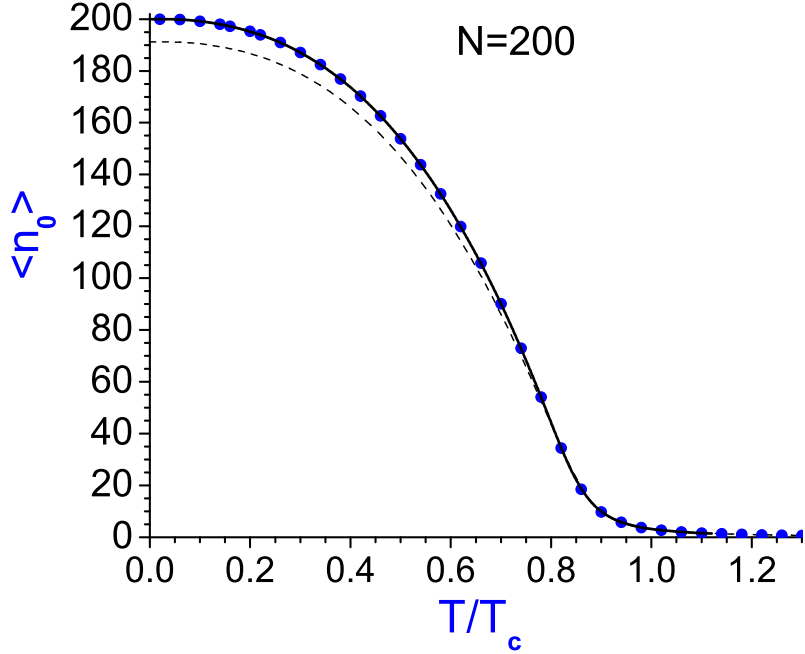}
\caption{Canonical occupation number of the ground state as a function of
temperature for $N=200$ atoms in a harmonic isotropic trap. Dashed and solid
curves are obtained by the conventional and the refined saddle-point method,
respectively. Dots are ``exact'' numerical answers obtained for the canonical
ensemble.}
\label{sad1}
\end{figure}

\begin{figure}[tbp]
\center \epsfxsize=12cm\epsffile{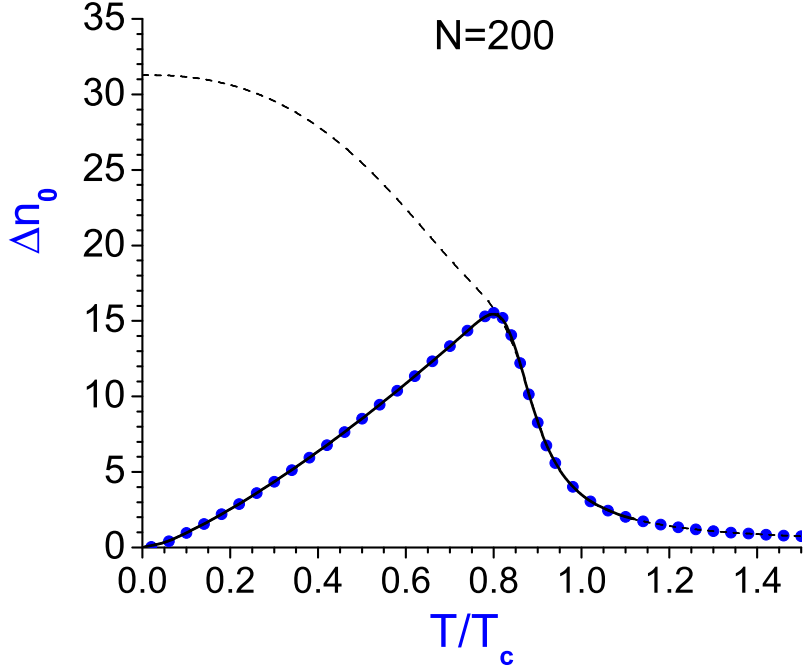}
\caption{Variance in the condensate particle number as a function of
temperature for $N=200$ atoms in a harmonic isotropic trap. Dashed and solid
curves are obtained by the conventional and the refined saddle-point method,
respectively. Dots are ``exact'' numerical answers in the canonical ensemble.}
\label{sad2}
\end{figure}

In the conventional saddle-point method, the whole function $F(z)$ is taken in
the saddle-point approximation
\begin{equation}
F(z)\approx F(z_{0})+\frac{1}{2}F^{\prime \prime }(z_{0})(z-z_{0})^{2}.
\label{x3}
\end{equation}
Doing the remaining Gaussian integral yields
\begin{equation}
Z_{N}(\beta )\approx \frac{\exp (-F(z_{0}))}{\sqrt{-2\pi F^{\prime \prime
}(z_{0})}}.
\end{equation}
The canonical occupation number of the ground state, and its mean-square
fluctuations, are obtained by differentiating the canonical partition
function:
\begin{equation}
\bar{n}_{0}=\frac{\partial \ln Z_{N}(\beta )}{\partial (-\beta \varepsilon
_{0})}=\frac{1}{Z_{N}(\beta )}\frac{1}{2\pi i}\oint dz\frac{1}{z^{N+1}}\frac{%
\partial \Xi (\beta ,z)}{\partial (-\beta \varepsilon _{0})},  \label{x1}
\end{equation}
\begin{equation}
\Delta n_{0}=\frac{\partial ^{2}\ln Z_{N}(\beta )}{\partial (-\beta
\varepsilon _{0})^{2}}=-\bar{n}_{0}^{2}+\frac{1}{Z_{N}(\beta )}\frac{1}{2\pi
i}\oint dz\frac{1}{z^{N+1}}\frac{\partial ^{2}\Xi (\beta ,z)}{\partial
(-\beta \varepsilon _{0})^{2}}.  \label{x2}
\end{equation}
The saddle-point approximation is then applied to the integrands of
Eqs.~(\ref{x1}) and (\ref{x2}). Figs.~\ref{sad1} and \ref{sad2} (dashed curves)
show results for $\bar{n}_{0}$ and $\Delta n_{0}$ obtained by the
conventional saddle-point method for a Bose gas with $N=200$ atoms in a
harmonic isotropic trap. In the condensate regime there is a substantial
deviation of the saddle-point curves from the ``exact'' numerical answer
obtained by solution of the recursion equations for the canonical statistics
(dots).

The reason for this inaccuracy is that in the condensate region the
saddle-point $z_{0}$ lies close to the singular point $z=\exp (\beta
\varepsilon _{0})$ of the function $F(z)$. As a result, the approximation
(\ref{x3}) becomes invalid in the condensate region. To improve the method,
Dingle~\cite{Dinglebook} proposed to treat the potentially dangerous term
in (\ref{x4}) as it is, and represent $Z_{N}(\beta )$ as
\begin{equation}
Z_{N}(\beta )=\frac{1}{2\pi i}\oint dz\frac{\exp (-F_{1}(z))}{1-z\exp
(-\beta \varepsilon _{0})},  \label{x5}
\end{equation}
where
\begin{equation}
F_{1}(z)=(N+1)\ln z+\sum_{\nu =1}^{\infty }\ln [1-z\exp (-\beta \varepsilon
_{\nu })]
\end{equation}
has no singularity at $z=\exp (\beta \varepsilon _{0})$. The singular point
to be watched now is the one at $z=\exp (\beta \varepsilon _{1})$. Since $%
z_{0}<\exp (\beta \varepsilon _{0}),$ the saddle-point remains separated
from that singularity by at least the $N$-independent gap $\exp (\beta
\varepsilon _{1})-\exp (\beta \varepsilon _{0})\simeq \hbar \omega /k_{B}T$.
This guarantees that the amputated function $F_{1}(z)$ remains
singularity-free in the required interval around $z_{0}$ for sufficiently
large $N$. Then the Gaussian approximation to $\exp (-F_{1}(z))$ is safe.
The subsequently emerging saddle-point integral for the canonical partition
function can be done exactly, yielding~\cite{Holt99}
\begin{equation}
Z_{N}(\beta )\approx \frac{1}{2}\exp [\beta \varepsilon
_{0}-F_{1}(z_{0})-1+\eta ^{2}/2] \text{erfc}\left( \frac{\eta _{1}}{\sqrt{2}}%
\right) ,
\end{equation}
where erfc$(z)=2/\sqrt{\pi }\int_{z}^{\infty }\exp (-t^{2})dt$ is the
complementary error function, $\eta =(\exp (\beta \varepsilon _{0})-z_{0})%
\sqrt{-F_{1}^{\prime \prime }(z_{0})}$, and $\eta _{1}=\eta -1/\eta $.

\begin{figure}[tbp]
\center \epsfxsize=12cm\epsffile{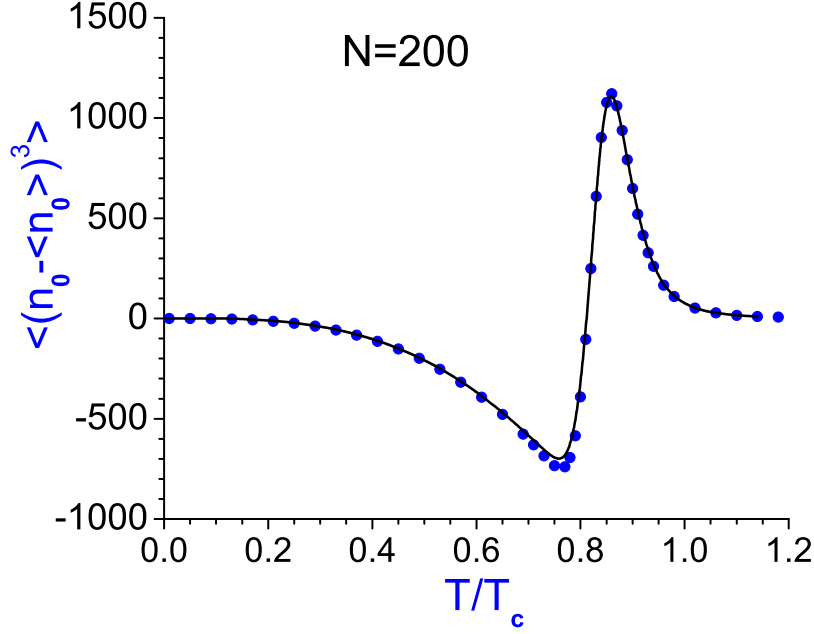}
\caption{The third central moment $\langle (n_{0}-\bar{n}_{0})^{3}\rangle $
for $N=200$ atoms in a harmonic isotropic trap obtained using the refined
saddle-point method (solid curve). Exact numerical results for the canonical
ensemble are shown by dots.}
\label{sad3}
\end{figure}

Calculation of occupation numbers and their fluctuations deals with
integrals from derivatives of $\Xi (\beta ,z)$ with respect to $-\beta
\varepsilon _{0}$. In such expressions the factors singular at $z=\exp
(\beta \varepsilon _{0})$ should be taken exactly. This leads to the
integrals of the following form:
\begin{equation}
\frac{1}{2\pi i}\oint dz\frac{\exp [-f_{1}(z)-(\sigma -1)\beta \varepsilon
_{0}]}{(1-z\exp (-\beta \varepsilon _{0}))^{\sigma }}\approx \frac{1}{\sqrt{%
2\pi }}(-f_{1}^{\prime \prime }(z_{\ast }))^{(\sigma -1)/2}\exp (\beta
\varepsilon _{0}-f_{1}(z_{\ast })-\sigma +\eta ^{2}/2-\eta
_{1}^{2}/4)D_{-\sigma }(\eta _{1})\text{,}
\end{equation}
where $\eta =(\exp (\beta \varepsilon _{0})-z_{\ast })\sqrt{-f_{1}^{\prime
\prime }(z_{\ast })}$, $\eta _{1}=\eta -\sigma /\eta$, and $z_{\ast }$ is a
saddle-point of the function
\begin{equation}
f(z)=f_{1}(z)+(\sigma -1)\beta \varepsilon _{0}+\sigma \ln (1-z\exp (-\beta
\varepsilon _{0}));
\end{equation}
$D_{-\sigma }(z)$ is a parabolic cylinder function, which can be expressed in
terms of hypergeometric functions as
\begin{equation}
D_{s}(z)=2^{s/2}e^{-z^{2}/4}\sqrt{\pi }\left[ \frac{%
_{1}F_{1}(-s/2,1/2,z^{2}/2)}{\Gamma \lbrack (1-s)/2]}-\frac{\sqrt{2}z\cdot
_{1}F_{1}((1-s)/2,3/2,z^{2}/2)}{\Gamma \lbrack -s/2]}\right] .
\end{equation}
Figs.~\ref{sad1} and \ref{sad2} (solid curves) show $\bar{n}_{0}(T)$ and
$\Delta n_{0}(T)$ obtained by the refined saddle-point method. These results
are in remarkable agreement with the exact dots. Fig.~\ref{sad3} demonstrates
that this refined method also provides good accuracy for the third central
moment of the number-of-condensed-atoms fluctuations.

\begin{thebibliography}{999}
\bibitem{exceptions} The only exception known to the authors is the paper
\cite{HKK} where the third moment of the ground-state occupation for the
ideal gas in a harmonic trap was discussed in the Maxwell's demon ensemble
approximation. Higher moments were also discussed on the basis of the master
equation approach in Refs. \cite{MOS99,KSZZ}.
\end{thebibliography}
\end{document}